\documentclass[a4paper,onecolumn,11pt,unpublished,accepted=2025-12-13]{quantumarticle}
\pdfoutput=1
\usepackage[utf8]{inputenc}
\usepackage[english]{babel}
\usepackage[T1]{fontenc}
\usepackage{amsmath, amsthm, thmtools}
\usepackage{booktabs}
\usepackage[pagebackref]{hyperref}
\usepackage[nameinlink,capitalise]{cleveref}

\newcommand{\sans}[1]{\mathsf{#1}}
\usepackage{cite}
\usepackage[dvipsnames]{xcolor}
\usepackage[pagebackref]{hyperref}
\hypersetup{
    colorlinks=true,      %
    linkcolor=Green,       %
    citecolor=NavyBlue,      %
    filecolor=Plum,    %
    urlcolor=Plum         %
}

\usepackage{mathtools, amsfonts, amssymb, mathrsfs}
\usepackage{enumerate,enumitem}
\usepackage{subcaption}
\usepackage{algorithm, algpseudocode, algorithmicx}
\usepackage{comment}
\usepackage{xspace}
\usepackage{mleftright}
\usepackage{xparse}

\newcommand{\complex}{\mathbb{C}}

\DeclareMathOperator{\orth}{orth}

\newcommand{\norm}[1]{\mleft\| #1 \mright\|}

\newcommand{\QR}{\textsf{QR}\xspace}

\newcommand{\expmat}[1]{\begin{bmatrix} #1 \end{bmatrix}}
\newcommand{\twobytwo}[4]{\expmat{#1 & #2 \\ #3 & #4}}

\newcommand{\onebytwo}[2]{\expmat{#1 & #2}}

\newcommand{\Id}{\mathrm{I}}

\DeclareMathOperator{\expect}{\mathbb{E}}

\newcommand{\order}{\mathcal{O}}

\newcommand{\ENE}[1]{}
\newcommand{\CC}[1]{}

\newcommand{\e}{\mathrm{e}}
\newcommand{\iu}{\mathrm{i}}

\renewcommand{\hat}[1]{\widehat{#1}}

\usepackage{tcolorbox}
\newcommand{\actionbox}[1]{\begin{tcolorbox}[colback=white,colframe=black,width=\columnwidth,boxsep=2pt,arc=4pt]
    #1
\end{tcolorbox}}

\NewDocumentCommand{\simplepicture}{O{0.35} m o}{%
  \centering
  \includegraphics[scale=#1]{#2}
  \IfNoValueTF{#3}{}{\captionof{figure}{#3}}
}

\NewDocumentCommand{\mypicture}{O{0.35} m o}{%
  \IfNoValueTF{#3}{\begin{center}
     \includegraphics[scale=#1]{#2}
   \end{center}}{\begin{figure}[h]
    \centering
    \includegraphics[scale=#1]{#2}
  \end{figure}
  \caption{#3}}
}
\newcommand{\conj}{\dagger}
\newcommand{\QB}{\textsf{QB}\xspace}
\DeclareMathOperator{\rank}{rank}

\usepackage[framemethod=tikz,xcolor=true]{mdframed}
\newmdenv[%
    leftmargin=0.5cm,
    roundcorner=5pt,%
    tikzsetting={draw=black, line width=2.0pt}%
    ]{specialtext}%

\DeclarePairedDelimiter{\ket}{|}{\rangle}
\DeclarePairedDelimiterX{\braket}[2]{\langle}{\rangle}{#1 \,\delimsize|\, #2}
\DeclarePairedDelimiterX{\braopket}[3]{\langle}{\rangle}{#1 \,\delimsize|\, #2 \,\delimsize|\, #3}

\renewcommand*{\backref}[1]{}
\renewcommand*{\backrefalt}[4]{%
	\ifcase #1 %
	(No citations.)% use this if no citations
	\or
	(Cited on page #2.)% use this if only one citation
	\else
	(Cited on pages #2.)% use this if multiple citations
	\fi
}
\makeatletter
\def\th@plain{%
  \thm@notefont{}% same as heading font
  \itshape % body font
}
\def\th@definition{%
  \thm@notefont{}% same as heading font
  \normalfont % body font
}
\makeatother

\crefname{equation}{}{}
\crefname{section}{section}{sections}
\crefname{appendix}{appendix}{appendices}
\crefname{figure}{Figure}{Figures}

\declaretheorem[name=Theorem]{theorem}

\declaretheorem[name=Proposition,numberlike=theorem]{proposition}

\theoremstyle{remark}

\theoremstyle{definition}

\usepackage{mdframed}
\newmdtheoremenv{question}{Question}

\numberwithin{equation}{section}

\begin{document}

\title{Successive randomized compression: A randomized algorithm for the compressed MPO-MPS product}

\author{Chris Cama\~no}
\affiliation{Department of Computing and Mathematical Sciences, California Institute of Technology, Pasadena, CA, 91125, USA}
\orcid{}
\author{Ethan N.\ Epperly}
\affiliation{Department of Computing and Mathematical Sciences, California Institute of Technology, Pasadena, CA, 91125, USA}
\orcid{0000-0003-0712-8296}
\author{Joel A.\ Tropp}
\affiliation{Department of Computing and Mathematical Sciences, California Institute of Technology, Pasadena, CA, 91125, USA}
\orcid{0000-0003-1024-1791}

\maketitle

\begin{abstract}
    Tensor networks like matrix product states (MPSs) and matrix product operators (MPOs) are powerful tools for representing exponentially large states and operators, with applications in quantum many-body physics, machine learning, numerical analysis, and other areas.
    In these applications, computing a compressed representation of the MPO--MPS product is a fundamental computational primitive.
    For this operation, this paper introduces a new single-pass, randomized algorithm, called successive randomized compression (SRC), that improves on existing approaches in speed or in accuracy.
    The performance of the new algorithm is evaluated on synthetic problems and unitary time evolution problems for quantum spin systems.
\end{abstract}

\section{Introduction}\label{sec:intro}
In the past three decades, tensor network methods \cite{Fannes92,V2004,Shi_2006,Verstraete_2006,Vidal2010,BC17} have established themselves as some of the most effective techniques for classical simulation of strongly interacting quantum many-body systems \cite{Evenbly_2011,Oru14,Oru19}.
Tensor networks have also shown promise for problems in computational mathematics and machine learning, for instance in numerical analysis \cite{oseQuant,Khoromskij2011OdlogNA,LIN24}, generative modeling \cite{Han_2018,HHL+22}, and compression of neural networks \cite{novikov2015tensorizing,garipov2016,yang2017,Memmel2022PositionTN,tomut2024compactifai}.
Given this diverse array of applications, the development of robust, accurate, and scalable tensor network algorithms is of wide interest.

This paper is concerned with matrix product states and matrix product operators, two of the most widely used tensor network formats:
\begin{itemize}
    \item \textbf{Matrix product states (MPSs).} Throughout this article, we will assume basic familiarity with tensor diagram notation; see \cite{Oru14,BC17,BB17} for a refresher if needed.
    Using this notation, an MPS is defined to be a tensor network of the form
    \mypicture{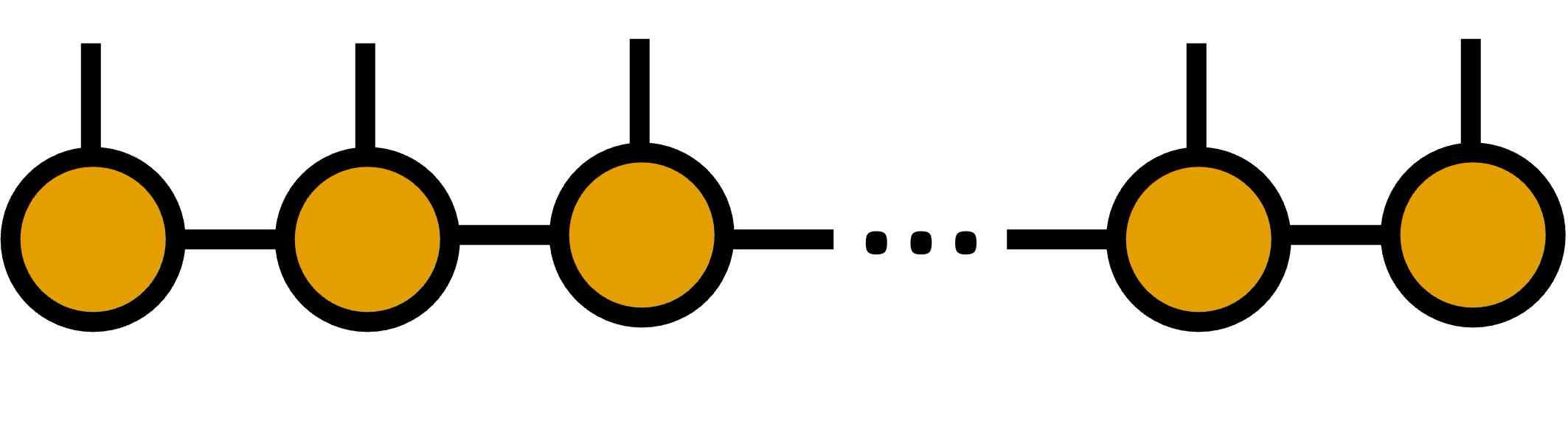}
    MPSs are used to represent \emph{states} of one-dimensional quantum systems such as chains of interacting spin-$S$ particles.
    In other applications, they are used to represent \emph{vectors} or \emph{functions}.
    MPSs are also called \emph{tensor trains} (TTs), particularly in the applied mathematics literature.
    \item \textbf{Matrix product operators (MPOs).}
    An MPO is a tensor network of the form
    \mypicture{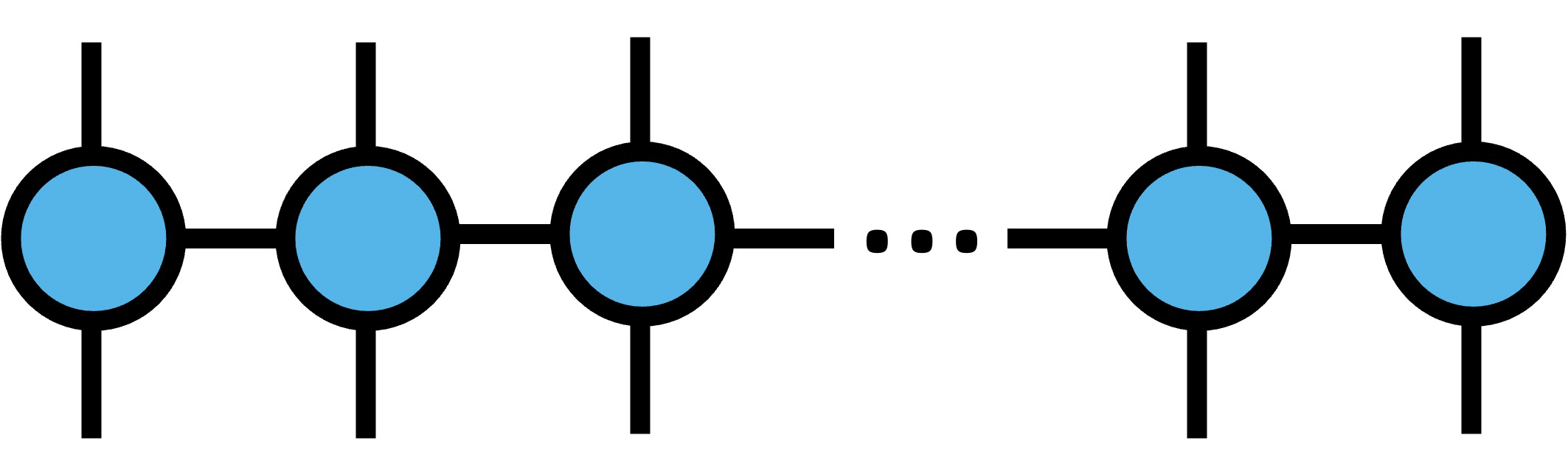}
    MPOs are used to represent \emph{operators} acting on one-dimensional quantum systems, such as local Hamiltonians \cite{V2004,Crosswhite_2008,Hubig_2017}, time evolution \cite{V2004,Paeckel_2019,vandamme2023}, or computational primitives like the quantum Fourier transform \cite{CSW23,chen2024}.
    In other applications, MPOs are used to represent \emph{matrices} or \emph{operators on functions}.
\end{itemize}
In an MPS or MPO, the horizontal and vertical edges are referred to as \emph{bond indices} and \emph{physical indices}, respectively.
We denote the number of sites $n$, the physical dimension $d$, the MPS bond dimension $\chi$, and the MPO bond dimension $D$.

This paper is concerned with the following basic and widely applicable problem:
\begin{specialtext}
\textbf{Compressed MPO--MPS multiplication problem.} Given an MPS $\ket{\psi}$ and an MPO $H$, compute an MPS representation $\ket{\eta} \approx H\ket{\psi}$ of the MPO--MPS product:
\begin{equation} \label{eq:mpo_times_mps}
    \begin{gathered}
    \includegraphics[scale=0.35]{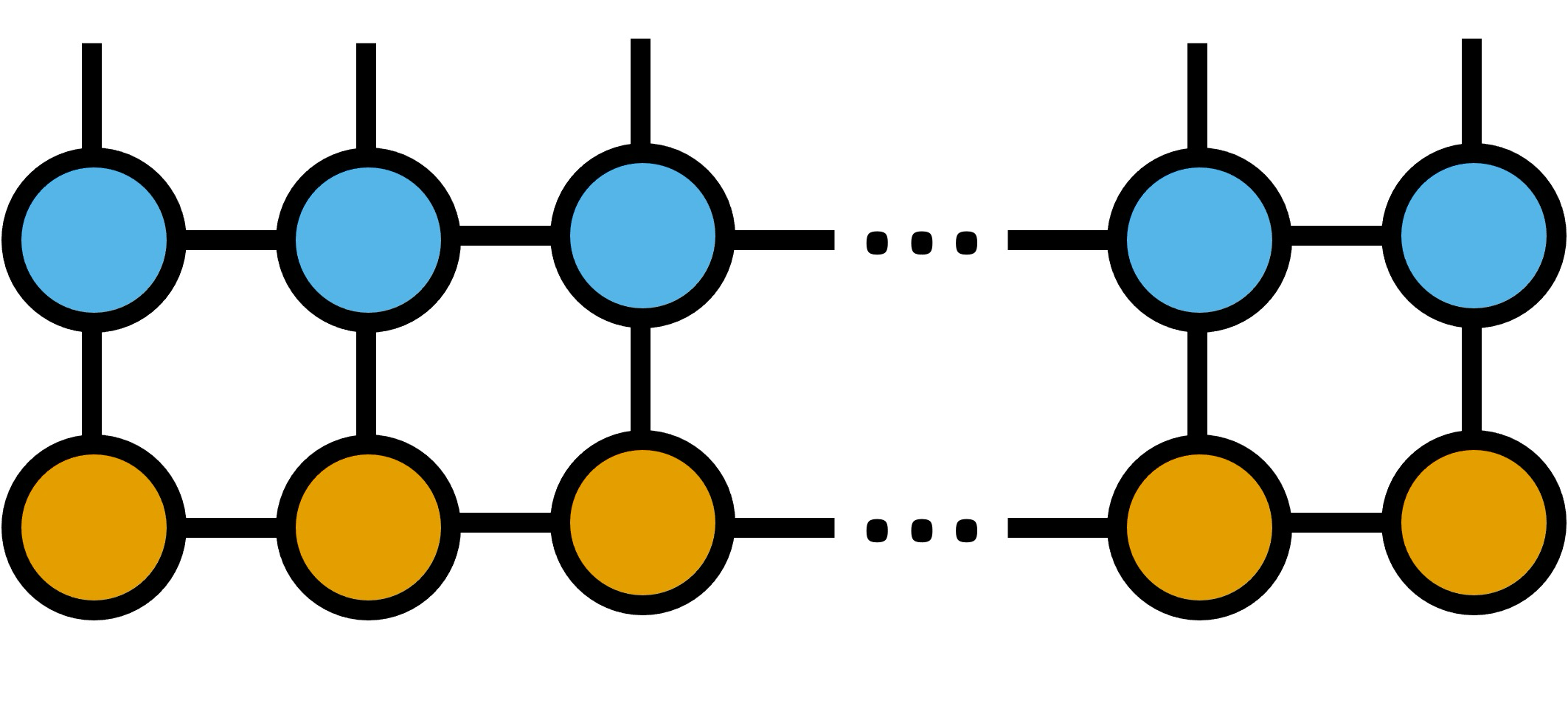}
    \end{gathered}
         \approx 
    \begin{gathered}     
       \includegraphics[scale=0.4]{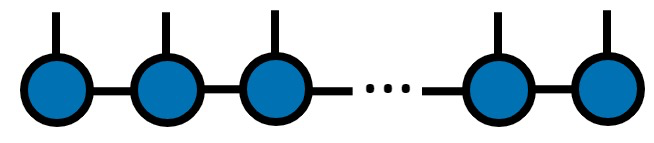}
    \end{gathered}.
\end{equation}
\end{specialtext}
The MPO--MPS product is a basic primitive which comes up in many computational problems.
In quantum physics, MPO--MPS products are used in algorithms for simulating real and imaginary time evolution of 1D quantum states \cite{Paeckel_2019,YA20}, the ``boundary MPS'' method for contracting tensor networks on higher-dimensional PEPS tensor networks \cite{VC04,JOR08}, among other use cases \cite{GTG24}.

\begin{figure}[t]
    \centering
    \includegraphics[width=0.98\textwidth]{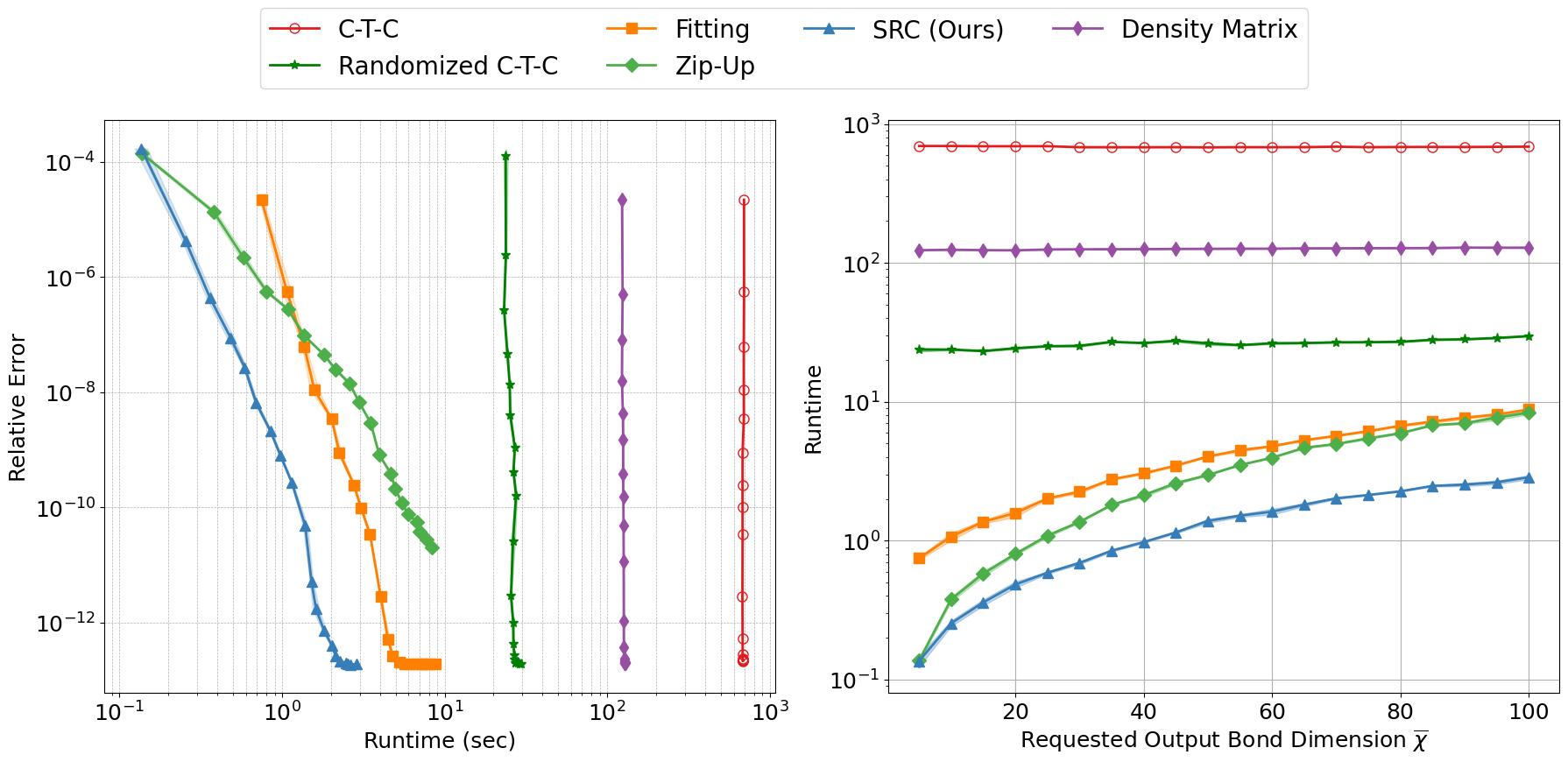}
    \caption{(\textbf{MPO--MPS contraction algorithm comparison}). Comparison of relative error versus computation time of several algorithms for the MPO--MPS product of  $n = 100$  tensor networks with bond dimension  $D = \chi = 50$. The maximum bond dimension  $\overline{\chi}$  was varied between $5$ and $100$, with the compressibility difficulty parameter  $\alpha$  set to $-0.5$.
    The fitting (variational) method was run for a single left-to-right sweep; more sweeps led to higher computational cost at no appreciable gain in accuracy.
    C-T-C is an abbreviation for ``contract-then-compress''.
    Randomized C-T-C refers to the contract-then-compress method with randomized MPS rounding (see \cref{sec:RCTC}).
    Each data point represents the mean of five independent runs.
    } \label{fig:intro}
\end{figure}

Given the importance of the compressed MPO--MPS multiplication problem, many algorithms have been designed for the task; see \cref{sec:previouswork} for a description of existing methods.
A comparison of these methods appears in \cref{fig:intro}, which shows the relative error $\|H\ket{\psi}-\ket{\eta}\|_{\rm F}/\|H\ket{\psi}\|_{\rm F}$ versus the runtime of several methods for computing the product of a random MPO and a random MPS.
The Frobenius norm $\norm{\cdot}_{\rm F}$ is defined as the square root of the sum of squared entries in any tensor. 
We set both bond dimensions to $D=\chi=50$, use $n=100$ sites, and consider a physical dimension $d=2$.
Each tensor is populated with uniformly random entries in the range $[\alpha, 1]$ for $\alpha=-0.5$.
The data for this problem is real, but we store it using complex data types (which is the required data format for most quantum tensor network problems). 
The parameter $\alpha \in [-1,1]$ sets the difficulty of the problem, with higher $\alpha$ making the problem easier \cite{gray2024,JCSH24,chen2025}; we choose $\alpha = -0.5$ for an intermediate level of difficulty.
This is a non-physical example purely intended to demonstrate the performance and accuracy characteristics of the existing methods.

\Cref{fig:intro} demonstrates limitations of the existing MPO--MPS algorithms.
The contract-then-compress (C-T-C) algorithm (also called the ``na\"ive'' algorithm), and the randomized contract-than-compress algorithm, and the density matrix algorithm are all accurate but slow (at least on this problem with $D,\chi \gg 1$).
The zip-up method, by contrast, is fast but less accurate.
On this example, the fitting (i.e., variational) method is accurate but slower than the zip-up method.
As we will see in \cref{sec:fitting}, the fitting method can require many sweeps to converge (making it significantly slower), or it may fail to converge at all.

For faster and more reliable MPO--MPS multiplication, this paper proposes a new \emph{randomized algorithm} for the compressed MPO--MPS product called \textit{successive randomized compression} (SRC).
The SRC method has the following desirable properties:
\begin{itemize}
    \item \textbf{Fast.} In terms of both empirical runtime and asymptotic operation count, our algorithm is competitive with the fastest available algorithms for MPO--MPS multiplication.
    In particular, the SRC method is asymptotically faster than the na{\"i}ve (contract-then-compress) and density matrix methods.
    \item \textbf{Accurate.}
    For a given bond dimension, the output $\ket{\eta}\approx H\ket{\psi}$ of the SRC algorithm achieves accuracy comparable to the contract-then-compress method, which is known to achieve near-optimal truncation error \cite[Cor.~2.4]{Ose11}.
    In particular, the accuracy of SRC is often significantly better than the zip-up algorithm.
    \item \textbf{One-shot.}
    SRC proceeds in a single shot with no need to iterate until convergence.
    This compares favorably to the fitting method and to nonlinear optimization methods, which can require many iterations to achieve convergence or can fail to converge.
\end{itemize}
We believe this suite of benefits make the SRC method a compelling candidate for deployment in applications.
Indeed, in \cref{fig:intro}, SRC achieves higher accuracy in less time than other randomized MPO--MPS methods.

\subsection{Outline for paper}

After introducing background material on randomized matrix approximation and Khatri--Rao products in \cref{sec:background}, we present our algorithm in \cref{sec:alg}.
Description of and comparison with existing MPO--MPS product methods appears in \cref{sec:previouswork}, and an application to unitary time evolution using the TDVP1-GSE algorithm appears in \cref{sec:time-evolution}.

\subsection{Notation}

The Frobenius norm of a matrix or tensor is $\norm{\cdot}_{\rm F}$.
We denote the Hermitian adjoint of a matrix by ${}^\dagger$.
The symbol $\otimes$ refers to the Kronecker product of vectors or matrices, while $\odot$ denotes the Khatri--Rao product of matrices (see \cref{sec:krp}).
We express tensors $\ket{\psi} \in \complex^{d\times \cdots \times d}$ using bra--ket notation.
The constituent tensors of an MPS $\ket{\psi}$ are denoted $\psi^{(1)},\ldots,\psi^{(n)}$ (similarly, $H^{(i)}$ for an MPO $H$).

\subsection{Source code}

Code for our algorithm, other MPO--MPS multiplication algorithms, and to reproduce the experiments in this paper can be found at the following URL: 
\actionbox{\centering \url{https://github.com/chriscamano/RandomMPOMPS}}

\section{Background} \label{sec:background}
This section reviews background material for our randomized method for the compressed MPO--MPS multiplication algorithm, including randomized low-rank approximation (\cref{sec:randomized-qb}) and the Khatri--Rao product (\cref{sec:krp}).

\subsection{Randomized \QB approximation} \label{sec:randomized-qb}

\emph{Randomized low-rank approximation algorithms} \cite{HMT11,Woo14,TW23a,MDM+23} are a class of fast, robust methods for approximating a large matrix $A\in \complex^{M\times N}$ as a product $A\approx QB$ of thin matrix $Q \in \complex^{M\times p}$ times a wide matrix $B \in \complex^{p\times N}$.
In this section, we review the simplest of such methods, randomized \QB approximation.
In the next section (\cref{sec:alg}), we will show how the randomized \QB approximation can be used to give a fast algorithm for the MPO--MPS product.

In its most basic form, the randomized \QB approximation consists of four steps:
\actionbox{\textbf{Randomized \QB approximation.} On input $A \in \complex^{M\times N}$, 
\begin{enumerate}[label=(\roman*)]
    \item \textbf{Generate randomness.} Generate a random matrix $\Omega \in \complex^{N\times p}$. \label{step:qb_random}
    \item \textbf{Collect information.} Compute the matrix product $Y \coloneqq A \Omega$. \label{step:qb_collect}
    \item \textbf{Orthonormalize.} Orthonormalize the columns of $Y$, yielding $Q \coloneqq \orth(Y)$. \label{step:qb_orth}
    \item \textbf{Project.} Set $B \coloneqq Q^\conj A$, defining an approximation $\hat{A} = QB$. \label{step:qb_proj}
\end{enumerate}
}
\noindent The randomized \QB approximation $\hat{A} = QB$ is the same as the approximation produced by the randomized SVD algorithm \cite{HMT11}, but presented in the form $\hat{A} = QB$ rather than as an SVD.
The matrix $Q$ has orthonormal columns, and the matrix $B$ has no special structure.
When the randomized \QB approximation is exact ($A=\hat{A}$), we call the expression $A = QB$ a (randomized) \QB \emph{decomposition}.

The accuracy of randomized \QB approximation can be understood using the following result \cite[Thm~10.5]{HMT11} (see also \cite[Fact~A.2]{TYUC17a}):

\begin{theorem}[Randomized \QB approximation: Gaussian] \label{thm:gaussian}
    Let $\hat{A}$ be the rank-$p$ randomized \QB approximation for $A$, formed using a random matrix $\Omega \in \mathbb{C}^{N \times p}$ whose entries are independent draws from either the real or complex standard normal distribution.
    Then
    \begin{itemize}
        \item \textbf{Exactly low-rank, exact recovery.} Suppose that $\rank A \le p$.
        With probability one, randomized \QB approximation recovers $A$ exactly (that is, $\hat{A} = A$).
        \item \textbf{Approximately low-rank, near-optimal approximation.} Fix a parameter $r \le p-1-\alpha$ and let $A_r$ denote the best rank-$r$ approximation to $A$ with respect to the Frobenius norm.
        Then 
        \begin{equation} \label{eq:randomized_qb_error}
            \expect \left[\norm{\smash{A - \hat{A}}}_{\rm F}^2\right] \le \left( 1 + \frac{r}{p-r-\alpha} \right) \norm{A - A_r}_{\rm F}^2.
        \end{equation}
        The number $\alpha = 0,1$ for the complex or real normal distributions, respectively.
    \end{itemize}
\end{theorem}

The conclusion of this result is that randomized \QB approximation exactly recovers genuinely low-rank matrices and, with modest oversampling such as $p=r+5$ or $p=2r$, produces an approximation that is comparable with the best rank-$r$ approximation.
See \cite[\S10]{HMT11}, \cite[\S8]{TW23a}, and the references therein for many more analyses of randomized \QB approximation.
While most of the \emph{analysis} of randomized \QB approximation focuses on the Gaussian case, the method usually works quite well \emph{in practice} for a much richer class of random matrices $\Omega$ (see \cite[\S11.5.2]{MT20a} and \cite[Fig.~1]{CEMT25a}).

The accuracy of randomized \QB approximation can be improved by using a random matrix $\Omega$ that is more aligned with the ``important directions'' (i.e., right singular vectors) of the matrix $A$.
Such a better $\Omega$ can be obtained systematically by applying subspace or Krylov iteration \cite{HMT11,Gu15,MM15,TW23a} or by using prior knowledge about the matrix \cite{BHOT24a}.
In the context of this work, subspace iteration and Krylov iteration are of prohibitively high cost, and we assume no prior information is available.
Therefore, a random $\Omega$, generated independently from the matrix $A$, is the best-available choice.

Randomized \QB approximation or the randomized SVD can be used as a replacement for a truncated full SVD in any algorithm, and there are several papers \cite{TRP15,MIZK18,KTK+18,MO24} that have proposed using the randomized SVD as a fast alternative to full SVDs in tensor network algorithms.
In this paper, we go further by applying the randomized \QB approximation directly to an entire tensor network.

\subsection{The Khatri--Rao product} \label{sec:krp}

In this work, we will apply randomized \QB approximations to \emph{matrix unfoldings of tensor networks}.
For instance, the following MPS 
\mypicture{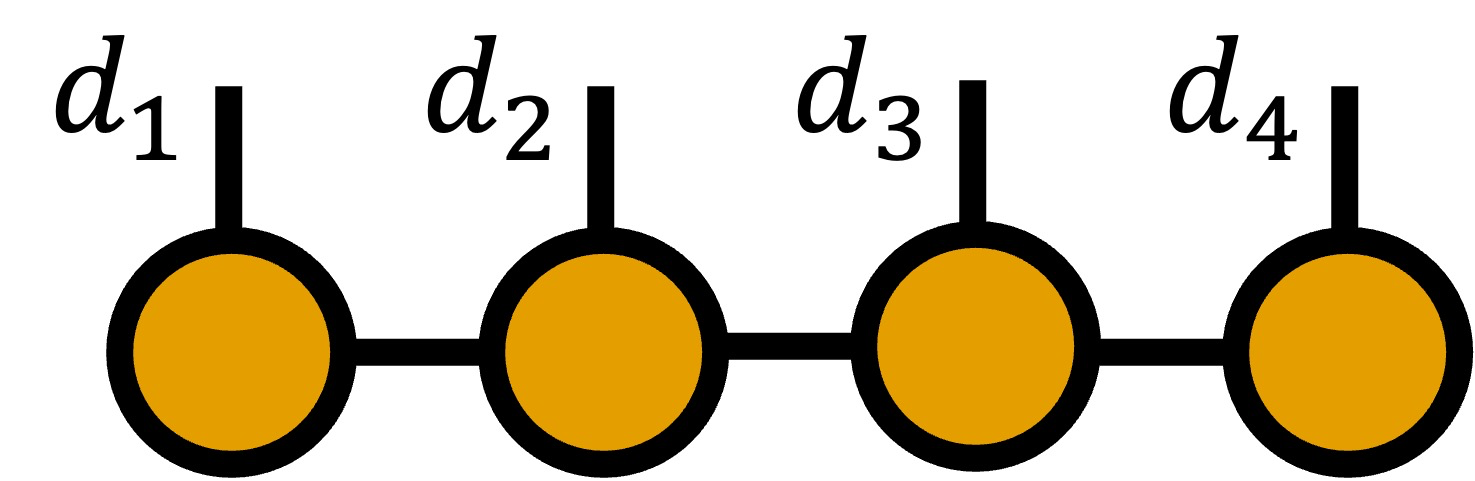}
represents a tensor of size $d_1\times d_2 \times \cdots \times d_4$, i.e., $\ket{\psi} \in \complex^{d_1\times d_2\times d_3\times d_4}$.
The tensor $\ket{\psi}$ can be viewed as a matrix in several different ways.
For instance, if we choose the third and fourth indices to be rows and the first and second to be columns, we can treat $\ket{\psi}$ as a $(d_3d_4)\times (d_1d_2)$ matrix.

To perform \QB approximation efficiently for matrix unfoldings of tensor networks, we can use a \emph{tensor-structured} random matrix $\Omega$.
There are many different types of tensor-structured random matrices that have been proposed for linear algebraic computations; see \cite[Ch.~7]{MDM+23} for a survey.
In this paper, we will use the simplest example, a Khatri--Rao product of random matrices.

The Khatri--Rao product \cite[\S2.6]{KB09} is defined as follows.
Let $\Omega^{(1)},\ldots,\Omega^{(n)} \in \complex^{d\times p}$ be matrices, where each matrix $\Omega^{(j)}$ has columns
\begin{equation*}
    \Omega^{(j)} = \begin{bmatrix} \omega_1^{(j)} & \cdots & \omega_p^{(j)} \end{bmatrix}.
\end{equation*}
The \emph{Khatri--Rao product} of $\Omega^{(1)},\ldots,\Omega^{(n)}$ is defined as the columnwise Kronecker product
\begin{equation*}
    \Omega^{(1)} \odot \cdots \odot \Omega^{(n)} = \begin{bmatrix}
        (\omega_1^{(1)} \otimes \omega_1^{(2)} \otimes \cdots \otimes \omega_1^{(n)}) & \cdots & (\omega_p^{(1)} \otimes \omega_p^{(2)} \otimes \cdots \otimes \omega_p^{(n)})
    \end{bmatrix} \in \complex^{d^n\times p}.
\end{equation*}
To apply randomized \QB approximation to MPO--MPS products in this paper, we will define $\Omega \coloneqq \Omega^{(1)}\odot \cdots \odot \Omega^{(n)}$ to be a Khatri--Rao product of random matrices $\Omega^{(1)},\ldots,\Omega^{(n)}$ with standard (real or complex) normal entries, all mutually independent.
The resulting \emph{Khatri--Rao random matrix} $\Omega$ consists of a collection of random product states $\omega_j^{(1)} \otimes \omega_j^{(2)} \otimes \cdots \otimes \omega_j^{(n)}$, stacked columnwise.
We can compute the action $A\Omega$ of a matrix $A$ on $\Omega$ by applying $A$ to each product state, one at a time.

In tensor diagram notation, the Khatri--Rao product can be denoted as
\mypicture{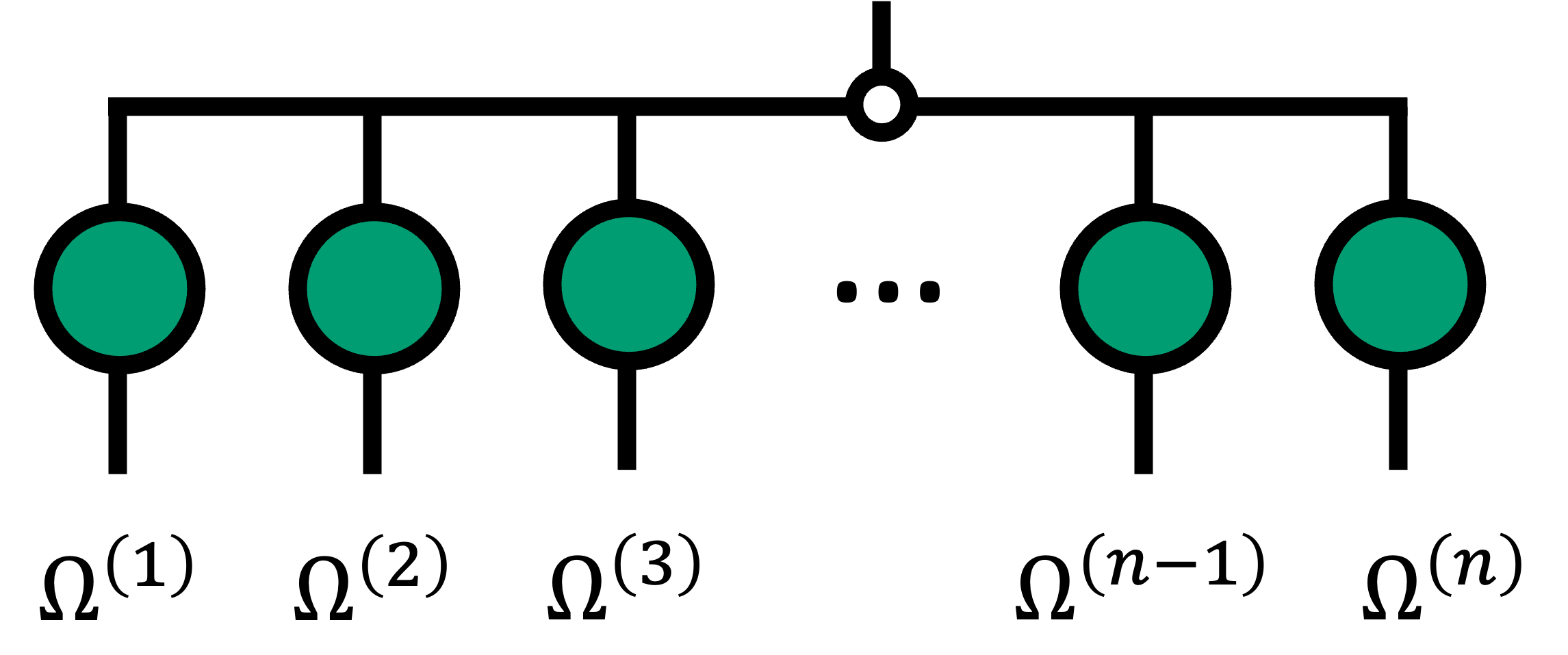}
\noindent We emphasize in this diagram that the white circle at the top of this diagram indicates multiplication of the corresponding indices entrywise.
See \cite[\S1.2 \& \S3.7]{Ahl25} for more details on denoting Khatri--Rao products using tensor diagrams.

At present, our theoretical understanding for randomized \QB approximation with a Khatri--Rao test matrix $\Omega = \Omega^{(1)} \odot \cdots \odot \Omega^{(n)}$ remains limited.  It is known that this method supports exact recovery of a low-rank matrix $A$:

\begin{theorem}[Randomized \QB approximation: Khatri--Rao] \label{thm:khatri-rao}
    Let $\hat{A}$ be the rank-$p$ randomized \QB approximation of $A$ where $\Omega = \Omega^{(1)} \odot \cdots \odot \Omega^{(n)} \in\mathbb{C}^{N\times p}$ formed as a Khatri--Rao product of matrices $\Omega^{(i)}$ with independent standard normal entries.
    Then
    \begin{itemize}
        \item \textbf{Exactly low-rank, exact recovery.} Suppose that $\rank A \le p$.
        With probability one, randomized \QB approximation recovers $A$ exactly (that is, $\hat{A} = A$).
    \end{itemize}
\end{theorem}

\Cref{app:khatri-rao-exact} contains a proof of
\cref{thm:khatri-rao}.
We expect that this result
appears in the literature, but we were unable to locate a citation.
After the initial release of this paper, the authors developed a quantitative error bound for randomized \QB approximation, similar to \cref{eq:randomized_qb_error}, in collaboration with Raphael Meyer \cite[\S5]{CEMT25a}.

The random test matrices with Khatri--Rao product structure used in this work may be contrasted with the framework of ``tensor train sketching'' (e.g., see \cite{ABC+21,HHL+22,KVV23,CK23,YZK25}).
In the latter approach, one chooses $\Omega$ to be a random MPS with a single extra exposed index at the right-most site, which has dimension $p$:
\mypicture{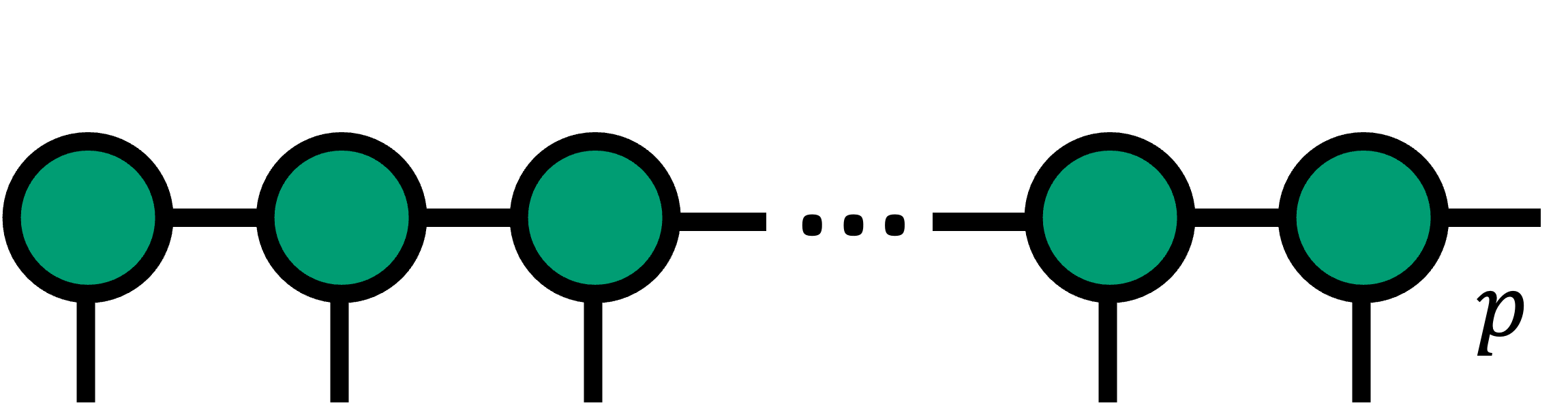}
The tensor train sketching framework has advantages and drawbacks over the random Khatri--Rao product matrices used in this work.
As advantages, the user has more freedom, as they can tune the bond dimensions of this MPS as a way of trading off the accuracy of the algorithm against computational cost.
As disadvantages, the computational effort of tensor train sketching is larger than random Khatri--Rao test matrices, and setting the bond dimensions of the tensor train sketching matrix generally requires either problem-dependent tuning or prior knowledge about the problem.

\section{Successive randomized compression} \label{sec:alg}

In this section, we introduce a new randomized algorithm, \textit{successive randomized compression} (SRC), for the compressed MPO--MPS product.
This algorithm constructs an MPS representation of $\ket{\eta}\approx H\ket{\psi}$ one site at a time in a single pass from right-to-left.
The outcome is an MPS representation of $\ket{\eta}\approx H\ket{\psi}$ in right canonical form.  Here is an illustration for $n=4$ sites:

\mypicture{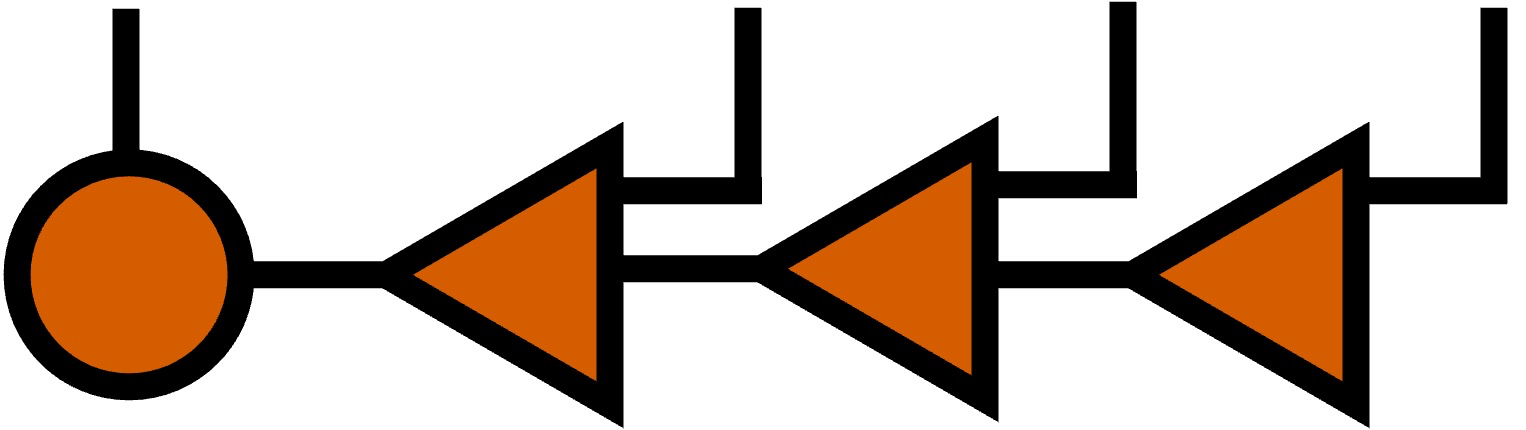}
In this network, triangular tensors denote partial isometries, satisfying
\begin{equation*}
    \begin{gathered}
    \includegraphics[scale=0.35]{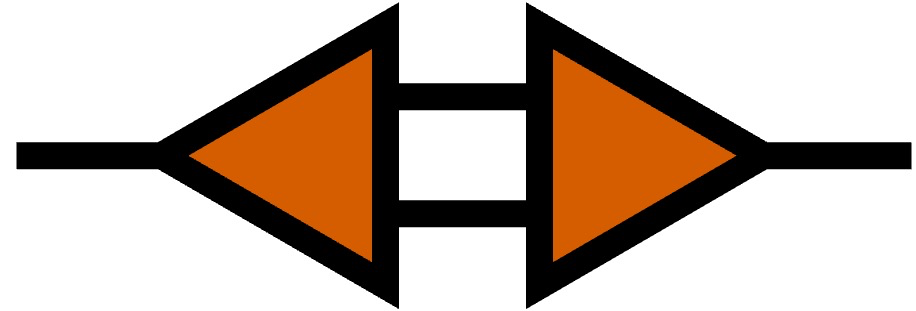}
    \end{gathered} = \begin{gathered}
    \includegraphics[scale=0.35]{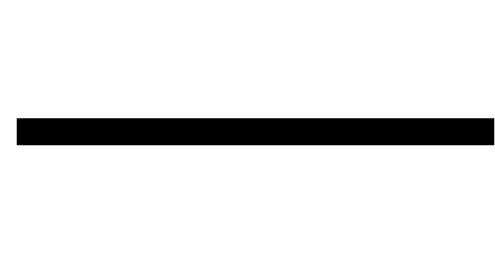}
    \end{gathered} = \mathrm{I} \qquad \text{(identity operator)}.
\end{equation*}

For clarity of exposition, we will describe the SRC algorithm under the following assumptions:
\begin{itemize}
    \item The product $\ket{\eta} = H\ket{\psi}$ is exactly representable as an MPS with bond dimension $\overline{\chi} < D \cdot \chi$.
    \item The output bond dimension $\overline{\chi}$ is provided as an input to the algorithm.
\end{itemize}
In practice, the goal is typically to compute an \emph{approximate} compression $\ket{\eta} \approx H\ket{\psi}$ and, often, to determine $\overline{\chi}$ adaptively to satisfy an error tolerance.
The algorithm works without modification for approximate contraction (though a final rounding step can be helpful, see \cref{sec:final_round}).
We describe the algorithm in detail in \cref{sec:Alg1,sec:Alg2,sec:Alg3}, and we analyze its computational cost in \cref{sec:pseudocode-complexity}.
Adaptive determination of bond dimension is deferred to \cref{sec:approx}.

\subsection{Step 1: The last site}\label{sec:Alg1}

Begin by viewing $H\ket{\psi}$ as a $d\times d^{n-1}$ matrix, with the last visible index as the rows and the first $n-1$ indices as the columns:
\mypicture{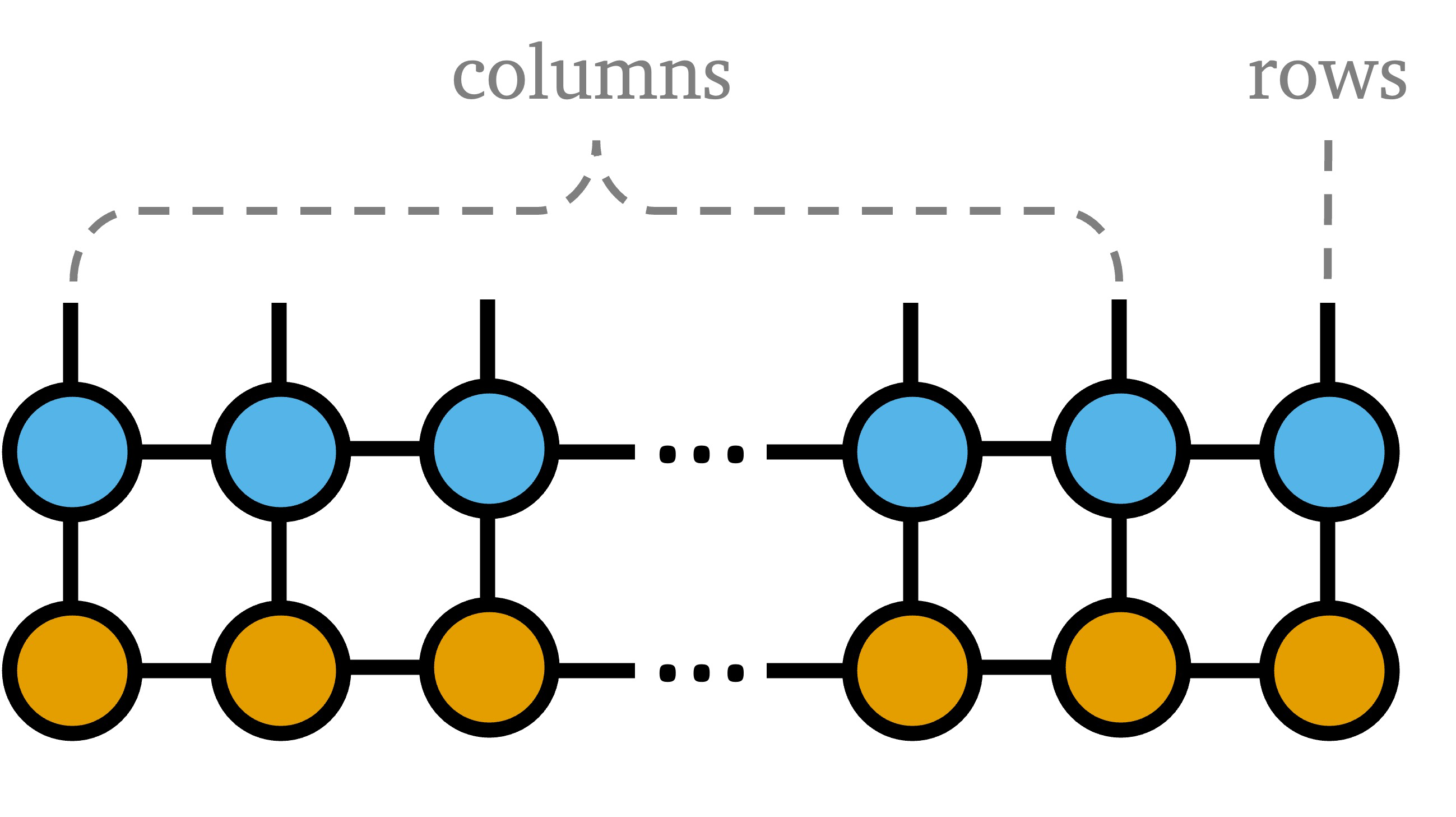}
We will compute a randomized \QB decomposition of this matrix.

Begin by drawing random matrices $\Omega^{(1)},\ldots,\Omega^{(n-1)} \in \complex^{d\times \overline{\chi}}$ with independent real or complex standard normal entries, instantiating the Khatri--Rao product $\Omega^{(1:n-1)} \coloneqq \Omega^{(1)} \odot \cdots \odot \Omega^{(n-1)}$, and applying it to $H\ket{\psi}$.
This process results in the tensor network 
\begin{align*}
    (H\ket{\psi}) \cdot \Omega^{(1:n-1)} &=
    \begin{gathered}
    \includegraphics[scale=0.35]{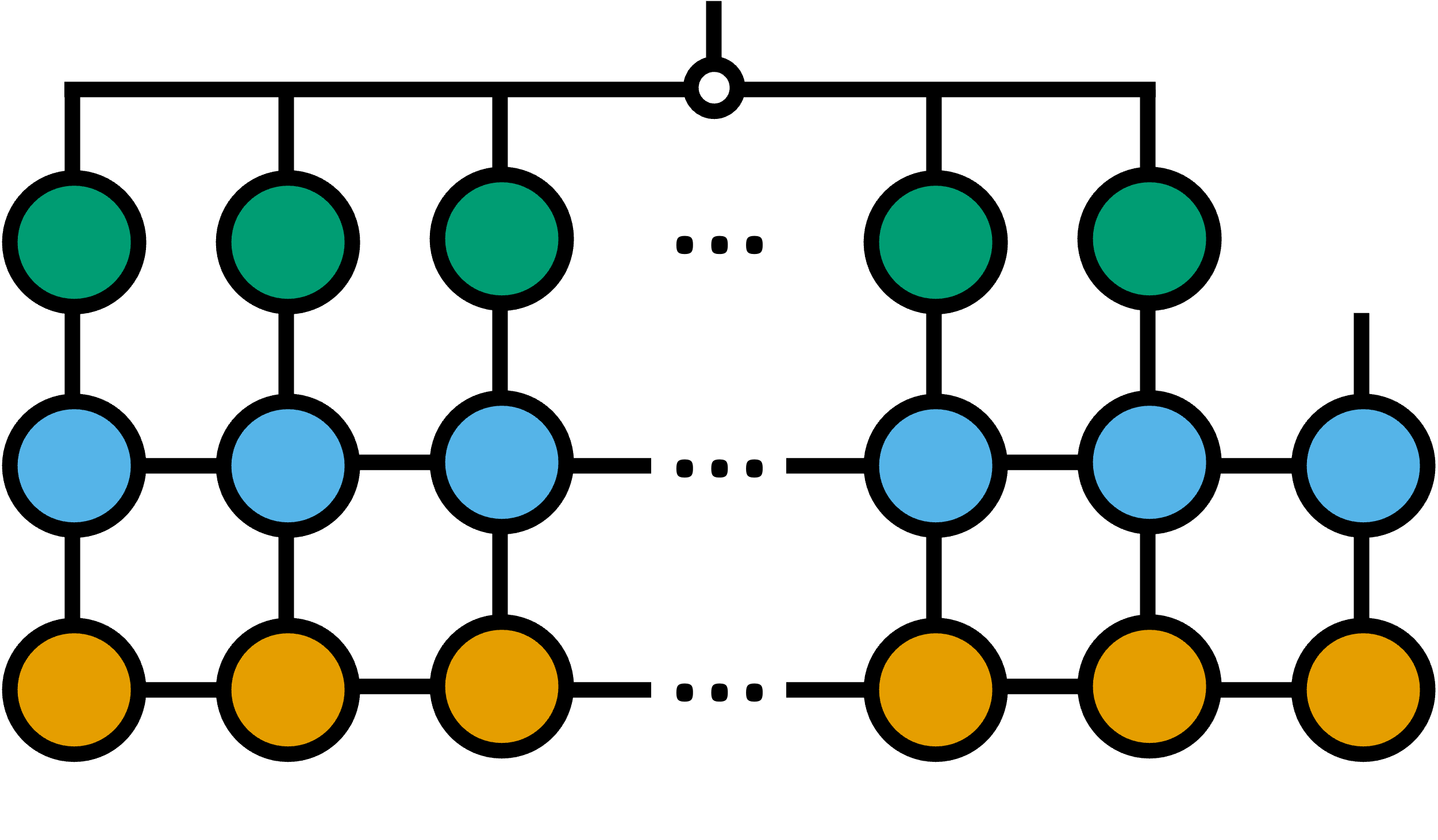}
    \end{gathered}\,.\\\intertext{We now contract this $(H\ket{\psi}) \cdot \Omega$ network.
First, contract the left-most stack of three tensors to obtain a new tensor $C^{(1)}$:}
    (H\ket{\psi}) \cdot \Omega^{(1:n-1)} &=
        \begin{gathered}
        \includegraphics[scale=0.35]{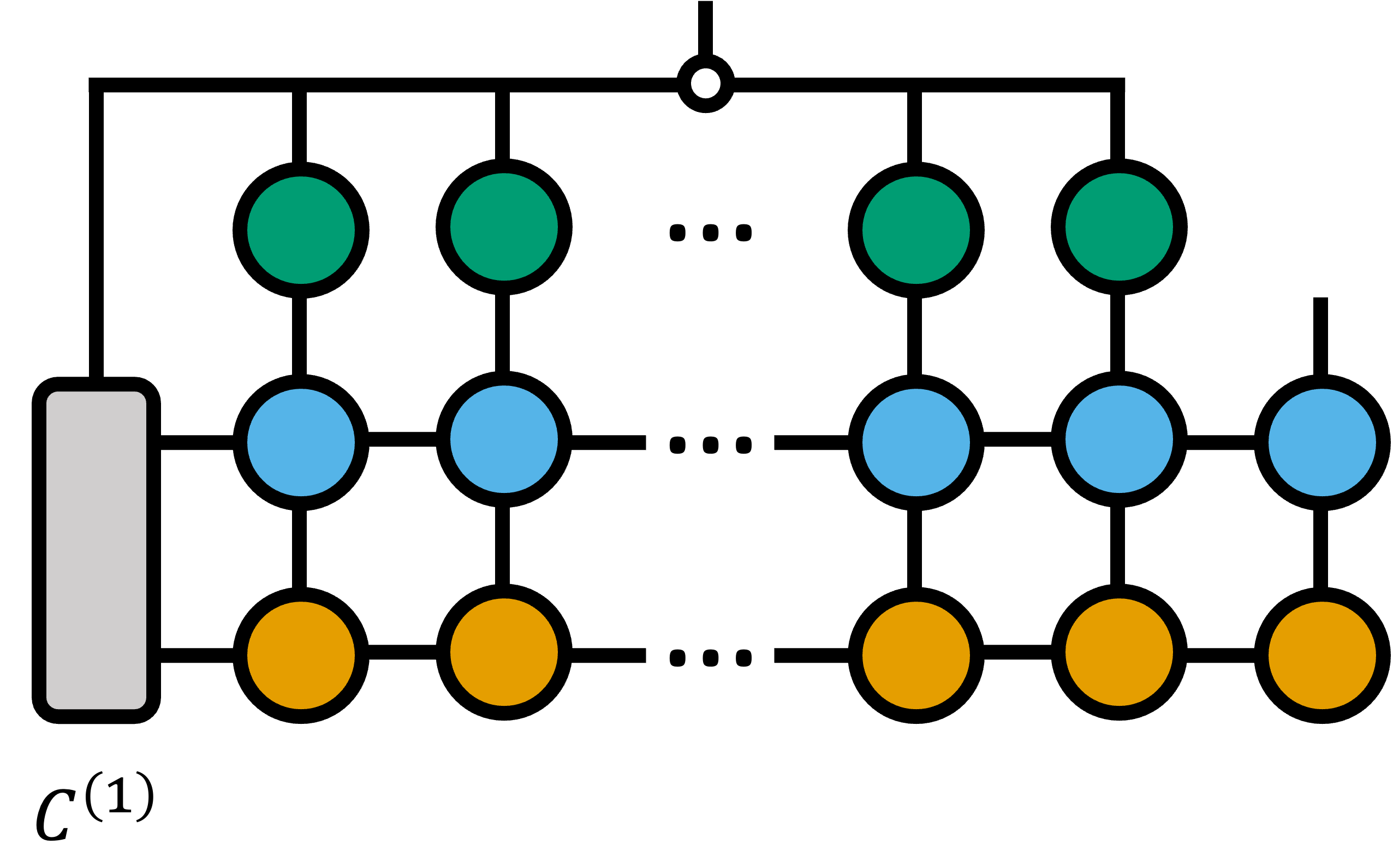}
        \end{gathered}\,. \\
    \intertext{Next, contract one step to the right, resulting in a new tensor $C^{(2)}$:}
    (H\ket{\psi}) \cdot \Omega^{(1:n-1)} &=
    \begin{gathered}
    \includegraphics[scale=0.35]{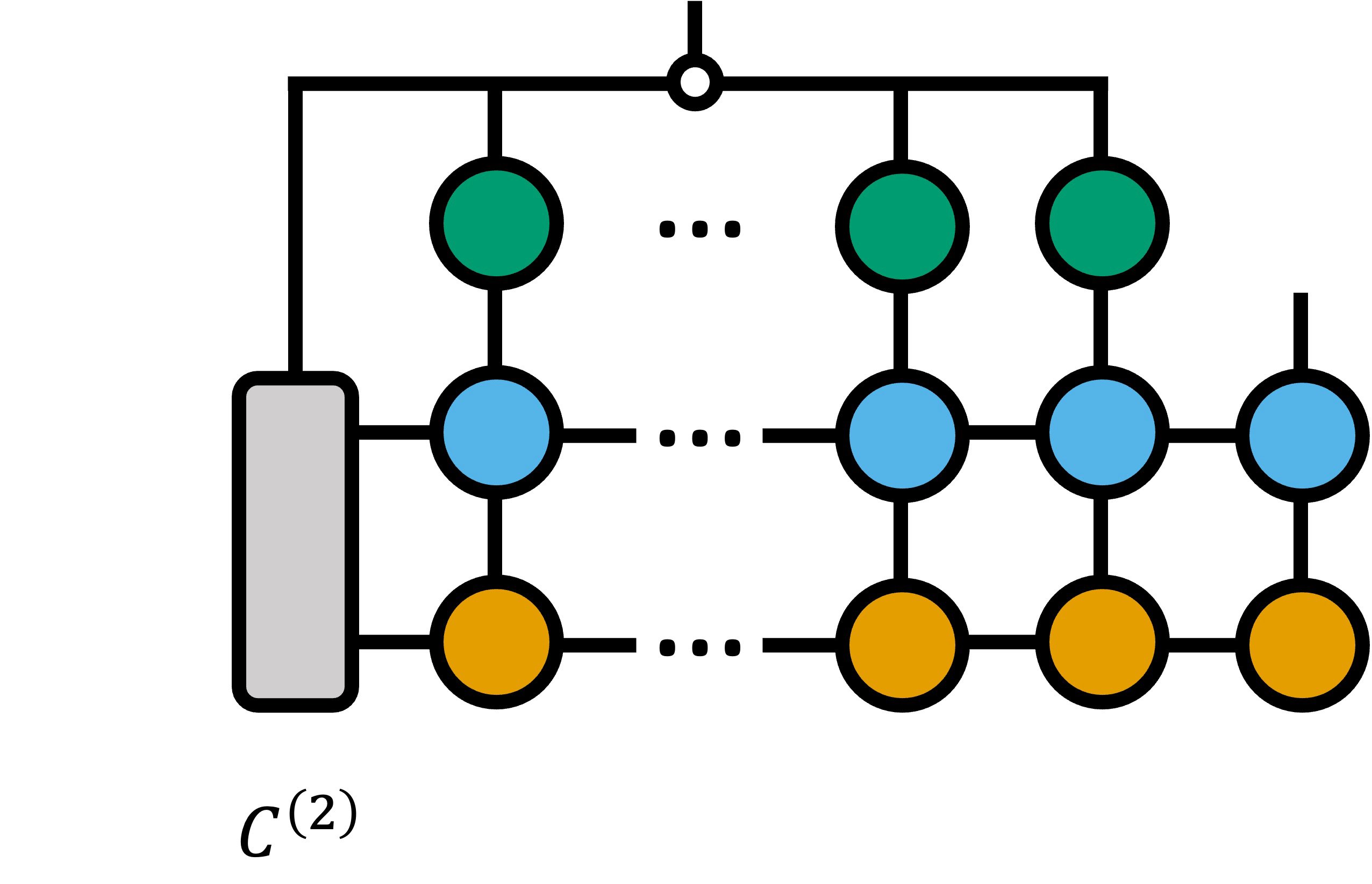}
    \end{gathered}. \\
    \intertext{We continue contracting left-to-right, producing tensors $C^{(3)},C^{(4)},\ldots, C^{(n-1)}$, which we contract down to a matrix:}
    (H\ket{\psi}) \cdot \Omega^{(1:n-1)} &=
    \begin{gathered}
    \includegraphics[scale=0.35]{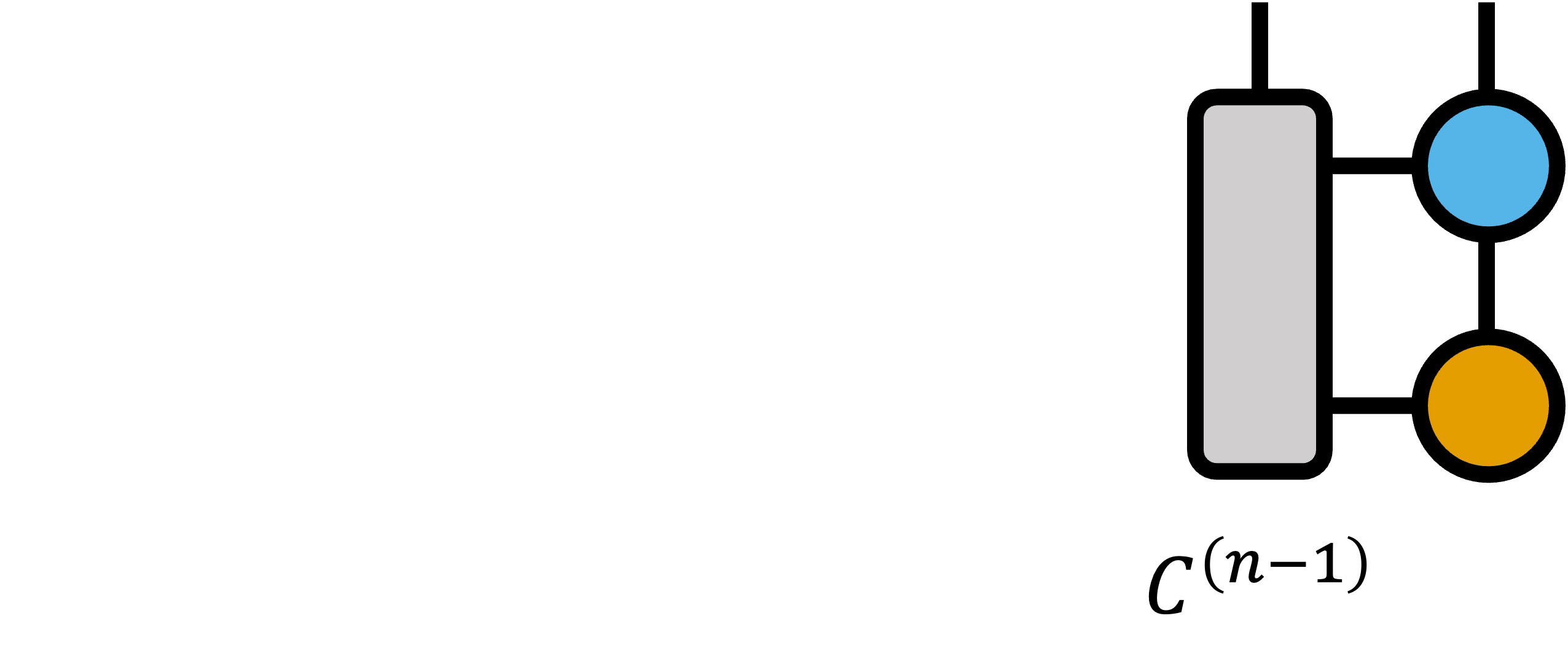}
    \end{gathered} \:\:= \:\:\: \begin{gathered}
    \includegraphics[scale=0.35]{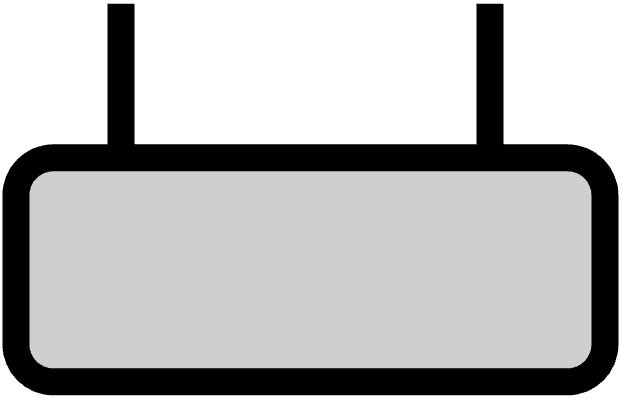}
    \end{gathered}\,.
\end{align*}
We have now collected $Y^{(n)} \coloneqq (H\ket{\psi}) \cdot \Omega^{(1:n-1)}$, which is step~\ref{step:qb_collect} of randomized \QB decomposition.

Next is step~\ref{step:qb_orth}, orthonormalization, which we accomplish by a \QR decomposition:
\begin{equation} \label{eqn:src-orth}
    Y^{(n)} = (H\ket{\psi}) \cdot \Omega^{(1:n-1)} = \begin{gathered}
    \includegraphics[scale=0.35]{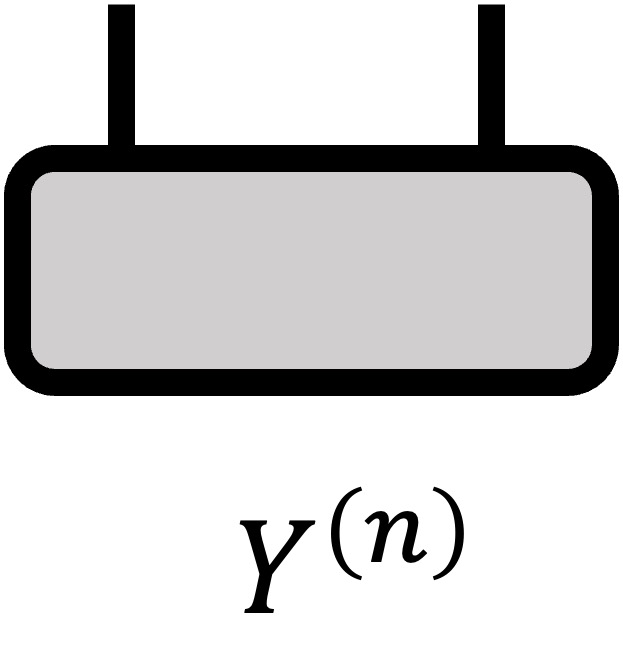}
    \end{gathered} = \begin{gathered}
    \includegraphics[scale=0.35]{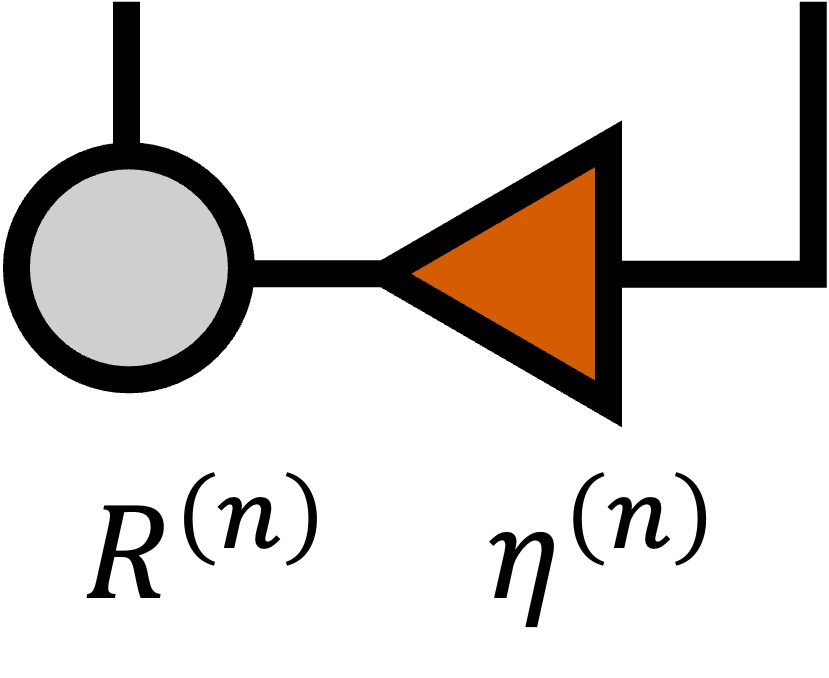}
    \end{gathered} =\eta^{(n)} \cdot R^{(n)}.
\end{equation}
Since the right index of $Y^{(n)}$ corresponds to the rows, the orthogonal ``$\sans{Q}$'' factor $\eta^{(n)}$ occurs to the right of the triangular ``$\sans{R}$'' factor $R^{(n)}$ in the tensor diagram~\cref{eqn:src-orth}, but the factorization is written $Y^{(n)} = \eta^{(n)} \cdot R^{(n)}$ in matrix notation.
The tensor $\eta^{(n)}$ will form the final site of the output $\ket{\eta}= H\ket{\psi}$.

Now, we perform step~\ref{step:qb_proj} of \QB decomposition, projection:
\begin{equation*}
    \begin{split}
    H\ket{\psi} &= \eta^{(n)} (\eta^{(n)})^\conj (H\ket{\psi})= \begin{gathered}
    \includegraphics[scale=0.35]{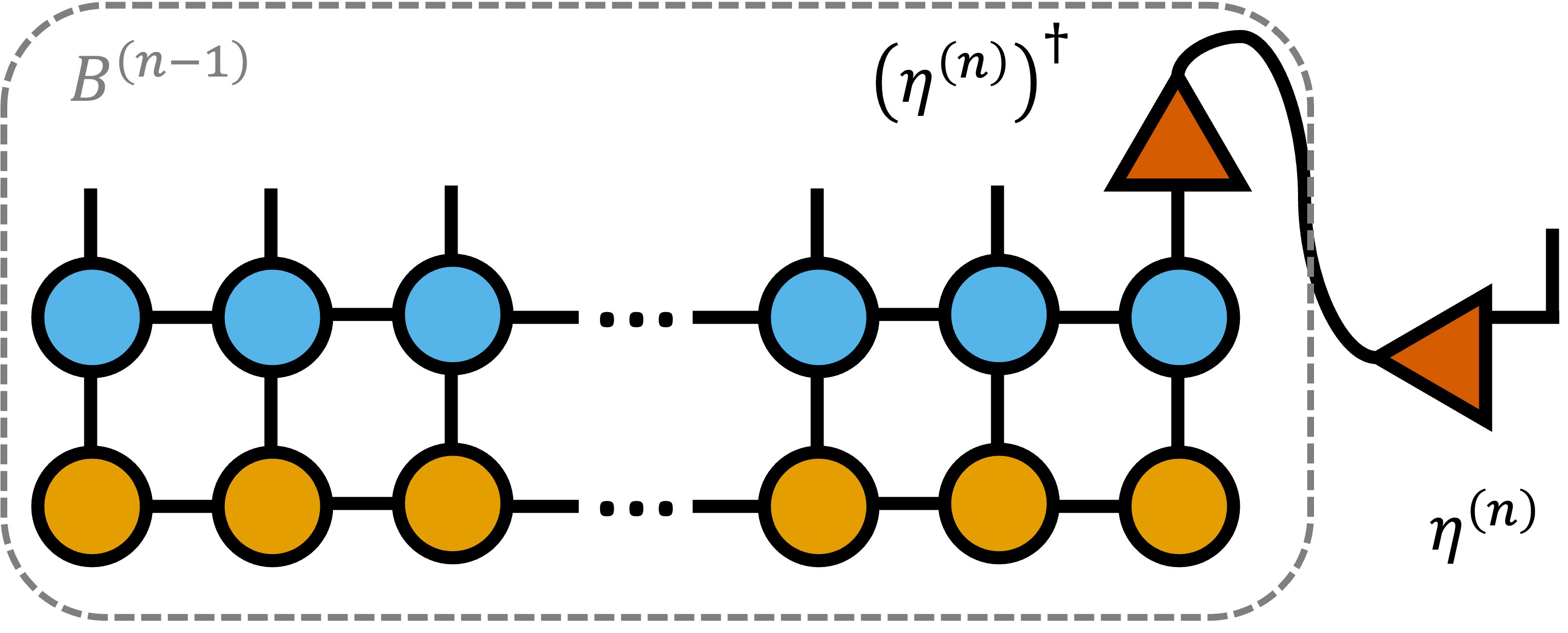}
    \end{gathered}\\
    &\eqqcolon \eta^{(n)}\cdot B^{(n-1)}\,.
    \end{split}
\end{equation*}
Remember that we are assuming, for now, that $H\ket{\psi}$ is exactly compressible to an MPS of bond dimension $\overline{\chi}$.
Thus, by \cref{thm:khatri-rao}, the randomized \QB decomposition is exact with probability one.

In the next step of the algorithm, we will produce site $n-1$ of $H\ket{\psi}$ by applying randomized \QB decomposition to $B^{(n-1)}$.
In preparation for this step, we contract $(\eta^{(n)})^\conj$ into the MPO and MPS sites below, obtaining 
\begin{equation}
    B^{(n-1)} = \begin{gathered}
    \includegraphics[scale=0.35]{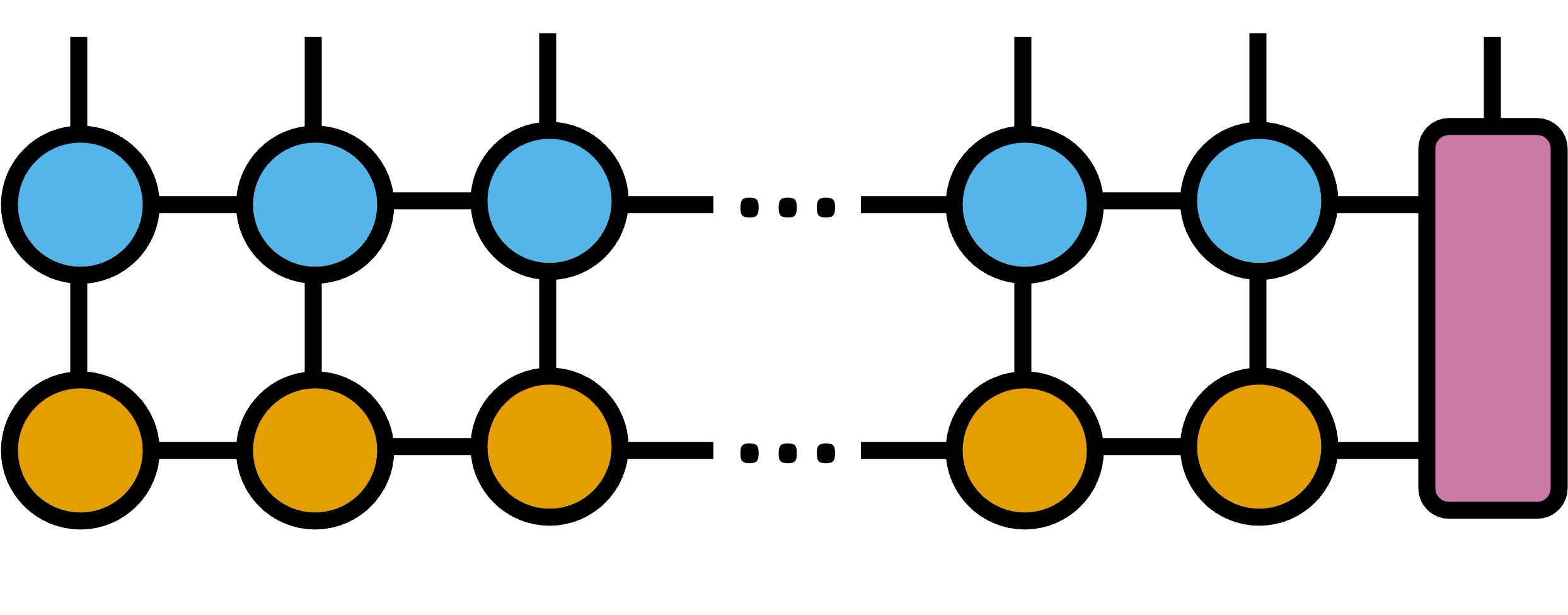}
    \end{gathered}\,.
\end{equation}

\subsection{Step 2: Second-to-last site}\label{sec:Alg2}

We now compute a randomized \QB decomposition of $B^{(n-1)}$.
We treat $B^{(n-1)}$ as a $\overline{\chi}d\times d^{n-2}$ matrix, with the last two exposed indices as rows and the first $n-2$ visible indices as columns: \mypicture{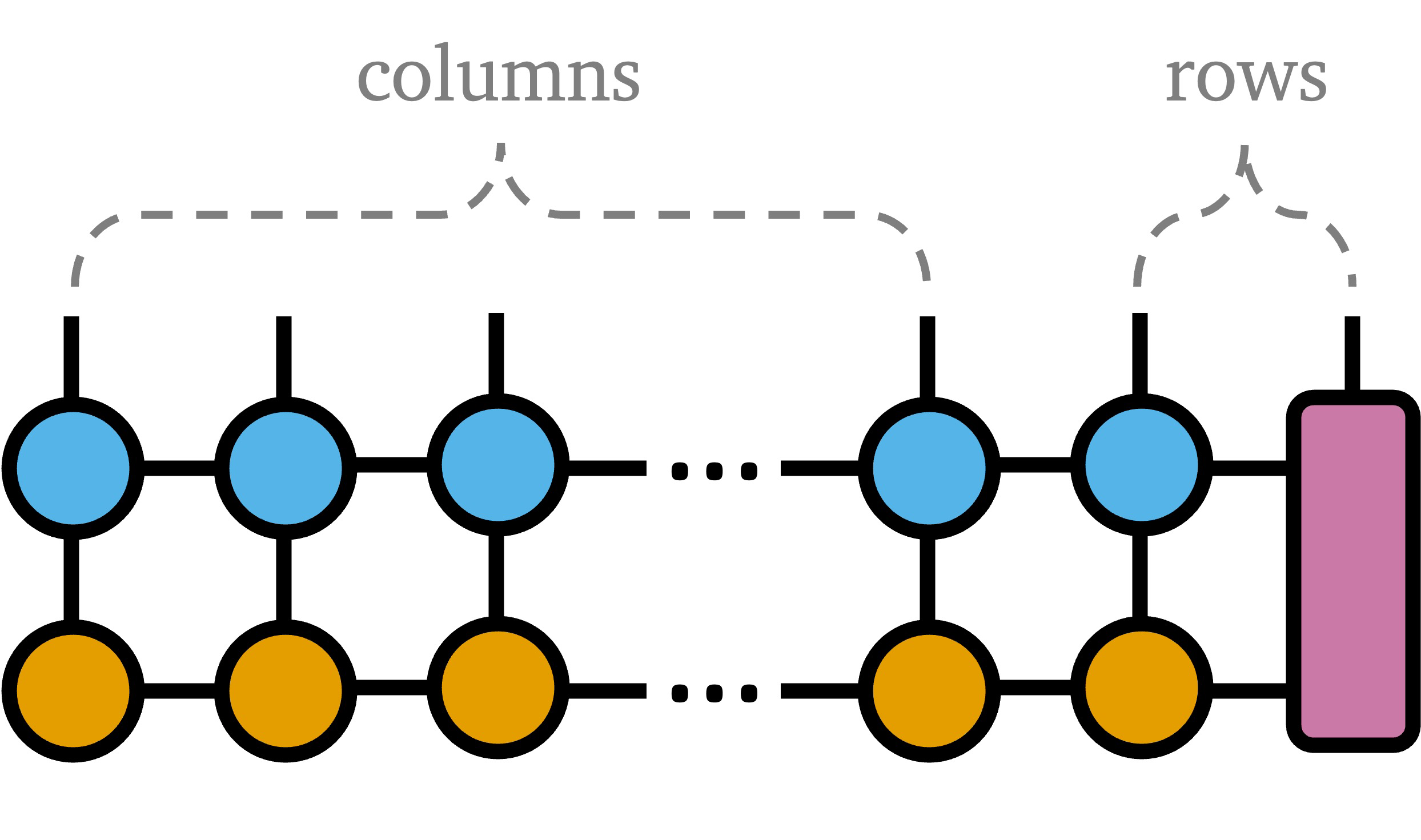}

The key step which allows our algorithm to be computationally efficient is \emph{reuse} of the same random matrices $\Omega^{(1)},\ldots,\Omega^{(n-2)}$ employed in the first step of the algorithm.
Specifically, we form the Khatri--Rao product $\Omega^{(1:n-2)} \coloneqq \Omega^{(1)} \odot \cdots \odot \Omega^{(n-2)}$ and apply it to $B^{(n-1)}$, resulting in the following tensor network:
\begin{equation*}
    B^{(n-1)} \cdot \Omega^{(1:n-2)} =
    \begin{gathered}
    \includegraphics[scale=0.35]{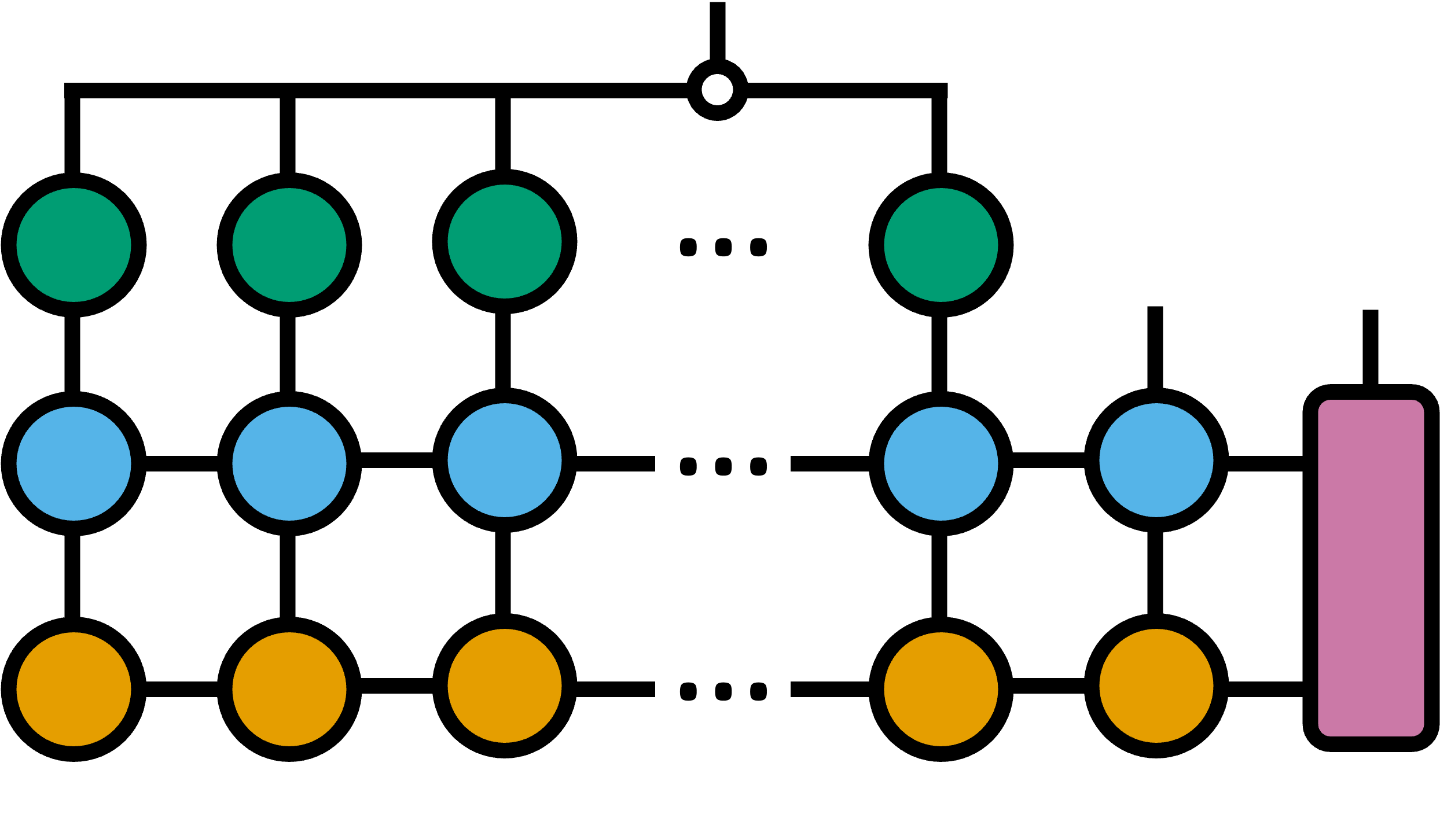}
    \end{gathered}\,.
\end{equation*}
Now, since we use the same random tensors $\Omega^{(1)},\ldots,\Omega^{(n-2)}$, we can reuse the work from step 1 by using the tensor $C^{(n-2)}$ we saved to obtain
\begin{equation*}
    B^{(n-1)} \cdot \Omega^{(1:n-2)} =
    \begin{gathered}
    \includegraphics[scale=0.35]{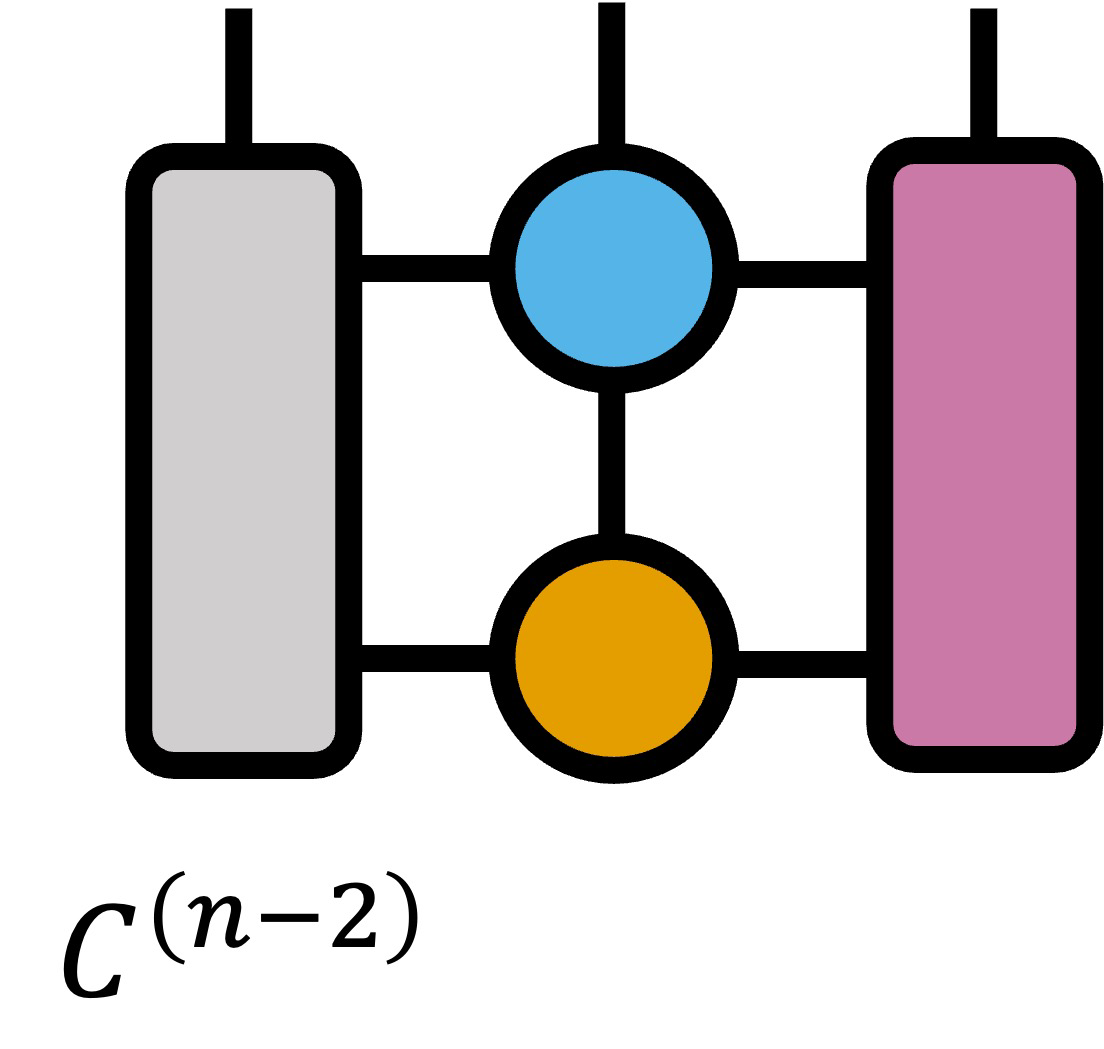}
    \end{gathered}\, \eqqcolon \,\begin{gathered}
    \includegraphics[scale=0.35]{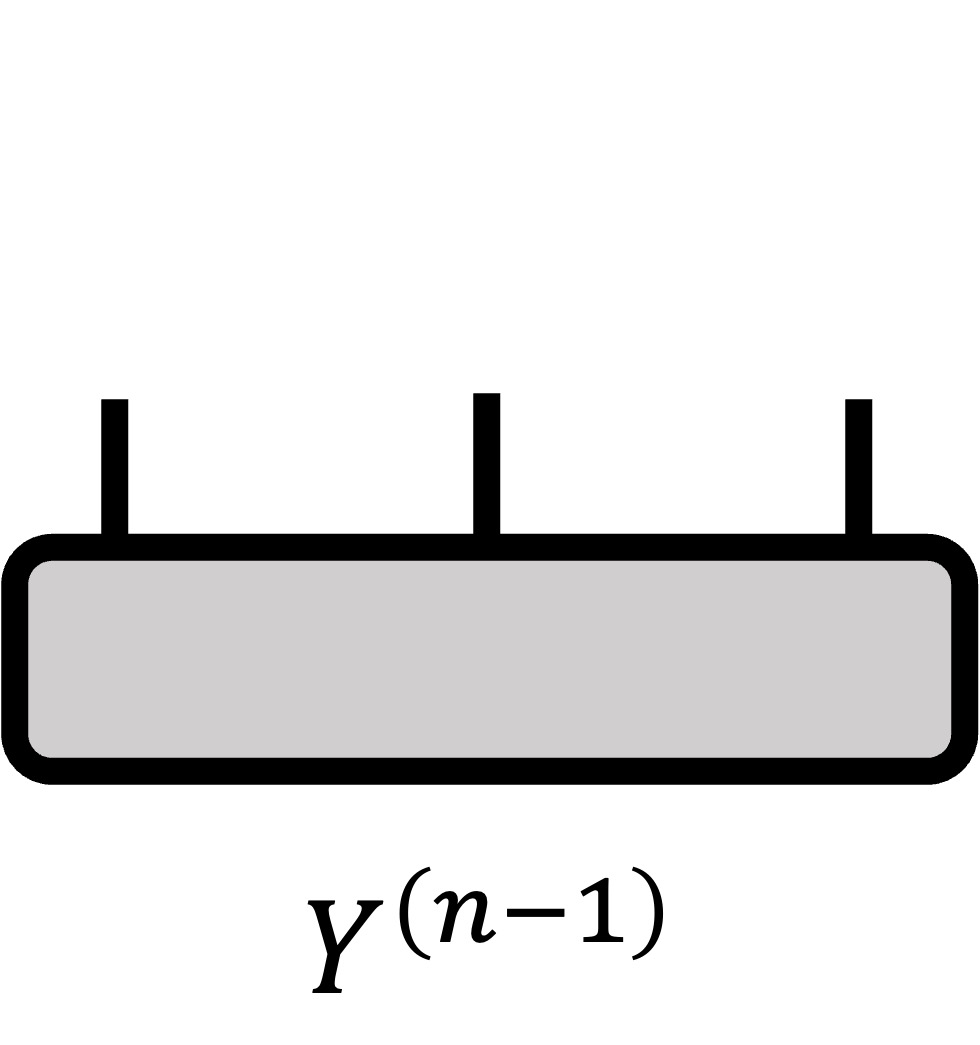}
    \end{gathered}\,.
\end{equation*}
We have further contracted down $B^{(n-1)} \cdot \Omega^{(1:n-2)}$ to a three-mode tensor $Y^{(n-1)}$.
This completes step~\ref{step:qb_collect} of \QB decomposition for $B^{(n-1)}$.

Next, we do step~\ref{step:qb_orth} of \QB decomposition, orthonormalization, by computing a \QR decomposition of $Y^{(n-1)}$:
\begin{equation*}
    Y^{(n-1)} = B^{(n-1)} \cdot \Omega^{(1:n-2)} = \begin{gathered}
    \includegraphics[scale=0.35]{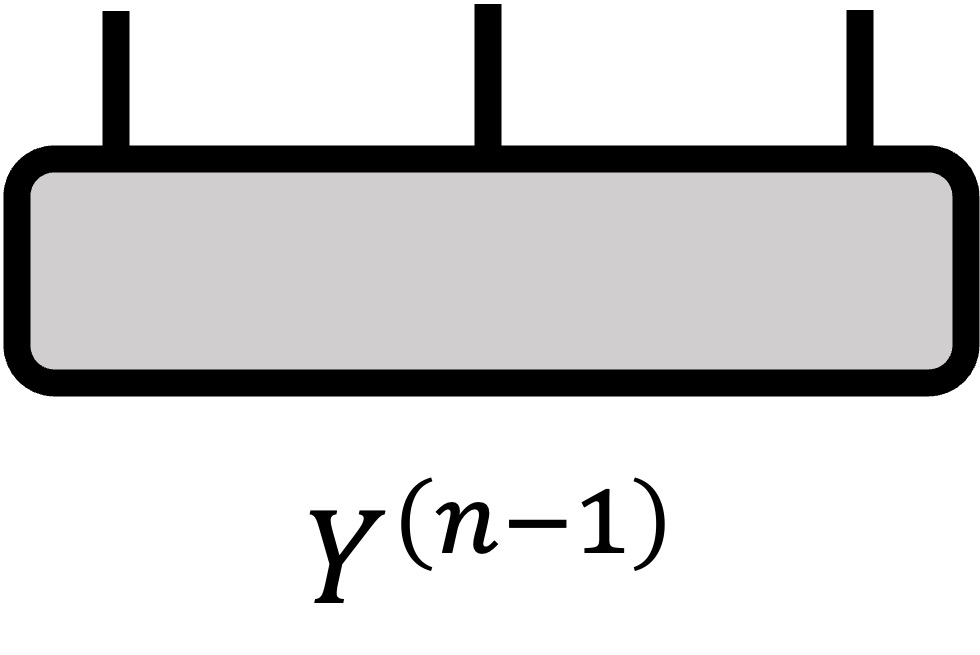}
    \end{gathered}= \begin{gathered}
    \includegraphics[scale=0.35]{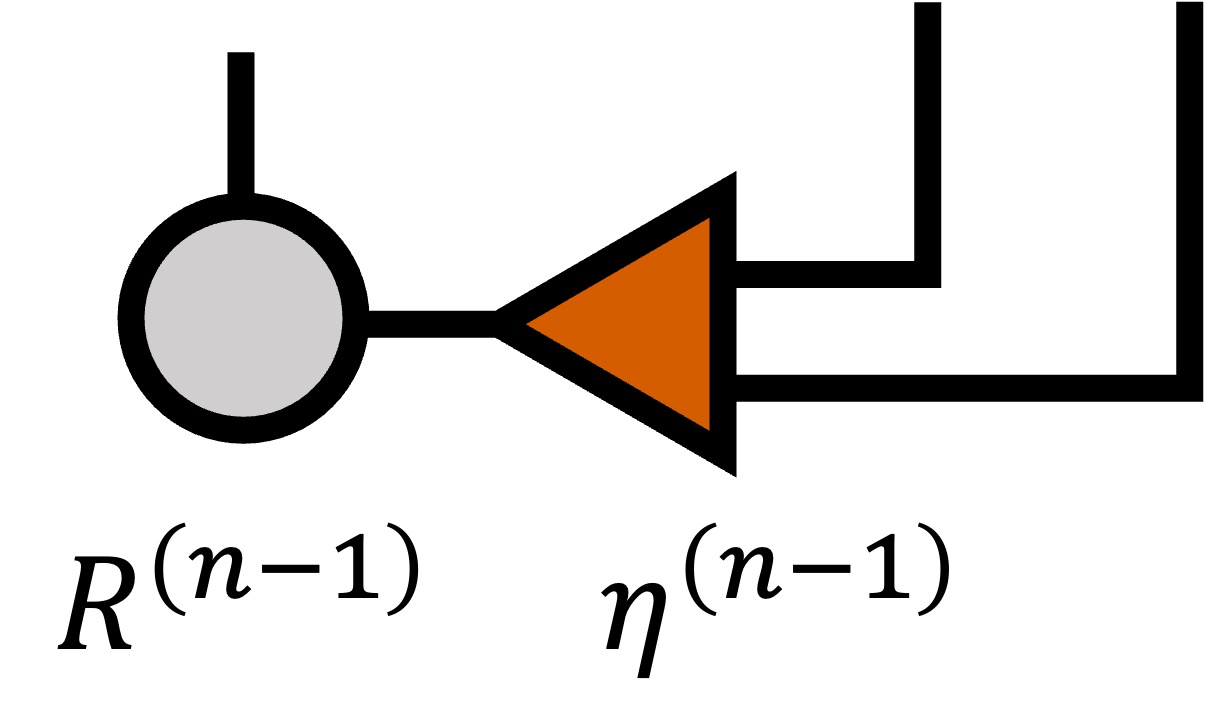}
    \end{gathered}=\eta^{(n-1)}\cdot R^{(n-1)}.
\end{equation*}

Now, we apply the step~\ref{step:qb_proj} of \QB decomposition, projection:
\begin{equation*}
    B^{(n-1)} = \eta^{(n-1)}(\eta^{(n-1)})^\conj B^{(n-1)} = \begin{gathered}
    \includegraphics[scale=0.35]{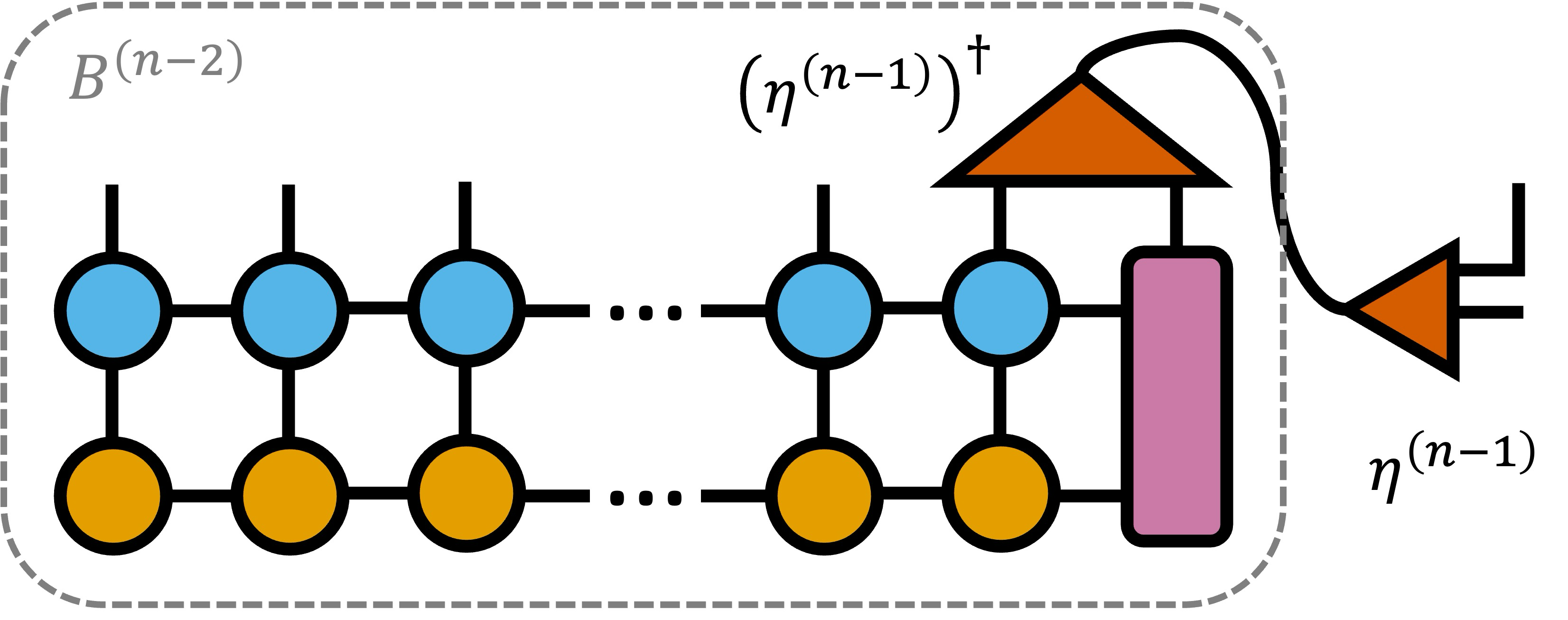}
    \end{gathered}\,.
\end{equation*}
Again, because we have assumed $H\ket{\psi}$ is exactly representable as an MPS with bond dimension $\overline{\chi}$, the \QB decomposition $B^{(n-1)}= \eta^{(n-1)} \cdot B^{(n-2)}$ is exact with probability one.
For the next step of the algorithm, we clean up the $B^{(n-2)}$ by contracting in $(\eta^{(n-1)})^\conj$:
\begin{equation*}
    B^{(n-2)} = \begin{gathered}
    \includegraphics[scale=0.35]{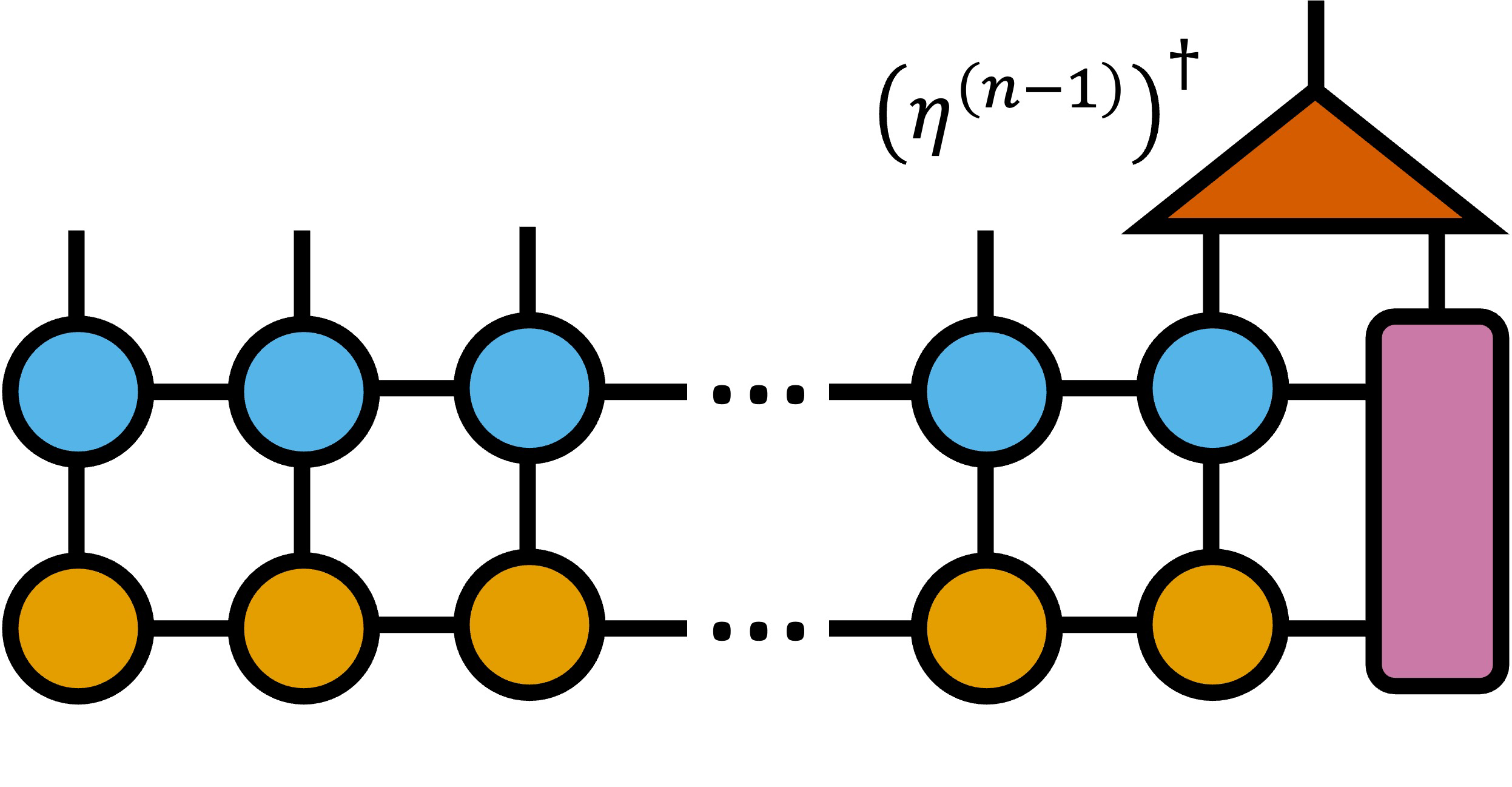}
    \end{gathered}=\begin{gathered}
    \includegraphics[scale=0.35]{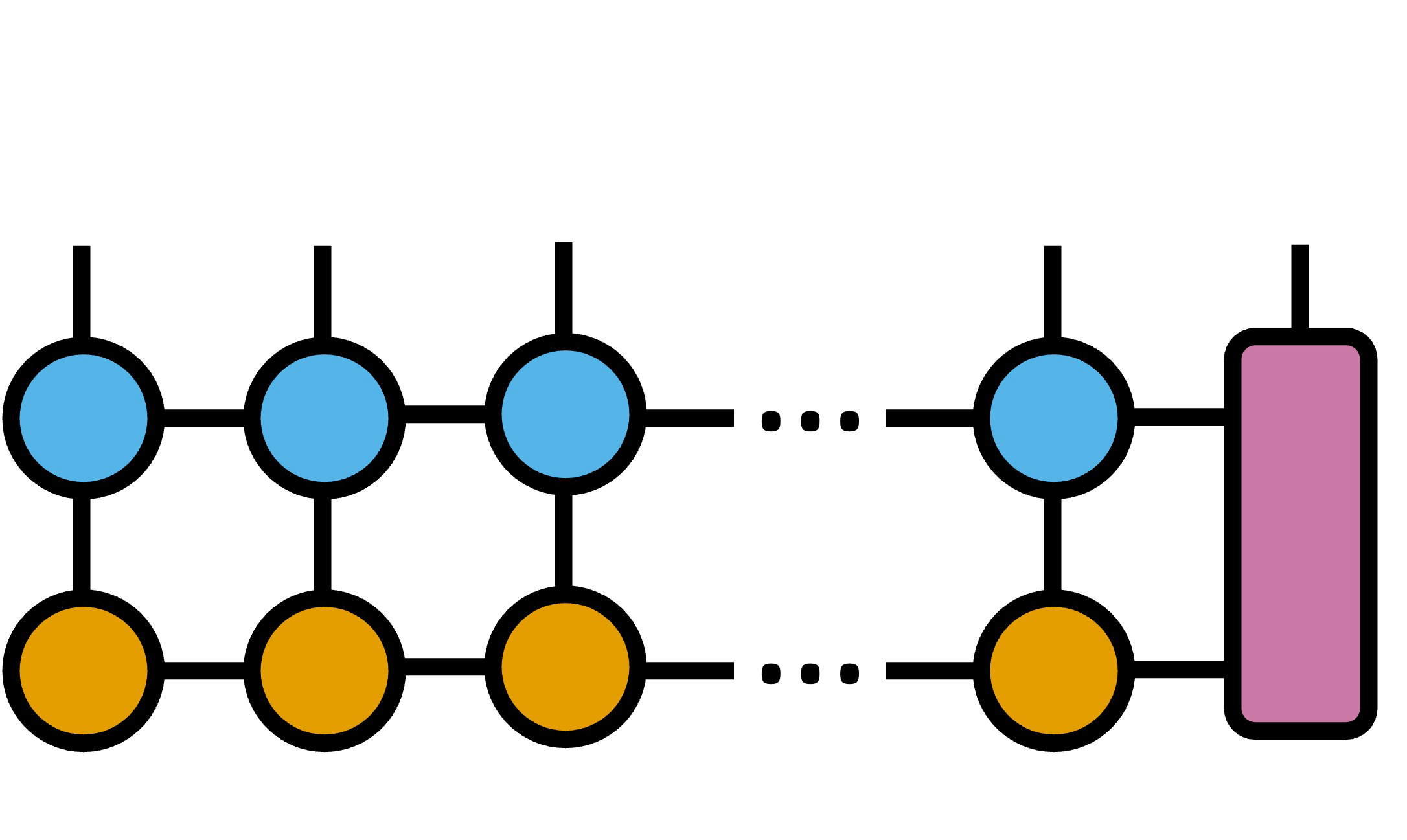}
    \end{gathered}\,.
\end{equation*}
Observe that $B^{(n-2)}$ now has $n-2$ exposed indices of dimension $d$ and $1$ exposed index of dimension $\overline{\chi}$; the total number of exposed indices, $n-1$, has been decreased by one.

To see how the algorithm is making progress in representing $H\ket{\psi}$ as an MPS $\ket{\eta}$, observe that in step one, we obtained a factorization $H\ket{\psi} = \eta^{(n)}\cdot B^{(n-1)}$:
\begin{equation*}
    H\ket{\psi} = \eta^{(n)}\cdot B^{(n-1)}=\begin{gathered}
    \includegraphics[scale=0.35]{figs/contract17.png}
    \end{gathered}.
\end{equation*}
Step two yielded a factorization $B^{(n-2)}= \eta^{(n-1)}\cdot B^{(n-2)}$. 
Thus, 
\begin{equation*}
    H\ket{\psi} = \eta^{(n)}\cdot (\eta^{(n-1)}\cdot B^{(n-2)}) = \begin{gathered}
    \includegraphics[scale=0.35]{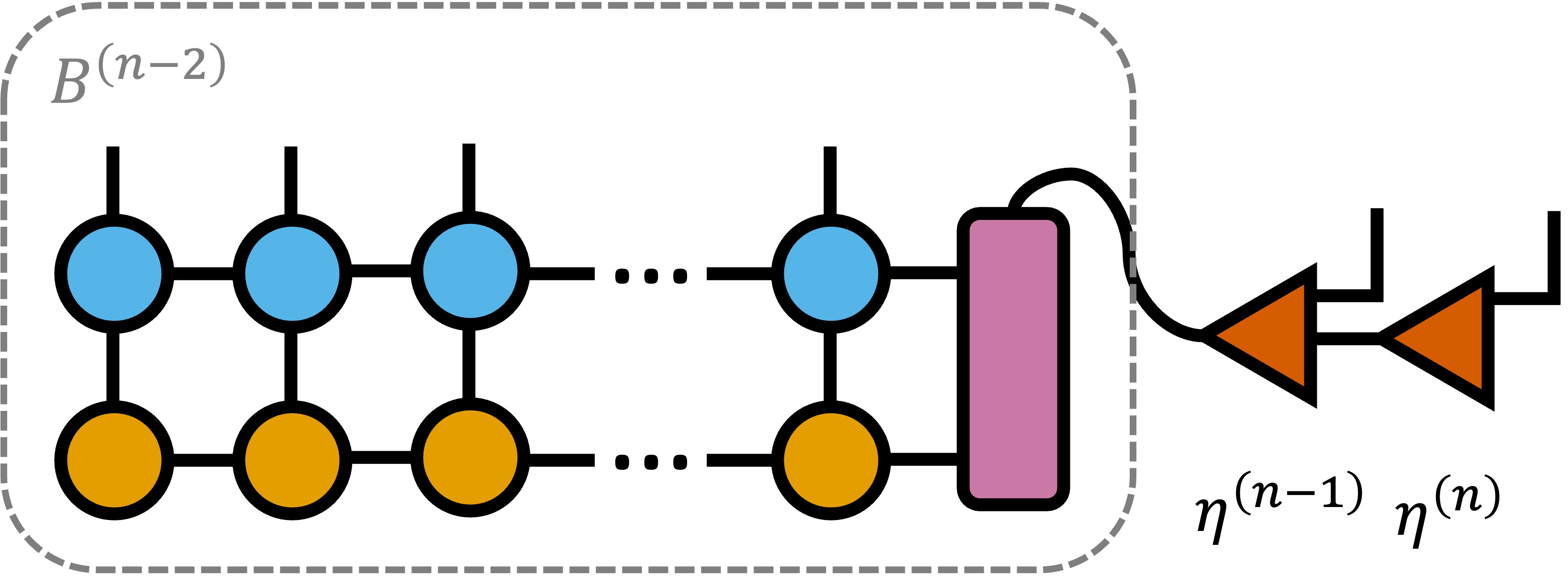}
    \end{gathered}.
\end{equation*}
After two steps of the algorithm, we have produced the right two sites $\eta^{(n-1)}$ and $\eta^{(n)}$ of an MPS $\ket{\eta} = H\ket{\psi}$.
When completed, the MPS $\ket{\psi}$ will be in left canonical form.

\subsection{Finishing up}\label{sec:Alg3}

Proceeding as in step two above, we continue right-to-left producing orthogonal tensors $\eta^{(n)},\eta^{(n-1)},\eta^{(n-2)},\eta^{(n-3)},\ldots$.
By the time we reach site two of $H\ket{\psi}$ and compute $\eta^{(2)}$, we have obtained a tensor network of the form
\begin{equation*}
    \begin{split}
    H\ket{\psi} &= \eta^{(n)}\times\eta^{(n-1)} \times \cdots \times \eta^{(2)}\times B^{(1)} \\ &= \begin{gathered}
    \includegraphics[scale=0.35]{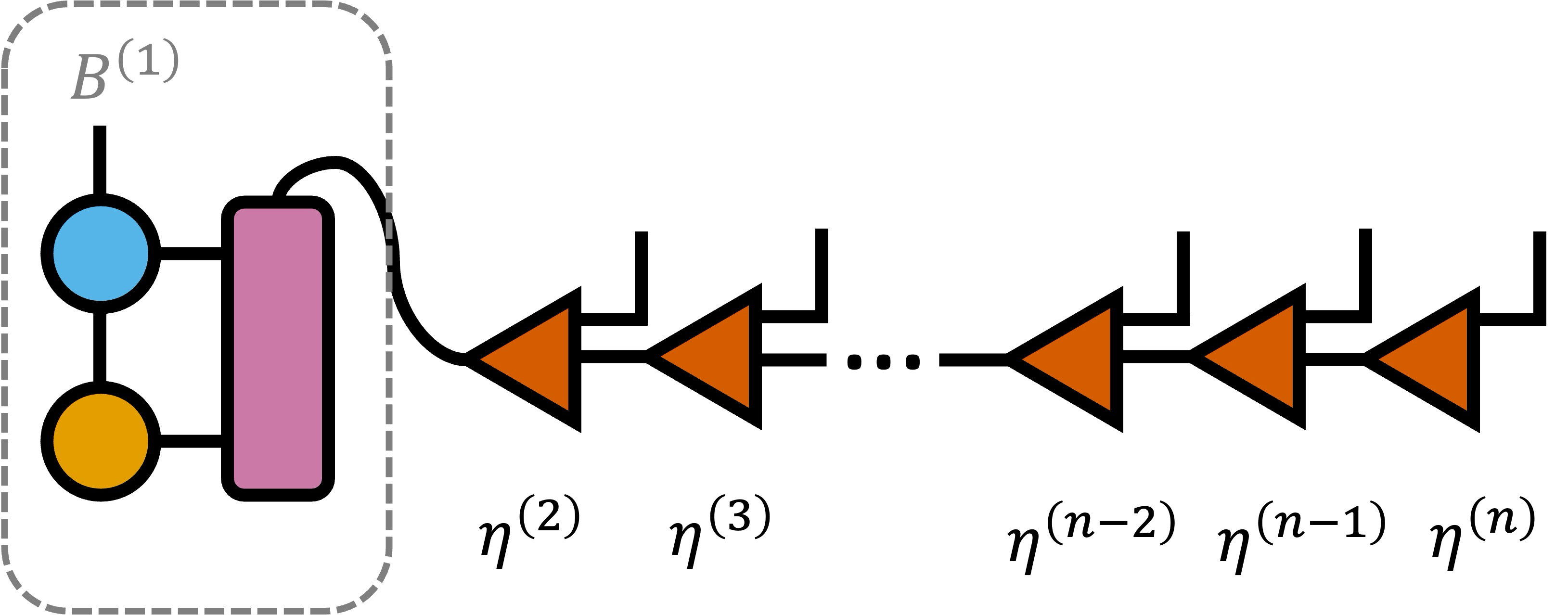}
    \end{gathered}.
    \end{split}
\end{equation*}
To obtain the first site tensor $\eta^{(1)}$, we take the $B^{(1)}$ tensor network and contract it down:
\begin{equation*}
    \eta^{(1)} \coloneqq \begin{gathered}
    \includegraphics[scale=0.35]{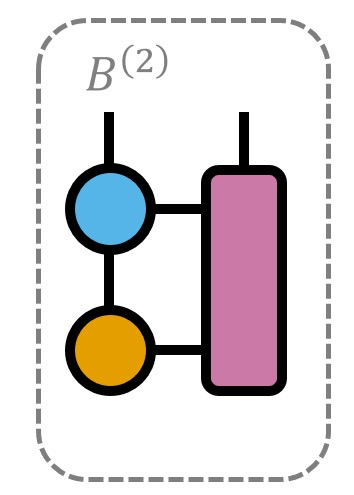}
    \end{gathered} = \begin{gathered}
    \includegraphics[scale=0.35]{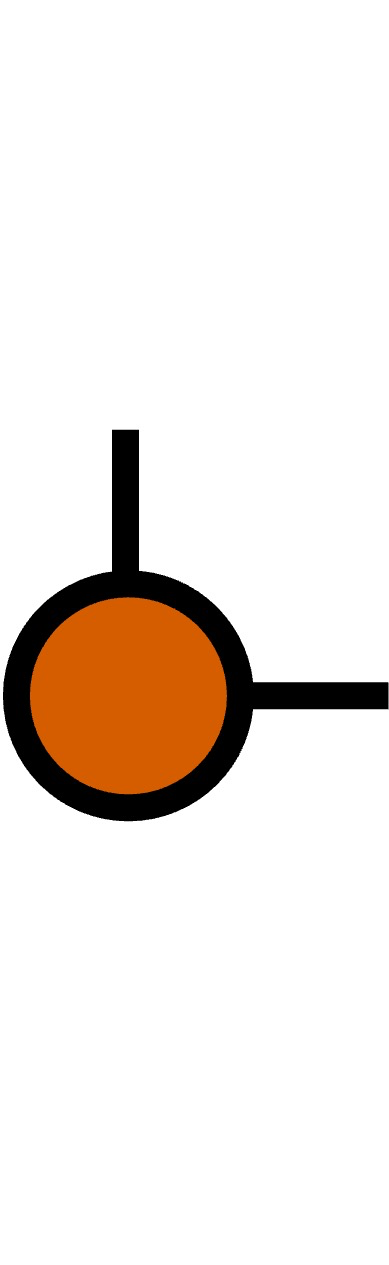}
    \end{gathered}\,.
\end{equation*}
We conclude, having computed the MPO--MPS product $\ket{\eta} = H\ket{\psi}$ as an MPS in right canonical form:
\begin{multline*}
    \begin{gathered}
    \includegraphics[scale=0.35]{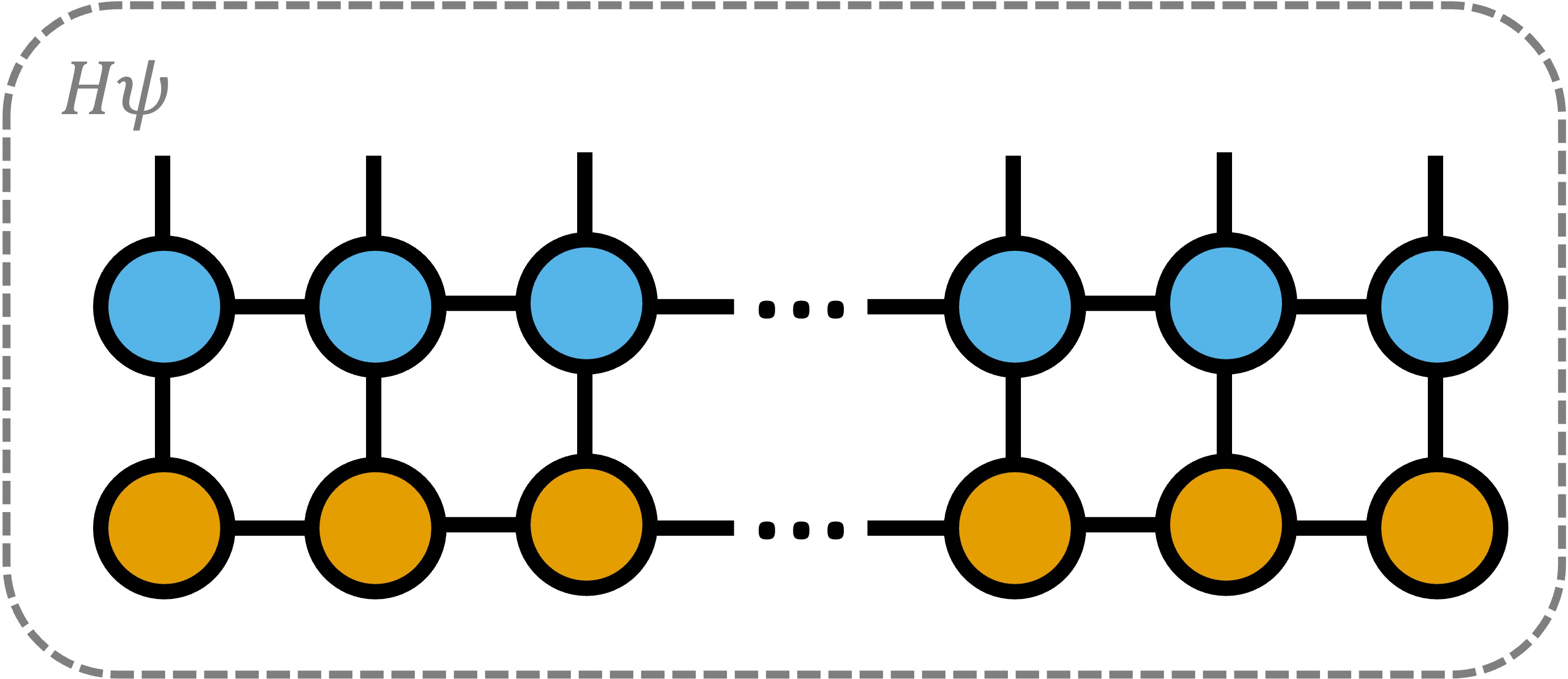}
    \end{gathered} \\ = \begin{gathered}
    \includegraphics[scale=0.35]{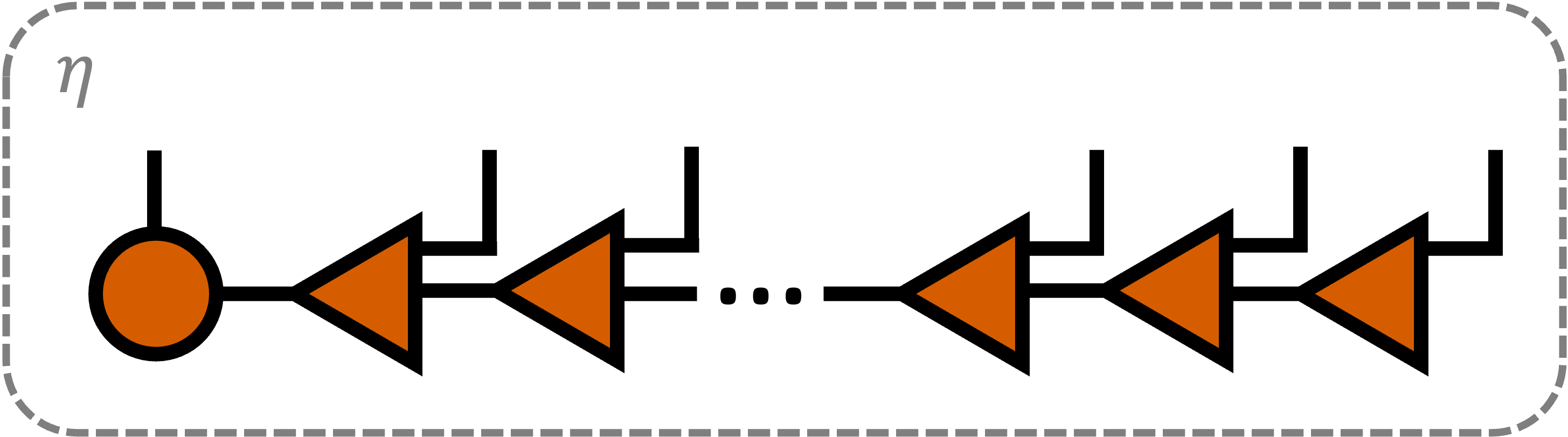}
    \end{gathered}\,.
\end{multline*}

\subsection{Optional step: Oversampling and final round} \label{sec:final_round}

In \cref{sec:Alg1,sec:Alg2,sec:Alg3}, we introduced the SRC algorithm under the assumption that the product $H\ket{\psi}$ is exactly representable as an MPS of known bond dimension $\overline{\chi}$.
However, the algorithm works without modification for computing a compressed approximation $H\ket{\psi}$ as an MPS with specified bond dimension $\overline{\chi}$.

When using the SRC method for approximate MPO--MPS multiplication, \emph{oversampling} can be used to achieve accuracy comparable with the best MPS representation of the product $H\ket{\psi}$.
To compute a compressed representation $\ket{\eta}\approx H\ket{\psi}$ of bond dimension $\overline{\chi}$, we do the following.
First, run SRC with a larger bond dimension $\overline{\chi}' > \overline{\chi}$, obtaining an MPS in right canonical form.
Then, use a standard MPS compression algorithm to truncate down to an MPS in left canonical form with desired output bond dimension $\overline{\chi}$.
As a sensible default, we recommend $\overline{\chi}' = \max(\lceil1.5\overline{\chi}\rceil,\overline{\chi}+10)$.

\subsection{Summary: Pseudocode and time complexity} \label{sec:pseudocode-complexity}

\begin{algorithm}[t]
	\caption{successive randomized compression for MPO--MPS multiplication} \label{alg:rand-MPO--MPS}
	\begin{algorithmic}[1]
        \algblock{Begin}{End}
		\Require MPO $H$, MPS $\ket{\psi}$, and output bond dimension $\overline{\chi}$ \textcolor{gray}{(or tolerance $\tau$, see \cref{sec:adaptivity})}
		\Ensure MPS $\ket{\eta} \approx H\ket{\psi}$ of bond dimension $\overline{\chi}$ 
        \Statex
        \Statex \textbf{Multiply with random matrix}
        \State Generate standard Gaussian matrices $\Omega^{(1)},\ldots,\Omega^{(n-1)}\in\mathbb{C}^{d\times \overline{\chi}}$ 
        \State Compute tensor $C^{(1)}(a,b,c) \gets \sum_{d,e} \Omega^{(1)}(a,d) H^{(1)}(d,b,e) \psi^{(1)}(e,c)$
        \For{$i= 2,\ldots,n-1$}
            \State Compute tensor $C^{(i)}(a,b,c) \gets \text{\small$\sum_{d,e,f,g} C^{(i-1)}(a,d,e) \Omega^{(i)}(a,f) H^{(i)}(d,f,b,g) \psi^{(i)}(e,g,c)$}$
        \EndFor
        \Statex
        \Statex \textbf{Determine the last site $\eta^{(n)}$}
        \State Compute tensor $Y^{(n)}(a,b) \gets \sum_{c,d,e} C^{(n-1)}(a,c,d) H^{(n)}(b,c,e) \psi^{(n)}(d,e)$
        \State Compute \QR decomposition $Y^{(n)}(a,b) = \sum_{c} \eta^{(n)}(c,b) R^{(n)}(a,c)$
        \State \textcolor{gray}{[\textit{\textbf{Optional:} Compute error estimate and increase $\overline{\chi}$ if needed; see \cref{sec:adaptivity}}]}
        \State Compute tensor $S^{(n)}(a,b,c) \gets \sum_{d,e} \overline{\eta^{(n)}(c,d)} H^{(n)}(b,d,e) \psi^{(n)}(c,e)$
        \Statex
        \Statex \textbf{Determine sites $\eta^{(n-1)},\ldots,\eta^{(2)}$}
        \For{$j = n-1,n-2,\ldots,2$}
        \State Compute tensor $Y^{(j)} \gets \text{\small $\sum_{d,e,f,g,h} C^{(j-1)}(a,d,e) H^{(j)}(d,b,f,g) \psi^{(j)}(e,g,h) S^{(j+1)}(h,f,c)$}$
        \State Compute \QR decomposition $Y^{(j)}(a,b,c) = \sum_d \eta^{(j)}(d,b,c) R^{(j)}(a,d)$ 
        \State \textcolor{gray}{[\textit{\textbf{Optional:} Compute error estimate and increase $\overline{\chi}$ if needed; see \cref{sec:adaptivity}}]}
        \State Compute tensor $S^{(j)}(a,b,c) \gets \text{\footnotesize $\sum_{d,e,f,g,h}\overline{\eta^{(j)}(c,d,e)}H^{(j)}(b,d,f,g)\psi^{(j)}(a,g,h) S^{(j+1)}(h,f,e)$}$
        \EndFor
        \Statex
        \Statex \textbf{Determine the first site $\eta^{(1)}$}
        \State Compute tensor $\eta^{1}(a,b) \gets \sum_{c,d,e} H^{(1)}(a,c,d) \psi^{(1)}(d,e) S^{(2)}(e,d,b)$
        \State \textcolor{gray}{[\textit{\textbf{Optional:} Run an MPS truncation algorithm on $\ket{\eta}$}]}
	\end{algorithmic}
\end{algorithm}

As we have seen, SRC produces an MPS representation $\ket{\eta}$ of the MPO--MPS product $H\ket{\psi}$ using a sequence of \QB decompositions or approximations.
For ease of presentation, we first presented the algorithm for the case in which $H\ket{\psi}$ is exactly representable as an MPS of specified bond dimension $\overline{\chi}$.
In this case, SRC produces an exact representation $\ket{\eta} = H\ket{\psi}$:
\begin{theorem}[Successive randomized compression: Exact recovery] \label{thm:src-exact}
    If the MPO--MPS product $H\ket{\psi}$ is exactly representable as an MPS of bond dimension $\overline{\chi}$, then SRC recovers $\ket{\eta} = H\ket{\psi}$ with probability one.
\end{theorem}
This result follows directly from \cref{thm:khatri-rao}, though the fact that we use a common set of random matrices $\Omega^{(1)},\ldots,\Omega^{(n-1)}$ across all steps of the algorithms makes the result not entirely trivial.
\Cref{app:src-exact-proof} contains a proof of this result.

When $H\ket{\psi}$ is not exactly equal to an MPS of bond dimension $\overline{\chi}$, SRC returns a compressed approximation $\ket{\eta}\approx H\ket{\psi}$ to the product.
We present pseudocode for SRC in \cref{alg:rand-MPO--MPS}.
See \cref{sec:approx} for a discussion of determining the output bond dimension $\overline{\chi}$ adaptively to meet a tolerance.

The runtime of SRC depends on the number of sites $n$, the MPO and MPS bond dimensions $D$ and $\chi$, the physical dimension $d$, and the output bond dimension $\overline{\chi}$.
In terms of all of these parameters, SRC has a time complexity of 
\begin{subequations} \label{eq:runtimes}
\begin{equation} \label{eq:runtime-full}
    \order(ndD\chi \overline{\chi}(\chi+\overline{\chi}+dD)) \text{ operations}.
\end{equation}
Making a few assumptions about parameter values, we obtain a simplified runtime estimate
\begin{equation} \label{eq:runtime-simplified}
    \order(n D\chi^3) \text{ operations}, \quad \text{assuming } D \le \chi = \overline{\chi} \text{ and } d = \mathrm{const}.
\end{equation}
\end{subequations}
Using either the full runtime \cref{eq:runtime-full} or simplified runtime \cref{eq:runtime-simplified}, SRC is as fast as or faster than all existing MPO--MPS multiplication algorithms; see \cref{sec:previouswork} for a description of other MPO--MPS multiplication algorithms and a runtime comparison with SRC.

\subsection{Extension: Linear combinations of MPO--MPS products}

As an extension, this algorithm can also be used to form a compressed representation of a linear combination of MPO--MPS products:
\begin{equation*}
    \ket{\eta} \approx \sum_{i=1}^t \alpha H_i\ket{\psi_i}.
\end{equation*}
At each phase of the algorithm, we compute a ``local'' $Y^{(j)}_i$ tensor separately for each term, and compute the global $Y^{(j)} = \sum_{i=1}^t \alpha_i Y^{(j)}_i$ as a linear combination.
The global $Y^{(j)} = \eta^{(j)} \cdot R^{(j)}$ tensor is then split using a \QR factorization, and we use the same $\eta^{(j)}$ to define a local tensor $B^{(j-1)}_i = (\eta^{(j)})^\dagger B^{(j)}_i$ for each term.
We conclude the algorithm by setting $\eta^{(1)} \coloneqq \sum_{i=1}^t \alpha_i B^{(1)}_i$.
This randomized algorithm for compressing sums of MPO--MPS products is similar to the algorithm for compressing sums of MPSs in \cite{ABC+21}.

\subsection{Limitation: Physical symmetries}

In physical problems, MPOs and MPSs often obey group symmetry relations associated with physical conservation laws (e.g., symmetry under the group $U(1)$ arises from particle-number conservation).
When symmetry is present, it can be built into the tensor network representation by requiring that the constituent tensors $\psi^{(1)},\ldots,\psi^{(n)}$ or $H^{(1)},\ldots,H^{(n)}$ have a prescribed block-sparsity structure; see \cite{SPV10,SPV11} for details.

Unlike several other compressed contraction algorithms, the SRC algorithm presented in this paper is not symmetry-preserving: The intermediate tensors $C^{(i)}$ and $S^{(i)}$ computed by the algorithm do not possess sparsity structure, and the algorithm does not ensure the compressed output $\ket{\eta}$ has the appropriate block-sparsity structure.
(The algorithm can, however, exploit sparsity of the constituent tensors by using sparse tensor formats to perform intermediate contractions in the algorithm.)
Investigating whether the SRC algorithm can be modified to respect symmetries is a natural subject for future research.

\section{Existing MPO--MPS product methods}\label{sec:previouswork}

Existing methods for the MPO--MPS product can be divided into three groups: contract-then-compress methods, optimization methods, and explicit construction algorithms.
We review each of these classes of algorithms in turn.

\subsection{Contract-then-compress methods}

Contract then compress methods work by first contracting the MPO--MPS product \emph{exactly} as an MPS of bond dimension $D\chi$, followed by a compression step to truncate the bond dimension:
\begin{equation*} \label{eq:contract-then-compress}
    \begin{split}
    \begin{gathered}
        \includegraphics[scale=0.3]{figs/mpo_times_mps.png}
    \end{gathered}
    &= 
    \begin{gathered}
        \includegraphics[scale=0.35]{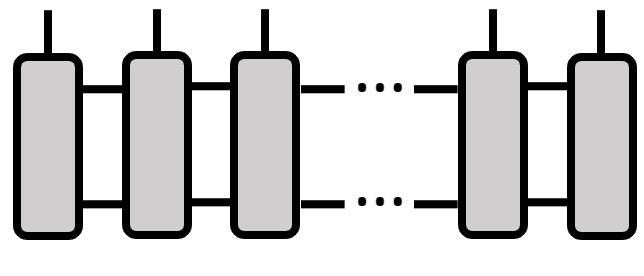} 
    \end{gathered}\\
     &\approx 
    \begin{gathered}     
       \includegraphics[scale=0.35]{figs/compress2.png}
    \end{gathered}
    \end{split}
\end{equation*}
For the second step, any MPS truncation algorithm can be used, leading to different variants of the contract-then-compress method.
We review these variants here.

\subsubsection{Basic contract-then-compress}
The simplest contract-then-compress method uses the standard SVD-based MPS compression algorithm (see, e.g., \cite[Alg.~2]{Ose11}).
This scheme is known to produce a near-optimal MPS approximation to $H\ket{\psi}$ \cite[Cor.~2.4 and sec.~3]{Ose11}, but it is slow, requiring
\begin{equation*}
    \order(n \cdot [dD^3\chi^3 + d^2D^2\chi^2]) \text{ operations}
\end{equation*}
since it requires converting $H\ket{\psi}$ to canonical form prior to compression. Basic contract-then-compress is typically about $D^2$ times slower than the SRC, in view of \cref{eq:runtime-simplified}.%\CC{With careful contraction it is possible to preform this process in $O(n(d D^3 \chi^3+d^2\chi^2+d^3D^2))$ time by sweeping left to right like we do our environment tensors absorbing one core of psi then one core of H  into the contracted left boundary. In fact this is precisely whats implemented in the library for both ctc and rctc. If we choose to update this bound the picture becomes less intutive, but it feels like an important attribution since this estimation is overemphasizing the cost of the only other randomized contraction algorithm.}

\subsubsection{Randomized contract-then-compress}\label{sec:RCTC}

The contract-than-compress method can be improved by using a \emph{randomized MPS compression method} \cite{ABC+21}.
Randomized contract-then-compress differs from the randomized method introduced in this paper in that it first represents $H\ket{\psi}$ exactly as an MPS and then compresses this MPS; by contrast, our algorithm avoids ever working with the exact MPS form of $H\ket{\psi}$, resulting in a faster algorithm.

The paper \cite{ABC+21} presents three different randomized MPS compression algorithms; for the purpose of this paper, we consider their \emph{randomize-then-orthogonalize} method \cite[Alg.~3.2]{ABC+21}.
Using this method, randomized contract-then-compress has a runtime of
\begin{equation*}
    \order(n \cdot [dD^2\chi^2\overline{\chi} + d^2D^2\chi^2]) \text{ operations}.
\end{equation*}
This improves on the basic contract-then-compress method, but is still typically a factor $D$ times slower than SRC. 
Similar to SRC, randomized contract-then-compress is typically a constant factor less accurate than the basic contract-then-compress method.
At present, randomized contract-then-compress is limited to the setting where the output bond dimension $\overline{\chi}$ is provided as an input to the algorithm (i.e., not determined adaptively).

\subsubsection{Variational contract-then-compress}

A third possible implementation of contract-then-compress uses \emph{variational compression} to compress the MPO--MPS product $H\ket{\psi}$.
Variational MPS is a cyclic minimization algorithm that attempts to minimize the error $\norm{\ket{\phi} - \ket{\eta}}_{\rm F}$ of approximating a provided MPS $\ket{\phi}$ by an MPS $\ket{\eta}$ of smaller dimension; see \cite[\S4.5.2]{Sch11} and \cite[\S2.6.2]{Paeckel_2019} for details.
As with DMRG (and fitting, below), there are one-site and two-site versions of the variational compression method, with the two-site version being somewhat more expensive but adaptive and more robust.
With the two-site procedure, variational contract-then-compress costs 
\begin{equation*}
    \mathcal{O}(n\cdot [d^2D^2\chi^2\overline{\chi}]) \text{ operations per sweep}.
\end{equation*}
Similar to the fitting algorithm (below), variational MPS can have slow convergence (requiring many sweeps) or fail to converge at all on some examples.

\subsection{Optimization methods}

Optimization methods treat the MPO--MPS multiplication problem as an optimization problem
\begin{equation} \label{eq:optimization}
    \operatorname*{minimize}_{\ket{\eta} = \operatorname{MPS}(\eta^{(1)},\ldots,\eta^{(n)})}\, \norm{\ket{\eta} - H\ket{\psi}}_{\rm F}^2.
\end{equation}
Here, the minimum is taken over all MPSs $\ket{\eta}$, say, of specified bond dimension $\overline{\chi}$.
There are two optimization methods, the fitting method and the global optimization method.

\subsubsection{Fitting} \label{sec:fitting}

The fitting method, proposed by Verstraete and Cirac \cite{VC04}, attempts to solve the optimization problem \cref{eq:optimization} by cyclic minimization, iteratively minimizing over each $\eta^{(i)}$ while holding the other sites.
This site-by-site minimization is performed in alternating sweeps left-to-right then right-to-left, \textit{\`a la} DMRG.
See \cite[\S 2.8.2]{Paeckel_2019} for details.
The fitting method is distinguished from the variational contract-then-compress method because contract-then-compress represents $H\ket{\psi}$ directly as an MPS of bond dimension $D\chi$, which it then compresses; the fitting method avoids ever forming the uncompressed MPO--MPS product $H\ket{\psi}$, leading to a fast algorithm.

The basic Verstraete--Cirac fitting method can suffer from slow convergence and does not admit a natural way to determine the bond dimension of $\ket{\eta}$ adaptively to meet a tolerance.
To mitigate these issues, Stoudenmire and White \cite{SW10} proposed a two-site version of the fitting method which minimizes a pair of sites $\eta^{(i)},\eta^{(i+1)}$ and then splits by SVD, allowing for adaptive determination of the bond dimension.
The relation of one-site and two-site fitting is analogous to one-site and two-site DMRG.
All experiments in this paper use the two-site fitting algorithm.

\begin{figure}
    \centering
\includegraphics[width=0.8\linewidth]{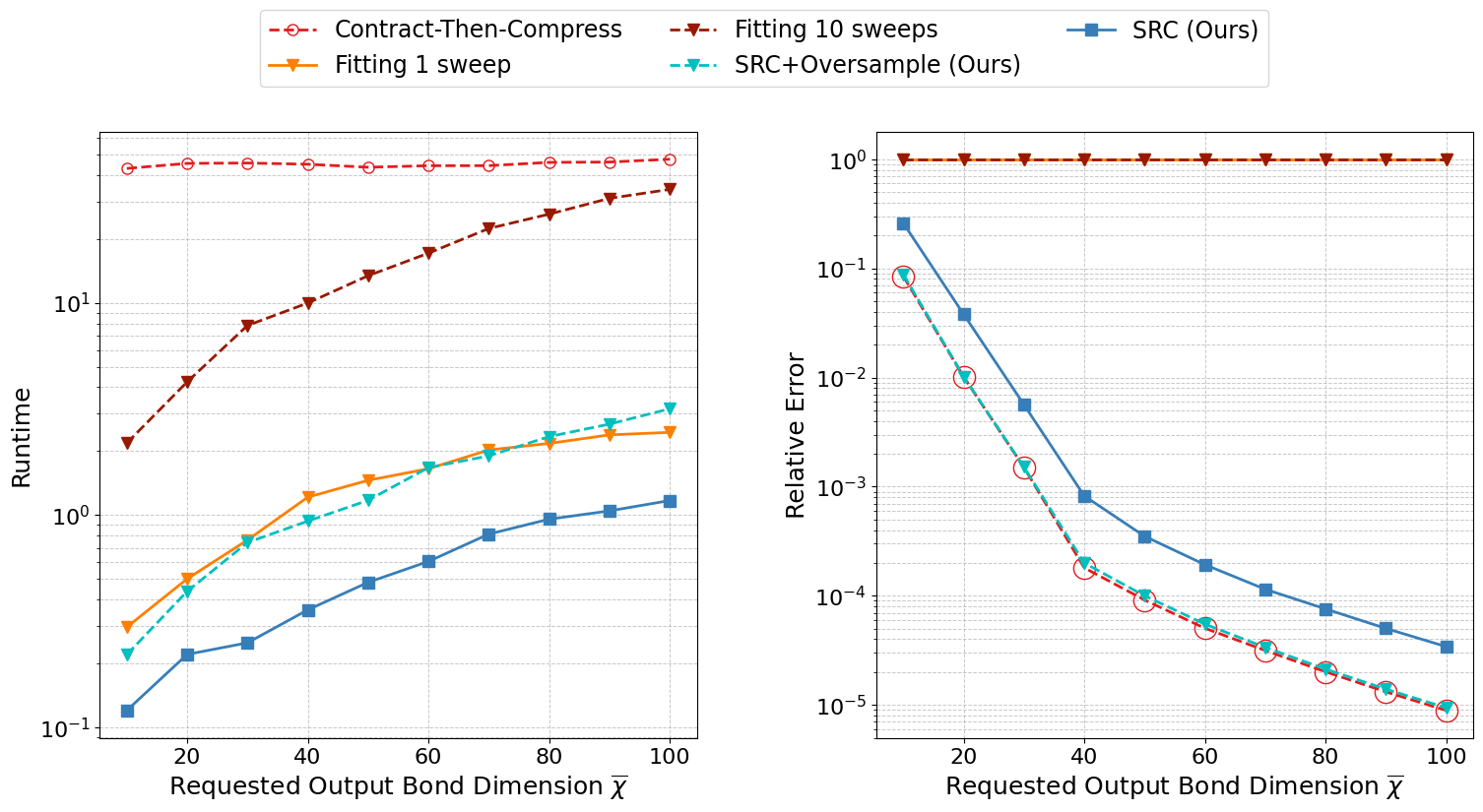}
    \caption{(\textbf{Fitting failure mode}). Runtime (\emph{left}) and relative error (\emph{right}) of the fitting algorithm, SRC, and contract-then-compress on the MPO--MPS product between the long range XY Hamiltonian given in \cref{sec:TDVP_exp} and its groundstate.}
    \label{fig:fittingexample}
\end{figure}

On a per-sweep basis, the fitting method is a fast, requiring $\order\left(n\cdot[ dD\chi\overline{\chi}^2+d^2D^2\chi^2]\right)$ operations.
However, as Stoudenmire and White \cite{SW10} warn, even the more robust two-site fitting algorithm can converge slowly or fail to converge at all, particularly for MPOs that capture long-range interactions. \cref{fig:fittingexample} illustrates this behavior for the long range $XY$ Hamiltonian $H$ used in \cref{sec:TDVP_exp}. In this example, we compute the MPO--MPS product of $H$ with its ground state $\ket{\psi_0}$, which, again, was obtained using DMRG2 with tolerance $10^{-8}$.
The fitting algorithm is supplied with the ground state $\ket{\psi_0}$ as the initial guess, which agrees with the output $H\ket{\psi_0}$ up to scaling by the energy $E_0 \neq 0$ and numerical errors.
Despite performing 10 sweeps, the fitting algorithm fails to converge, or make any progress towards solving the problem.
We observe the same behavior when we initialize the fitting algorithm with a random MPS with Gaussian random entries.

Similar to the fitting algorithm, the DMRG algorithm can also suffer from convergence to local minima.
Researchers have suggested adding random noise~\cite{white2005density,hubig2015strictly} or using randomized low-rank approximation algorithms \cite{MO24} as strategies to avoid this behavior, and these strategies can also be applied to the fitting algorithm.
We have not evaluated these more sophisticated versions of the fitting algorithm in this paper.

\subsubsection{Global optimization}

To overcome the shortcomings of cyclic minimization algorithms like fitting, Hauru, Van Damme, and Haegeman investigates global Riemmanian optimization algorithms for tensor network problems \cite{HVH21}.
In principle, these methods can have a runtime competitive with other fast MPO--MPS algorithms, but they are relatively complicated to implement.
We are unaware of a publicly available implementation of this method for the MPO--MPS product.

\subsection{Explicit construction methods}

The last family of algorithms, explicit construction methods, directly construct an approximate MPO--MPS product without either the initial representation of $H\ket{\psi}$ as an MPS of bond dimension $D\chi$ in contract-then-compress or the need to iterate until convergence required by optimization methods.

\subsubsection{Zip-up}
The zip-up method was proposed by Stoudenmire and White \cite{SW10} as an alternative to the potentially slow convergence of the fitting algorithm.
Zip-up begins by left-canonizing the input MPO and MPS at a cost of $\mathcal{O}(n\cdot [dD^3+d\chi^3])$.
Then, it proceeds left-to-right merging sites of the MPO and MPS, truncating as it proceeds.
See \cite[\S2.8.3]{Paeckel_2019}, \cite[\S3.2.2]{ma2024} for details.

As advantages, the zip-up algorithm has a fast $\mathcal{O}(n\cdot [dD\chi\overline{\chi}^2+d^2D^2\chi\overline{\chi}])$ runtime and is a single-shot procedure, immune from the slow or non-convergence than can affect fitting-type methods.
However, zip-up is less accurate than other MPO--MPS algorithm because the $i$th truncation step uses information from the first $i$ sites, rather than the full environment.
Describing their experience with the algorithm, Paeckel et al.\ \cite[\S2.8.3]{Paeckel_2019} write that the zip-up algorithm is ``\emph{typically} sufficiently accurate and fast'' (emphasis ours) and suggest combining it with a few sweeps of the fitting algorithm to improve accuracy when necessary.
By contrast, SRC is both fast and always achieves near-optimal truncation using the full environment.

\subsubsection{Density matrix}
The density matrix algorithm \cite{Tenc,MFSS24a} is conceptually similar to the randomized method developed in this work.
In SRC, we construct an MPS approximation $\ket{\eta}$ to $H\ket{\psi}$ by a sequence of randomized \QB approximations $B^{(n)} = H\ket{\psi} \approx \eta^{(n)} \cdot B^{(n-1)}$, $B^{(n-1)} \approx \eta^{(n-1)} \cdot B^{(n-2)}$, etc.; each $B^{(j)}$ is a matrix unfolding of an appropriate tensor network.
Instead of randomized \QB approximation, the density matrix algorithm constructs each $\eta^{(j)}$ as the leading eigenvectors of the \textit{density matrix} $\rho^{(j)} \coloneqq B^{(j)}(B^{(j)})^\dagger$.

The density matrix has advantages and disadvantages.
Neglecting rounding errors, it produces an identical output to the contract-then-compress method, thus achieving near-optimality accuracy.
Also, the density method is faster than contract-then-compress and more reliable than the fitting and zip-up methods.

The main disadvantage of the density matrix algorithm is speed, being the second slowest method in \cref{fig:intro}.
SRC is nearly as accurate as the density matrix algorithm and often much faster.
Another issue with the density matrix algorithm is numerical precision.
Forming density matrix $B^{(j)}(B^{(j)})^\dagger$ effectively squares the matrix $B^{(j)}$, potentially halving the effective numerical precision.
On certain examples, the relative error of the density matrix can stagnate near $10^{-8}$ in double precision ($10^{-4}$ in single), where other methods achieve full accuracy ($\approx 10^{-16}$ in double, $10^{-8}$ in single).
The density matrix algorithm faces these precision issues only on some examples, achieving full precision on others; we do not have an explanation for this inconsistency.

\subsection{Comparison}\label{sec:comp}

In this section, we compare the SRC method to these alternatives and provide guidance about which algorithm to use.
In all experiments, we use real standard Gaussian entries in random Khatri--Rao products for the SRC algorithm.

\begin{table}[t]
  \centering
  \caption{Simplified operation counts for methods to apply an MPO (bond dimension $D$) to an MPS (bond dimension $\chi$). For this table, we assume $d = \order(1)$ and $D \le \chi \le \overline{\chi} \le D\chi$; see \cref{tab:Complete_operation_counts} for a complete operation count that does not make these assumptions.} \label{tab:simplified_operation_counts}
  \begin{tabular}{ll}\toprule  
    Method & Op.\ Count (simplified) \\ \midrule
    Contract-then-compress \\
    \quad $\bullet$ Basic & $\order( nD^3\chi^3 )$ \\
    \quad $\bullet$ Randomized & $\order( nD^2\chi^2\overline{\chi} )$  \\
    Density Matrix & $\order( nD^2 \chi^2\overline{\chi} )$ \\
    Fitting & $\order( nD \chi\overline{\chi}^2)$ per iteration \\
    Zip-up &  $\order( nD\chi\overline{\chi}^2)$ \\
    SRC & $\order(nD\chi\overline{\chi}^2)$  
\\
    \bottomrule
  \end{tabular}
\end{table} 
\subsubsection{Complexity}
We first compare the methods based on their operation counts, which are shown in \cref{tab:simplified_operation_counts}.
This table displays simplified operation counts where the physical dimension $d$ is treated as a constant and the bond dimensions are ordered as $D\le \chi \le \overline{\chi}$; see \cref{tab:Complete_operation_counts} for more detailed operation counts in terms of all five parameters $n$, $d$, $\chi$, $D$, and $\overline{\chi}$.

In terms of pure operation count, we see that the randomized and zip-up methods are fastest followed by density matrix and randomized contract-then-compress, all of which are faster than basic contract-then-compress.
These operation counts are consistent with the runtime experiments provided in paper.
The \emph{per-iteration} cost of the fitting method is also comparable to the randomized and zip-up methods, but---as \cref{fig:fittingexample} shows---the fitting method can take many iterations to converge or fail to converge outright for difficult examples.
Finally, we notice that the operation counts of all methods are comparable when $\overline{\chi} \approx D\chi$; therefore, speedups are only possible for MPO--MPS multiplication when a significant amount of compression is used $\overline{\chi} \ll D\chi$.

\begin{figure}[t]
    \centering
\includegraphics[width=.8\textwidth]{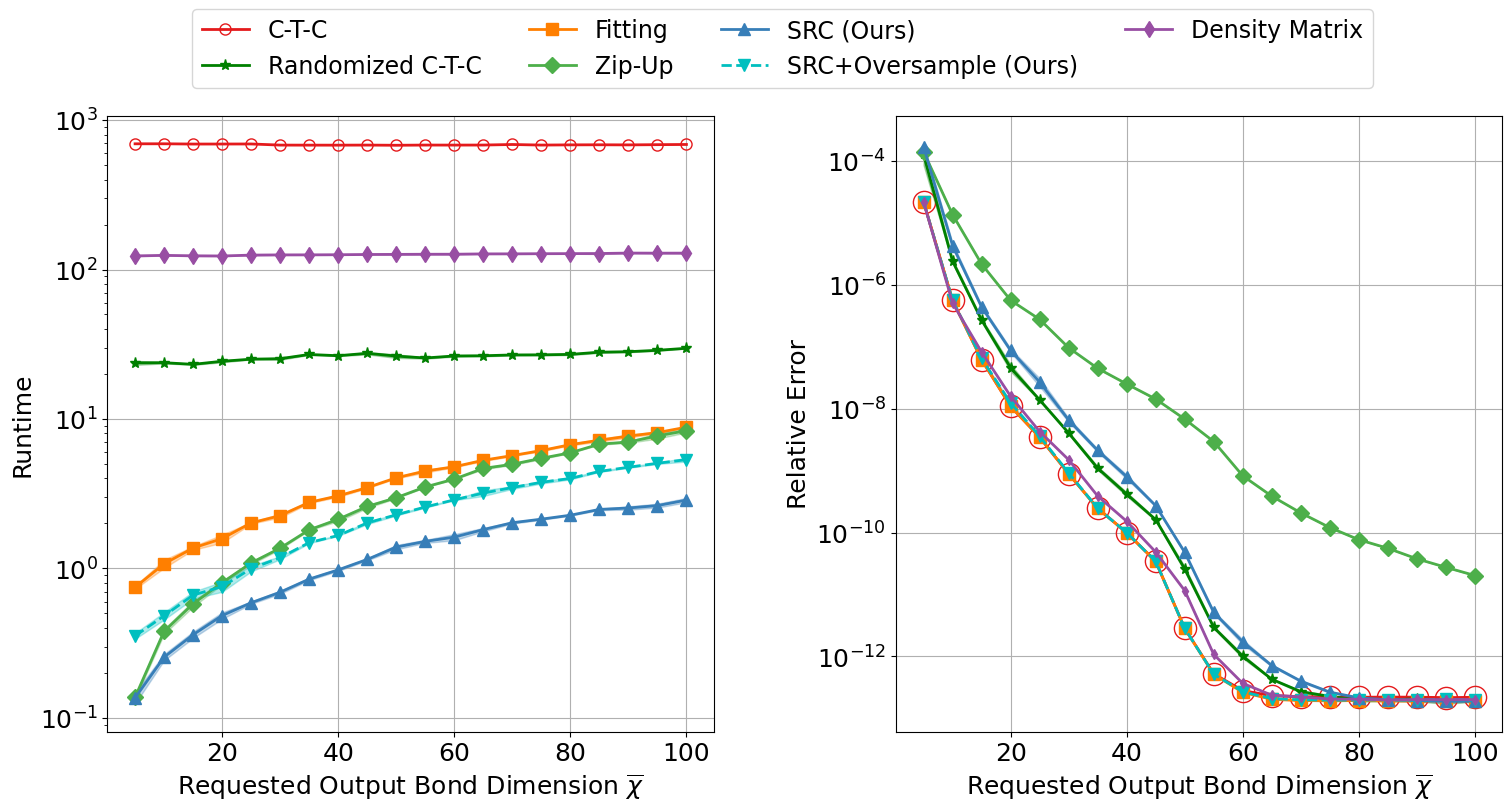}
    \caption{(\textbf{Fixed bond dimension}). Runtime (\emph{left}) and relative error (\emph{right}) of several MPO--MPS multiplication algorithms for a synthetic problem as a function of the output bond dimension $\overline{\chi}$, as described in \cref{sec:Fixed dimension}.}
    \label{fig:finalComparison}
\end{figure}
\subsubsection{Comparison experiment 1: Fixed output bond dimension} \label{sec:Fixed dimension}
First, we compare the methods when the output bond dimension $\overline{\chi}$ is fixed by the user.
We use the same problem setup as \cref{sec:intro}, where we generate random a random MPS and MPO with bond dimensions $D=\chi=50$ across $n=100$ sites and uniformly random entries $[\alpha, 1]$ with $\alpha=-0.5$.
As in the introduction, we use complex data types and apply the contract-then-compress method with a machine-precision tolerance to compute a baseline.

A comparison of existing MPO--MPS methods and SRC is presented in \cref{fig:finalComparison} for a range of requested bond dimensions $\overline{\chi}$.
We implement SRC both as-is and with the oversampling strategy presented in \cref{sec:final_round}.
For this problem, we see that SRC is the fastest method for all requested output bond dimensions $\overline{\chi}$ and, when run with oversampling, achieves comparable accuracy to contract-then-compress at a faster runtime than other accurate methods.

\begin{figure}
    \centering
\includegraphics[width=1\linewidth]{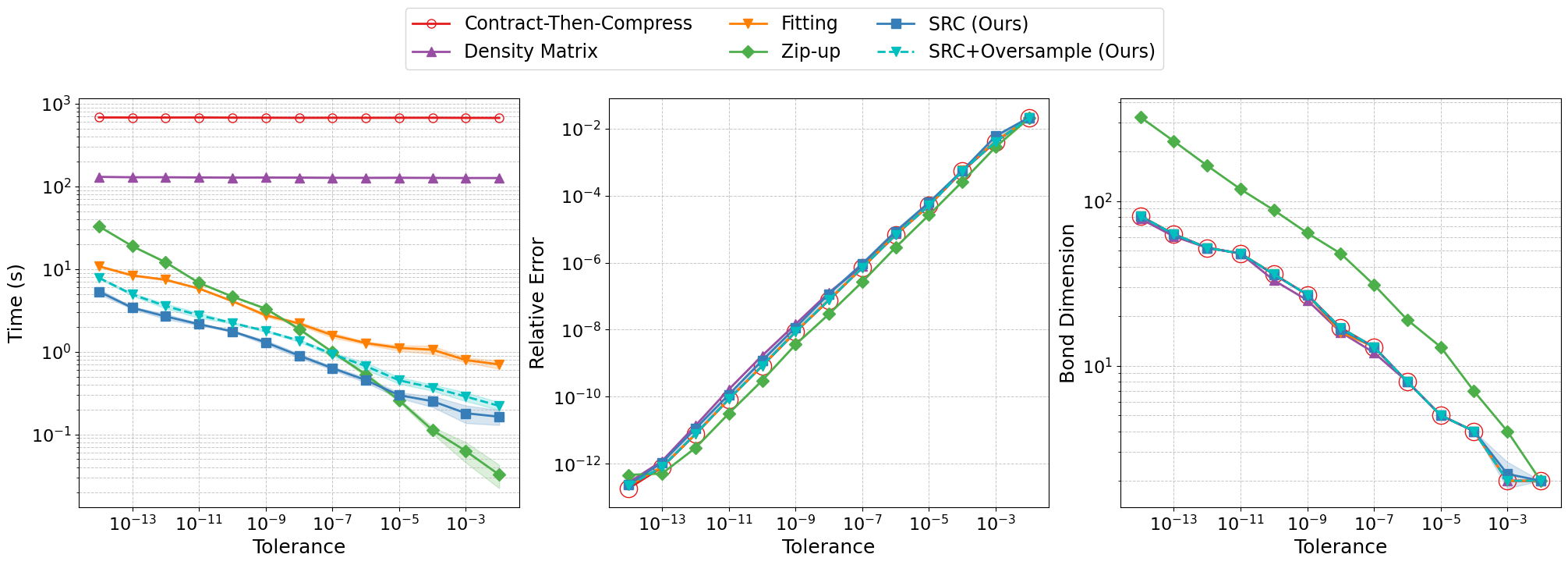}
    \caption{(\textbf{Accuracy threshold experiment}). Runtime (\emph{left}), relative error (\emph{center}), and final bond dimension (\emph{right}) of several MPO--MPS multiplication algorithms for a synthetic problem as a specified approximation threshold.}
    \label{fig:syntheticcutoff}
\end{figure}

\subsubsection{Comparison experiment 2: Adaptive determination of bond dimension}\label{sec:adaptive_experiments}
Next, we compare the the methods when the bond dimension is determined adaptively to meet a tolerance; see \cref{sec:approx} for how to implement SRC in this setting.
The runtime behavior of the algorithms in this setting is somewhat changed, as the adaptive determination of bond dimension places a different level of overhead on each algorithm.

\Cref{fig:syntheticcutoff} shows the results.
For both the density matrix and contract-then-compress methods, the runtime behavior remains largely unaffected by the requested tolerance.
SRC continues to be the fastest method in this setting, except at large values of the tolerance parameter where the zip-up method becomes fastest.
This speed of zip-up comes with a cost, with the zip-up method resulting in meaningfully larger bond dimensions than other algorithms.
Therefore, even for high values of the tolerance, SRC (with a final round) may be preferable to obtain an output of the smallest possible bond dimension, with the zip-up method being preferable for fast, approximate computations.

\subsubsection{Summary of comparison}
Both with a fixed bond dimension and with a tolerance, SRC achieves the greatest speed among the existing methods. When run with oversampling, it achieves comparable accuracy to the best method.
We believe these experiments support the deployment of the randomized MPO--MPS multiplication as a method-of-choice in applications.\CC{consider for example as a segue into dynamics result see checklist.}

We conclude with two important caveats and limitations.
First, as we already saw, SRC can be slower than the zip-up method when used with a loose tolerance, though the zip-up method results in higher bond dimensions.
The zip-up method may still be the method of choice for coarse approximate MPO--MPS multiplication. 
Second, the speedups of SRC require having a problem with large MPS \emph{and MPO} bond dimensions $D,\chi \gg 1$ and significant compression of the output bond dimension below its maximum value $\overline{\chi} \ll D\chi$.
This makes SRC a powerful tool for physical problems with complicated or long-range interactions, but it may not be the best tool for simple Hamiltonians of bond dimension $D\le 5$.

\section{Application: Unitary time evolution} \label{sec:time-evolution}

To demonstrate the effectiveness of the randomized contraction algorithm, we compare the performance of several compressed MPO--MPS multiplication algorithms for computing the unitary time evolution $\ket{\psi(t)} = \e^{-\iu t H} \ket{\psi(0)}$ of an initial state $\ket{\psi(0)}$ under the influence of a Hamiltonian $H$.
\Cref{sec:tdvp} describes the TDVP-GSE algorithm, a leading algorithm for time evolution that requires compressed MPO--MPS products, and \cref{sec:TDVP_exp} provides a comparison of different MPO--MPS algorithms for use with this algorithm.

\subsection{The time-dependent variational principle with global subspace expansion} \label{sec:tdvp}

The time-dependent variational principle (TDVP) algorithm \cite{HA11} and its variants are currently among the best-performing methods for simulating time-evolution.
In its basic form (``TDVP1''), TDVP simulates time evolution by updating individual site of the MPS in sequential left-to-right, right-to-left sweeps.
This basic one-site procedure can have issues (similar to one-site DMRG), and even two-site versions of TDVP (``TDVP2'', \cite[App.~5]{Haeg16}) can struggle on difficult examples.
To address this issues, Yang and White \cite{YA20} proposed an enhancement to TDVP by enriching the state at each time step with a set of global Krylov vectors.
Given a current state $\ket{\psi(t)}$ and timestep $\delta$, a single step of Yang and White's method proceeds as follows:
\begin{enumerate}
    \item Generate the Krylov vectors
    \begin{equation*}
        \mathcal{K} = \{ \ket{\psi(t)}, (\Id - \delta H)\ket{\psi(t)}, \ldots, (\Id - \delta H)^{k-1} \ket{\psi(t)} \}.
    \end{equation*}
    Here, $k$ sets the number of the Krylov vectors, which Yang and White recommend setting to a small value, e.g., $k=3$.
    \item Perform \emph{basis expansion} on the MPS $\ket{\psi(t)}$, increasing its bond rank by incorporating information from the Krylov space $\mathcal{K}$.
    \item Perform a single step of TDVP1 with step size $\delta$.
\end{enumerate}
We refer to \cite{YA20} for a justification and a description of the basis expansion step.
Yang and White refer to their procedure as \emph{global subspace expansion} TDVP (GSE-TDVP1).
In our implementation of their method, we use an SVD-based version of the basis expansion procedure, which differs from but is equivalent to the eigendecomposition-based procedure presented in \cite{YA20}.

The GSE-TDVP1 method requires (compressed) MPO--MPS products to form the Krylov space $\mathcal{K}$, and these MPO--MPS products can take up a large fraction of the method's runtime depending on the algorithm used.
To respond to this, Yang and White recommend using the density matrix algorithm or the fitting algorithm, depending on the bond dimension of $H$.
In the example documented in next section, we found the density matrix algorithm was too slow and the fitting algorithm was not accurate enough, leading to an explosion of the bond dimension of $\ket{\psi(t)}$ as time evolution progressed.
Responding to these issues, we propose the SRC method as an alternative for computing MPO--MPS products in GSE-TDVP1.

\subsection{Comparison of MPO--MPS methods for GSE-TDVP1}\label{sec:TDVP_exp}
To compare the performance of GSE-TDVP1 with different MPO--MPS methods, we emulate an experiment performed in \cite{Haeg16}.
This experiment considers an XY model with power-law interactions
\begin{equation*}
    H=\frac{1}{2} \sum_{1\le i<j\le n} \frac{J}{|i-j|^\alpha}\left(X_iX_j+Y_iY_j\right) \quad \text{with } \alpha = 1.5, \: n = 101.
\end{equation*}
Here, $X_i$ and $Y_i$ denote Pauli matrices acting on the $i$th site.
We construct the initial state $\ket{\psi(0)} = \e^{\iu (\pi/4) Y_{\lceil N/2\rceil}}\ket{\psi_{\rm min}}$ by applying a local operation to the middle site of the ground state $\ket{\psi_{\rm min}}$.
As demonstrated in previous studies, the difference in magnetization
\begin{equation} \label{eq:magnetization_difference}
    \braopket{\psi(t)}{X_i}{\psi(t)} - \braopket{\psi_{\rm min}}{X_i}{\psi_{\rm min}}
\end{equation}
between the ground state $\ket{\psi_{\rm min}}$ and the time-evolved perturbed state $\ket{\psi(0)}$ shows a light cone emanating from the middle cite $i = \lceil N/2\rceil$ with power law leakage outside the cone.

To simulate this system, we represent $H$ as an MPO using the linear-combination-of-exponentials method of \cite[App]{PMCV10} with tolerance $10^{-8}$, and we compute the ground state $\ket{\psi_{\rm min}}$ using DMRG2 with tolerance $10^{-8}$.
The bond dimensions of $H$ and $\ket{\psi(0)}$ are $D=26$ and $\chi = 32$.
Time evolution is performed with GSE-TDVP1 with a time step of $\delta = 0.2 /J$ to a final time $tJ = 5$, and all MPO--MPS products $(\Id - \delta H) \ket{\phi}$ are truncated to have the same bond dimension as $\ket{\phi}$.
Our purpose in presenting this experiment is to compare several MPO--MPS methods for use in GSE-TDVP1, not to claim that the GSE-TDVP1 method is necessary to resolve this particular physical system.
(Indeed, the qualitative features of this model can be resolved even using TDVP1.)

\begin{figure}[t]
    \centering
    \begin{subfigure}[b]{0.45\textwidth}
        \centering
        \includegraphics[width=\textwidth]{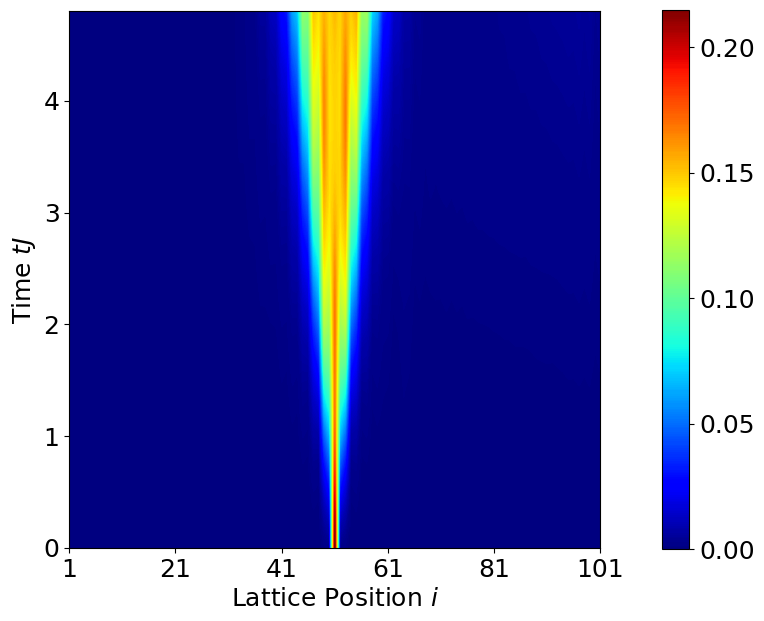}
        \label{fig:rand}
    \end{subfigure}
    \hfill
    \begin{subfigure}[b]{0.23\textwidth}
        \centering
        \includegraphics[width=\textwidth]{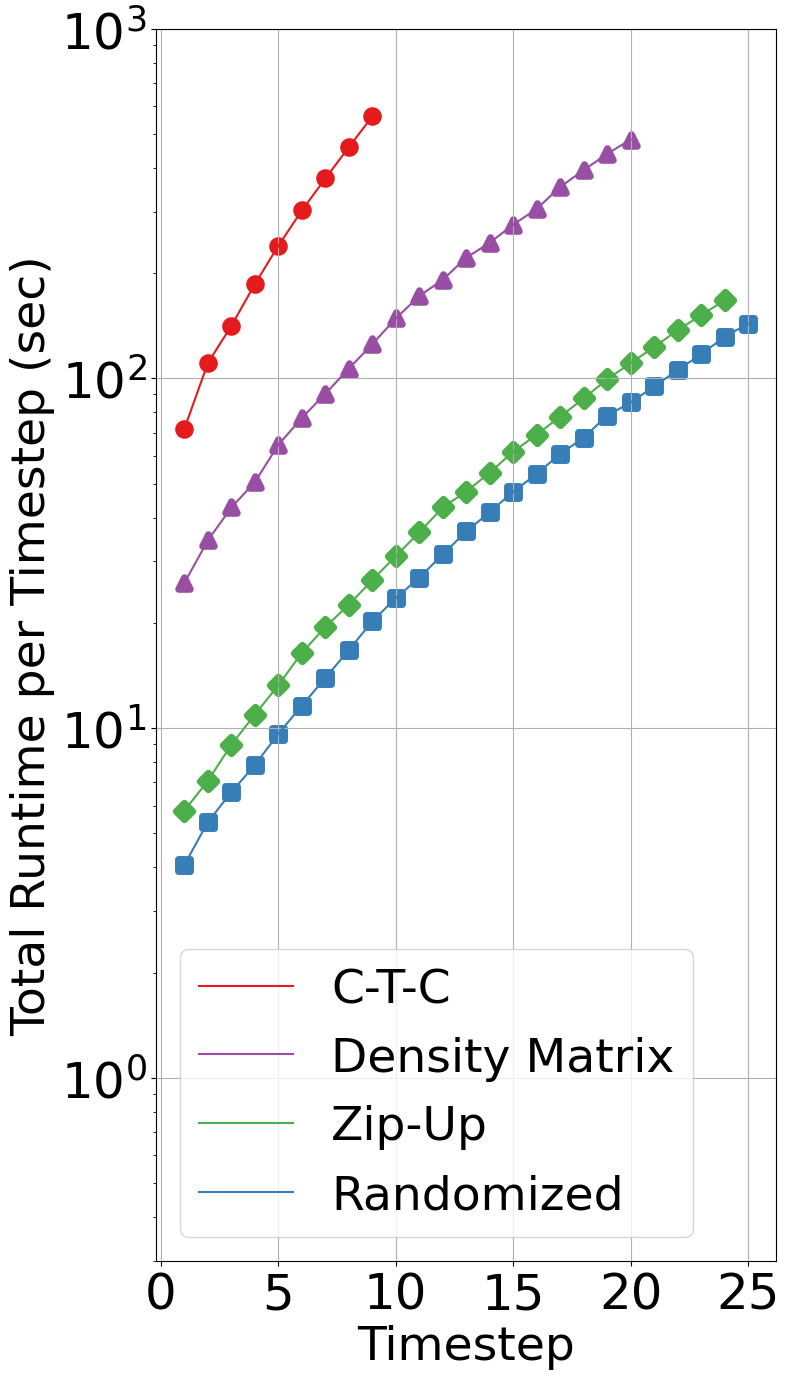}
        \label{fig:time}
    \end{subfigure}
    \hfill
    \begin{subfigure}[b]{0.23\textwidth}
        \centering
        \includegraphics[width=\textwidth]{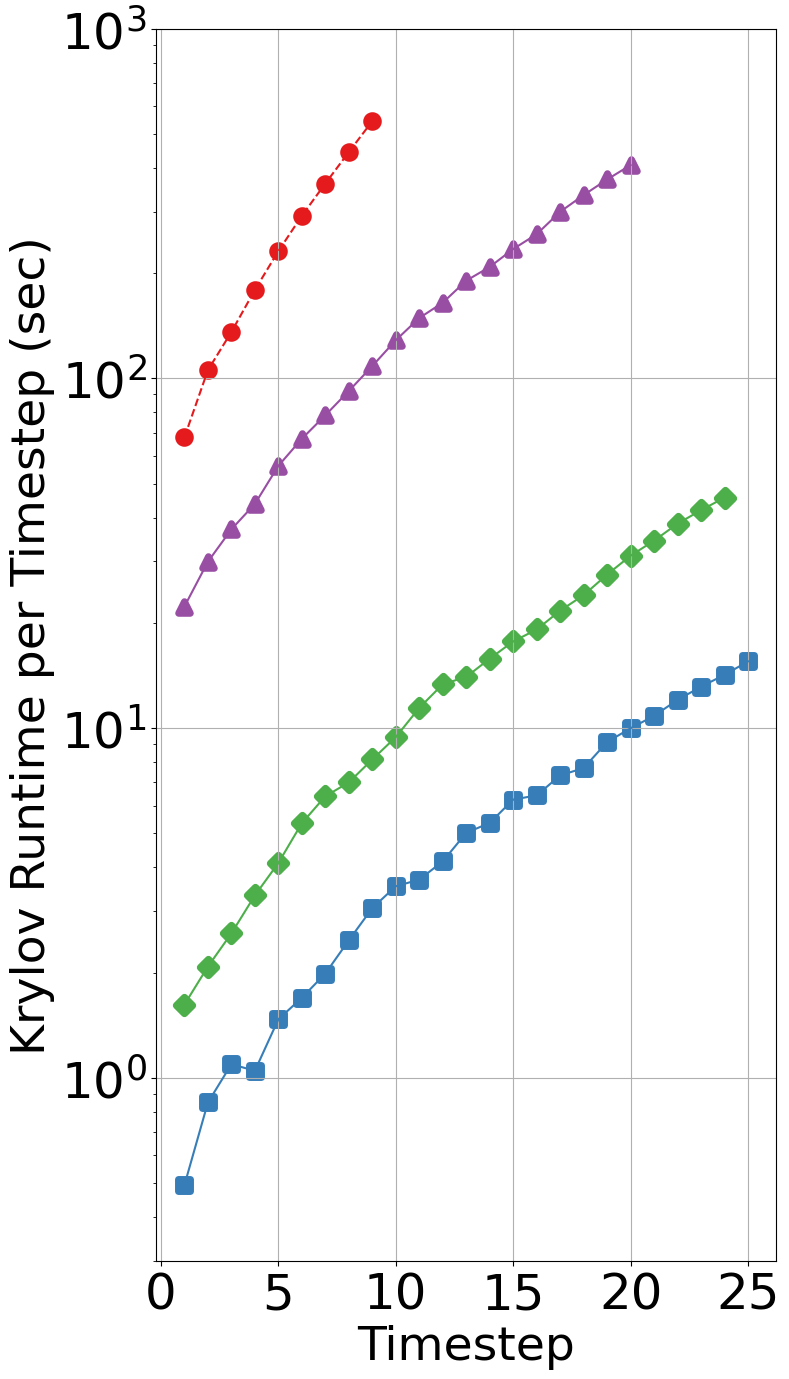}
        \label{fig:kry}
    \end{subfigure}
    \caption{(\textbf{TDVP unitary time evolution}). \emph{Left:} Magnetization difference \cref{eq:magnetization_difference} at each position $1\le i\le 101$ and time $0\le tJ\le 5$ computed by GSE-TDVP1 with randomized MPO--MPS multiplication, showing a light cone with power law leakage.
    \emph{Middle and right:} Total runtime (\emph{middle}) and runtime spent during Krylov vector construction (\emph{right}) for different MPO--MPS methods per timestep.\CC{move to pg 13}}
    \label{fig:time-evolution}
\end{figure}

\Cref{fig:time-evolution} shows the results.
The left panel shows that the magnetization difference \cref{eq:magnetization_difference} exhibits a light cone with power law leakage, in concurrence with the results of previous studies.
The middle and right panels show the runtime with various MPO--MPS methods.
(The fitting method yielded very poor accuracy and resulted in a rapid explosion in the bond dimension during the basis expansion step, so we omit it.)
The SRC method is definitely the fastest, achieving speedups of up to 181$\times$ over contract-then-compress, 45$\times$ over density matrix, and 3.2$\times$ over zip-up for the Krylov vector construction.

\section*{Acknowledgements}

We thank Jielun Chen, Garnet Chan, Johnnie Gray, Sandeep Sharma, and Robert Webber for helpful discussions.

\bibliographystyle{quantum}
\bibliography{refs}

\begin{thebibliography}{10}

\bibitem{Fannes92}
M.~{Fannes}, B.~{Nachtergaele}, and R.~F. {Werner}.
\newblock ``{Finitely correlated states on quantum spin chains}''.
\newblock \href{https://dx.doi.org/10.1007/BF02099178}{Communications in
  Mathematical Physics {\bf 144}, 443--490}~(1992).

\bibitem{V2004}
F.~Verstraete, J.~J. Garc\'{\i}a-Ripoll, and J.~I. Cirac.
\newblock ``Matrix product density operators: Simulation of finite-temperature
  and dissipative systems''.
\newblock \href{https://dx.doi.org/10.1103/PhysRevLett.93.207204}{Phys. Rev.
  Lett. {\bf 93}, 207204}~(2004).

\bibitem{Shi_2006}
Y.-Y. Shi, L.-M. Duan, and G.~Vidal.
\newblock ``Classical simulation of quantum many-body systems with a tree
  tensor network''.
\newblock \href{https://dx.doi.org/10.1103/physreva.74.022320}{Physical Review
  A{\bf 74}}~(2006).

\bibitem{Verstraete_2006}
F.~Verstraete, M.~M. Wolf, D.~Perez-Garcia, and J.~I. Cirac.
\newblock ``Criticality, the area law, and the computational power of projected
  entangled pair states''.
\newblock \href{https://dx.doi.org/10.1103/physrevlett.96.220601}{Physical
  Review Letters{\bf 96}}~(2006).

\bibitem{Vidal2010}
Guifre Vidal.
\newblock ``Entanglement renormalization: an introduction''~(2010).
\newblock  url:~\url{https://arxiv.org/abs/0912.1651v2}.

\bibitem{BC17}
Jacob~C. Bridgeman and Christopher~T. Chubb.
\newblock ``Hand-waving and interpretive dance: An introductory course on
  tensor networks''.
\newblock \href{https://dx.doi.org/10.1088/1751-8121/aa6dc3}{Journal of Physics
  A: Mathematical and Theoretical {\bf 50}, 223001}~(2017).

\bibitem{Evenbly_2011}
G.~Evenbly and G.~Vidal.
\newblock ``Tensor network states and geometry''.
\newblock \href{https://dx.doi.org/10.1007/s10955-011-0237-4}{Journal of
  Statistical Physics {\bf 145}, 891918}~(2011).

\bibitem{Oru14}
Rom{\'a}n Or{\'u}s.
\newblock ``A practical introduction to tensor networks: {{Matrix}} product
  states and projected entangled pair states''.
\newblock \href{https://dx.doi.org/10.1016/j.aop.2014.06.013}{Annals of Physics
  {\bf 349}, 117--158}~(2014).

\bibitem{Oru19}
Rom{\'a}n Or{\'u}s.
\newblock ``Tensor networks for complex quantum systems''.
\newblock \href{https://dx.doi.org/10.1038/s42254-019-0086-7}{Nature Reviews
  Physics {\bf 1}, 538--550}~(2019).

\bibitem{oseQuant}
I.~V. Oseledets.
\newblock ``Approximation of $2^d \times 2^d$ matrices using tensor
  decomposition''.
\newblock \href{https://dx.doi.org/10.1137/090757861}{SIAM Journal on Matrix
  Analysis and Applications {\bf 31}, 2130--2145}~(2010).

\bibitem{Khoromskij2011OdlogNA}
Boris~N. Khoromskij.
\newblock ``{$O(d \log N)$}-quantics approximation of {$N$-d} tensors in
  high-dimensional numerical modeling''.
\newblock \href{https://dx.doi.org/10.1007/s00365-011-9131-1}{Constructive
  Approximation {\bf 34}, 257--280}~(2011).

\bibitem{LIN24}
Michael Lindsey.
\newblock ``Multiscale interpolative construction of quantized tensor
  trains''~(2024).
\newblock  url:~\url{https://arxiv.org/abs/2311.12554v3}.

\bibitem{Han_2018}
Zhao-Yu Han, Jun Wang, Heng Fan, Lei Wang, and Pan Zhang.
\newblock ``Unsupervised generative modeling using matrix product states''.
\newblock \href{https://dx.doi.org/10.1103/physrevx.8.031012}{Physical Review
  X{\bf 8}}~(2018).

\bibitem{HHL+22}
YoonHaeng Hur, Jeremy~G. Hoskins, Michael Lindsey, E.~M. Stoudenmire, and
  Yuehaw Khoo.
\newblock ``Generative modeling via tensor train sketching''.
\newblock \href{https://dx.doi.org/10.1016/j.acha.2023.101575}{Applied and
  Computational Harmonic Analysis {\bf 67}, 101575}~(2023).

\bibitem{novikov2015tensorizing}
Alexander Novikov, Dmitry Podoprikhin, Anton Osokin, and Dmitry Vetrov.
\newblock ``Tensorizing neural networks''.
\newblock In Proceedings of the 29th International Conference on Neural
  Information Processing Systems.
\newblock \href{https://dx.doi.org/10.5555/2969239.2969289}{Volume~1, pages
  442--450}.
\newblock Cambridge, MA~(2015). MIT Press.

\bibitem{garipov2016}
Timur Garipov, Dmitry Podoprikhin, Alexander Novikov, and Dmitry Vetrov.
\newblock ``Ultimate tensorization: compressing convolutional and {FC} layers
  alike''~(2016).
\newblock  url:~\url{https://arxiv.org/abs/1611.03214v1}.

\bibitem{yang2017}
Yinchong Yang, Denis Krompass, and Volker Tresp.
\newblock ``Tensor-train recurrent neural networks for video classification''.
\newblock In Proceedings of the 34th International Conference on Machine
  Learning.
\newblock \href{https://dx.doi.org/10.5555/3305890.3306083}{Pages 3891--3900}.
\newblock Sydney, NSW~(2017). JMLR.

\bibitem{Memmel2022PositionTN}
Eva Memmel, Clara Menzen, Jetze Schuurmans, Frederiek Wesel, and Kim Batselier.
\newblock ``Position: Tensor networks are a valuable asset for green {{AI}}''.
\newblock In Proceedings of the 41st International Conference on Machine
  Learning.
\newblock \href{https://dx.doi.org/10.5555/3692070.3693509}{Volume 235, pages
  35340--35353}.
\newblock Vienna, Austria~(2024).

\bibitem{tomut2024compactifai}
Andrei Tomut, Saeed~S. Jahromi, Sukhbinder Singh, Faysal Ishtiaq, Cesar Muoz,
  Prabdeep~Singh Bajaj, Ali Elborady, Gianni del Bimbo, Mehrazin Alizadeh,
  David Montero, Pablo Martin-Ramiro, Muhammad Ibrahim, Oussama~Tahiri Alaoui,
  John Malcolm, Samuel Mugel, and Roman Orus.
\newblock ``Compactif{AI}: {Extreme} compression of large language models using
  quantum-inspired tensor networks''~(2024).
\newblock  url:~\url{arXiv:2401.14109v2}.

\bibitem{BB17}
Jacob Biamonte and Ville Bergholm.
\newblock ``Tensor networks in a nutshell''~(2017).
\newblock  url:~\url{https://arxiv.org/abs/1708.00006v1}.

\bibitem{Crosswhite_2008}
Gregory~M. Crosswhite, A.~C. Doherty, and Guifr Vidal.
\newblock ``Applying matrix product operators to model systems with long-range
  interactions''.
\newblock \href{https://dx.doi.org/10.1103/physrevb.78.035116}{Physical Review
  B{\bf 78}}~(2008).

\bibitem{Hubig_2017}
C.~Hubig, I.~P. McCulloch, and U.~Schollwck.
\newblock ``Generic construction of efficient matrix product operators''.
\newblock \href{https://dx.doi.org/10.1103/physrevb.95.035129}{Physical Review
  B{\bf 95}}~(2017).

\bibitem{Paeckel_2019}
Sebastian Paeckel, Thomas Khler, Andreas Swoboda, Salvatore~R. Manmana, Ulrich
  Schollwck, and Claudius Hubig.
\newblock ``Time-evolution methods for matrix-product states''.
\newblock \href{https://dx.doi.org/10.1016/j.aop.2019.167998}{Annals of Physics
  {\bf 411}, 167998}~(2019).

\bibitem{vandamme2023}
Maarten~Van Damme, Jutho Haegeman, Ian McCulloch, and Laurens Vanderstraeten.
\newblock ``Efficient higher-order matrix product operators for time
  evolution''~(2023).
\newblock  url:~\url{https://arxiv.org/abs/2302.14181v1}.

\bibitem{CSW23}
Jielun Chen, E.M. Stoudenmire, and Steven~R. White.
\newblock ``Quantum {{Fourier Transform Has Small Entanglement}}''.
\newblock \href{https://dx.doi.org/10.1103/PRXQuantum.4.040318}{PRX Quantum
  {\bf 4}, 040318}~(2023).

\bibitem{chen2024}
Jielun Chen and Michael Lindsey.
\newblock ``Direct interpolative construction of the discrete fourier transform
  as a matrix product operator''~(2024).
\newblock  url:~\url{https://arxiv.org/abs/2404.03182v1}.

\bibitem{YA20}
Mingru Yang and Steven~R. White.
\newblock ``Time-dependent variational principle with ancillary {Krylov}
  subspace''.
\newblock \href{https://dx.doi.org/10.1103/physrevb.102.094315}{Physical Review
  B{\bf 102}}~(2020).

\bibitem{VC04}
F.~Verstraete and J.~I. Cirac.
\newblock ``Renormalization algorithms for {Quantum-Many Body Systems} in two
  and higher dimensions''~(2004).
\newblock  url:~\url{https://arxiv.org/abs/cond-mat/0407066v1}.

\bibitem{JOR08}
J.~Jordan, R.~Or\'us, G.~Vidal, F.~Verstraete, and J.~I. Cirac.
\newblock ``Classical simulation of infinite-size quantum lattice systems in
  two spatial dimensions''.
\newblock \href{https://dx.doi.org/10.1103/PhysRevLett.101.250602}{Phys. Rev.
  Lett. {\bf 101}, 250602}~(2008).

\bibitem{GTG24}
Paula {Garc{\'i}a-Molina}, Luca Tagliacozzo, and Juan~Jos{\'e}
  {Garc{\'i}a-Ripoll}.
\newblock ``Global optimization of {{MPS}} in quantum-inspired numerical
  analysis''~(2024).
\newblock  url:~\url{https://arxiv.org/abs/2303.09430v2}.

\bibitem{gray2024}
Johnnie Gray and Garnet Kin-Lic Chan.
\newblock ``Hyperoptimized approximate contraction of tensor networks with
  arbitrary geometry''.
\newblock
  \href{https://dx.doi.org/https://doi.org/10.1103/PRXQuantum.6.010312}{Physical
  Review X {\bf 14}, 011009}~(2024).

\bibitem{JCSH24}
Jiaqing Jiang, Jielun Chen, Norbert Schuch, and Dominik Hangleiter.
\newblock ``Positive bias makes tensor-network contraction tractable''.
\newblock \href{https://dx.doi.org/10.48550/arXiv.2410.05414}{Foundation of
  Computer Science 2025, to appear}~(2024).

\bibitem{chen2025}
Jielun Chen, Jiaqing Jiang, Dominik Hangleiter, and Norbert Schuch.
\newblock ``Sign problem in tensor-network contraction''.
\newblock
  \href{https://dx.doi.org/https://doi.org/10.1103/PRXQuantum.6.010312}{PRX
  Quantum {\bf 6}, 010312}~(2025).

\bibitem{Ose11}
I.~V. Oseledets.
\newblock ``Tensor-train decomposition''.
\newblock \href{https://dx.doi.org/10.1137/090752286}{SIAM Journal on
  Scientific Computing {\bf 33}, 2295--2317}~(2011).

\bibitem{HMT11}
Nathan Halko, Per-Gunnar Martinsson, and Joel~A. Tropp.
\newblock ``Finding structure with randomness: {{Probabilistic}} algorithms for
  constructing approximate matrix decompositions''.
\newblock \href{https://dx.doi.org/10.1137/090771806}{SIAM Review {\bf 53},
  217--288}~(2011).

\bibitem{Woo14}
David~P. Woodruff.
\newblock ``Sketching as a tool for numerical linear algebra''.
\newblock \href{https://dx.doi.org/10.1561/0400000060}{Foundations and Trends
  in Theoretical Computer Science {\bf 10}, 1--157}~(2014).

\bibitem{TW23a}
Joel~A. Tropp and Robert~J. Webber.
\newblock ``Randomized algorithms for low-rank matrix approximation:
  {{Design}}, analysis, and applications''~(2023).
\newblock  url:~\url{https://arxiv.org/abs/2306.12418v1}.

\bibitem{MDM+23}
Riley Murray, James Demmel, Michael~W. Mahoney, N.~Benjamin Erichson, Maksim
  Melnichenko, Osman~Asif Malik, Laura Grigori, Piotr Luszczek, Micha{\l}
  Derezi{\'n}ski, Miles~E. Lopes, Tianyu Liang, Hengrui Luo, and Jack Dongarra.
\newblock ``Randomized numerical linear algebra : {{A}} perspective on the
  field with an eye to software''~(2023).
\newblock  url:~\url{https://arxiv.org/abs/2302.11474v2}.

\bibitem{TYUC17a}
Joel~A. Tropp, Alp Yurtsever, Madeleine Udell, and Volkan Cevher.
\newblock ``Practical sketching algorithms for low-rank matrix approximation''.
\newblock \href{https://dx.doi.org/10.1137/17M1111590}{SIAM Journal on Matrix
  Analysis and Applications {\bf 38}, 1454--1485}~(2017).

\bibitem{MT20a}
Per-Gunnar Martinsson and Joel~A. Tropp.
\newblock ``Randomized numerical linear algebra: {{Foundations}} and
  algorithms''.
\newblock \href{https://dx.doi.org/10.1017/S0962492920000021}{Acta Numerica
  {\bf 29}, 403--572}~(2020).

\bibitem{CEMT25a}
Chris Cama{\~n}o, Ethan~N. Epperly, Raphael~A. Meyer, and Joel~A. Tropp.
\newblock ``Faster linear algebra algorithms with structured random
  matrices''~(2025).
\newblock  \href{http://arxiv.org/abs/2508.21189}{arXiv:2508.21189}.

\bibitem{Gu15}
Ming Gu.
\newblock ``Subspace {{Iteration Randomization}} and {{Singular Value
  Problems}}''.
\newblock \href{https://dx.doi.org/10.1137/130938700}{SIAM Journal on
  Scientific Computing {\bf 37}, A1139--A1173}~(2015).

\bibitem{MM15}
Cameron Musco and Christopher Musco.
\newblock ``Randomized block {{Krylov}} methods for stronger and faster
  approximate singular value decomposition''.
\newblock In Proceedings of the 28th {{International Conference}} on {{Neural
  Information Processing Systems}}.
\newblock Pages 1396--1404.
\newblock MIT Press~(2015).
\newblock  url:~\url{https://dl.acm.org/doi/10.5555/2969239.2969395}.

\bibitem{BHOT24a}
Nicolas Boull{\'e}, Diana Halikias, Samuel~E. Otto, and Alex Townsend.
\newblock ``Operator learning without the adjoint''.
\newblock Journal of Machine Learning Research {\bf 25}, 1--54~(2024).
\newblock  url:~\url{https://jmlr.org/papers/v25/24-0162.html}.

\bibitem{TRP15}
D.~Tamascelli, R.~Rosenbach, and M.~B. Plenio.
\newblock ``Improved scaling of time-evolving block-decimation algorithm
  through reduced-rank randomized singular value decomposition''.
\newblock \href{https://dx.doi.org/10.1103/PhysRevE.91.063306}{Physical Review
  E {\bf 91}, 063306}~(2015).

\bibitem{MIZK18}
Satoshi Morita, Ryo Igarashi, Hui-Hai Zhao, and Naoki Kawashima.
\newblock ``Tensor renormalization group with randomized singular value
  decomposition''.
\newblock \href{https://dx.doi.org/10.1103/PhysRevE.97.033310}{Physical Review
  E {\bf 97}, 033310}~(2018).

\bibitem{KTK+18}
Lucas Kohn, Ferdinand Tschirsich, Maximilian Keck, Martin~B. Plenio, Dario
  Tamascelli, and Simone Montangero.
\newblock ``Probabilistic low-rank factorization accelerates tensor network
  simulations of critical quantum many-body ground states''.
\newblock \href{https://dx.doi.org/10.1103/PhysRevE.97.013301}{Physical Review
  E {\bf 97}, 013301}~(2018).

\bibitem{MO24}
Ian~P. McCulloch and Jesse~J. Osborne.
\newblock ``Comment on "{{Controlled Bond Expansion}} for {{Density Matrix
  Renormalization Group Ground State Search}} at {{Single-Site Costs}}"
  ({{Extended Version}})''~(2024).
\newblock  \href{http://arxiv.org/abs/2403.00562}{arXiv:2403.00562}.

\bibitem{KB09}
Tamara~G. Kolda and Brett~W. Bader.
\newblock ``Tensor decompositions and applications''.
\newblock \href{https://dx.doi.org/10.1137/07070111X}{SIAM Review {\bf 51},
  455--500}~(2009).

\bibitem{Ahl25}
Thomas~D. Ahle.
\newblock ``The tensor cookbook''.
\newblock Manuscript in progress, Version: February 2025. ~(2025).
\newblock  url:~\url{https://tensorcookbook.com}.

\bibitem{ABC+21}
Hussam Al~Daas, Grey Ballard, Paul Cazeaux, Eric Hallman, Agnieszka Mi{\k
  e}dlar, Mirjeta Pasha, Tim~W. Reid, and Arvind~K. Saibaba.
\newblock ``Randomized algorithms for rounding in the tensor-train format''.
\newblock \href{https://dx.doi.org/10.1137/21M1451191}{SIAM Journal on
  Scientific Computing {\bf 45}, A74--A95}~(2023).

\bibitem{KVV23}
Daniel Kressner, Bart Vandereycken, and Rik Voorhaar.
\newblock ``Streaming tensor train approximation''.
\newblock \href{https://dx.doi.org/10.1137/22M1515045}{SIAM Journal on
  Scientific Computing {\bf 45}, A2610--A2631}~(2023).

\bibitem{CK23}
Yian Chen and Yuehaw Khoo.
\newblock ``Combining {{Monte Carlo}} and {{Tensor-network Methods}} for
  {{Partial Differential Equations}} via {{Sketching}}''~(2023).
\newblock  \href{http://arxiv.org/abs/2305.17884}{arXiv:2305.17884}.

\bibitem{YZK25}
Ziang Yu, Shiwei Zhang, and Yuehaw Khoo.
\newblock ``Re-anchoring {{Quantum Monte Carlo}} with {{Tensor-Train
  Sketching}}''~(2025) \href{http://arxiv.org/abs/2411.07194}{math:2411.07194}.

\bibitem{SPV10}
Sukhwinder Singh, Robert N.~C. Pfeifer, and Guifr{\'e} Vidal.
\newblock ``Tensor network decompositions in the presence of a global
  symmetry''.
\newblock \href{https://dx.doi.org/10.1103/PhysRevA.82.050301}{Physical Review
  A {\bf 82}, 050301}~(2010).

\bibitem{SPV11}
Sukhwinder Singh, Robert N.~C. Pfeifer, and Guifre Vidal.
\newblock ``Tensor network states and algorithms in the presence of a global
  {{U}}(1) symmetry''.
\newblock \href{https://dx.doi.org/10.1103/PhysRevB.83.115125}{Physical Review
  B {\bf 83}, 115125}~(2011).

\bibitem{Sch11}
Ulrich Schollw{\"o}ck.
\newblock ``The density-matrix renormalization group in the age of matrix
  product states''.
\newblock \href{https://dx.doi.org/10.1016/j.aop.2010.09.012}{Annals of Physics
  {\bf 326}, 96--192}~(2011).

\bibitem{SW10}
E~M Stoudenmire and Steven~R White.
\newblock ``Minimally entangled typical thermal state algorithms''.
\newblock \href{https://dx.doi.org/10.1088/1367-2630/12/5/055026}{New Journal
  of Physics {\bf 12}, 055026}~(2010).

\bibitem{white2005density}
Steven~R White.
\newblock ``Density matrix renormalization group algorithms with a single
  center site''.
\newblock
  \href{https://dx.doi.org/https://doi.org/10.1103/PhysRevB.72.180403}{Physical
  Review B—Condensed Matter and Materials Physics {\bf 72}, 180403}~(2005).

\bibitem{hubig2015strictly}
Claudius Hubig, Ian~P McCulloch, Ulrich Schollw{\"o}ck, and F~Alexander Wolf.
\newblock ``Strictly single-site dmrg algorithm with subspace expansion''.
\newblock
  \href{https://dx.doi.org/https://doi.org/10.1103/PhysRevB.91.155115}{Physical
  Review B {\bf 91}, 155115}~(2015).

\bibitem{HVH21}
Markus Hauru, Maarten Van~Damme, and Jutho Haegeman.
\newblock ``Riemannian optimization of isometric tensor networks''.
\newblock \href{https://dx.doi.org/10.21468/SciPostPhys.10.2.040}{SciPost
  Physics {\bf 10}, 040}~(2021).

\bibitem{ma2024}
Linjian Ma, Matthew Fishman, Edwin~Miles Stoudenmire, and Edgar Solomonik.
\newblock ``Approximate contraction of arbitrary tensor networks with a
  flexible and efficient density matrix algorithm''.
\newblock \href{https://dx.doi.org/10.22331/q-2024-12-27-1580}{Quantum {\bf 8},
  1580}~(2024).

\bibitem{Tenc}
``{{MPO-MPS}} multiplication: Density matrix algorithm''.
\newblock Web page:
  \url{https://www.tensornetwork.org/mps/algorithms/denmat_mpo_mps/}.

\bibitem{MFSS24a}
Linjian Ma, Matthew Fishman, Edwin~Miles Stoudenmire, and Edgar Solomonik.
\newblock ``Approximate contraction of arbitrary tensor networks with a
  flexible and efficient density matrix algorithm''.
\newblock \href{https://dx.doi.org/10.22331/q-2024-12-27-1580}{Quantum {\bf 8},
  1580}~(2024).

\bibitem{HA11}
Jutho Haegeman, J.~Ignacio Cirac, Tobias~J. Osborne, Iztok Piorn, Henri
  Verschelde, and Frank Verstraete.
\newblock ``Time-dependent variational principle for quantum lattices''.
\newblock \href{https://dx.doi.org/10.1103/physrevlett.107.070601}{Physical
  Review Letters{\bf 107}}~(2011).

\bibitem{Haeg16}
Jutho Haegeman, Christian Lubich, Ivan Oseledets, Bart Vandereycken, and Frank
  Verstraete.
\newblock ``Unifying time evolution and optimization with matrix product
  states''.
\newblock \href{https://dx.doi.org/10.1103/physrevb.94.165116}{Physical Review
  B{\bf 94}}~(2016).

\bibitem{PMCV10}
B~Pirvu, V~Murg, J~I Cirac, and F~Verstraete.
\newblock ``Matrix product operator representations''.
\newblock \href{https://dx.doi.org/10.1088/1367-2630/12/2/025012}{New Journal
  of Physics {\bf 12}, 025012}~(2010).

\bibitem{Loj64}
Stanislaw Lojasiewicz.
\newblock ``Triangulation of semi-analytic sets''.
\newblock Annali della Scuola Normale Superiore di Pisa-Classe di Scienze {\bf
  18}, 449--474~(1964).
\newblock  url:~\url{http://www.numdam.org/item/ASNSP_1964_3_18_4_449_0/}.

\bibitem{1920527}
Nate Eldredge.
\newblock ``The {Lebesgue} measure of zero set of a polynomial function is
  zero''.
\newblock Math StackExchange post~(2016).
\newblock  url:~\url{https://math.stackexchange.com/q/1920527}.

\bibitem{ET24}
Ethan~N. Epperly and Joel~A. Tropp.
\newblock ``Efficient {{Error}} and {{Variance Estimation}} for {{Randomized
  Matrix Computations}}''.
\newblock \href{https://dx.doi.org/10.1137/23M1558537}{SIAM Journal on
  Scientific Computing {\bf 46}, A508--A528}~(2024).

\end{thebibliography}

\appendix

\section{\texorpdfstring{Proof of \cref{thm:khatri-rao}}{Proof of Theorem 2}} \label{app:khatri-rao-exact}

We may assume without loss of generality that $p = \rank A$.
The column space of the matrix $Y$ is a subset of the column space of $A$, and randomized \QB approximation is exact precisely when the column space of these two matrices is the same.
Thus, randomized \QB approximation is exact if and only if $\rank Y = p$, which occurs if and only if $\det(Y^\conj Y) \ne 0$, so that $Y^\dagger Y$ is nonsingular.

Since $Y = A\Omega$ and $\Omega = \Omega^{(1)}\odot \cdots \odot \Omega^{(n)}$, $p(\Omega^{(1)},\ldots,\Omega^{(n)}) = \det(Y^\dagger Y)$ is a polynomial in the entries of $\Omega^{(1)},\ldots,\Omega^{(n)}$.
The set of zeros of a nonzero multivariate polynomial is a null set.
(See \cite{Loj64} for a sophisticated generalization of this fact; an elementary argument is provided in \cite{1920527}.)
Therefore, provided $p(\cdot)$ is not identitically zero as a polynomial, $p(\Omega^{(1)},\ldots,\Omega^{(n)}) = \det(Y^\dagger Y) \ne 0$ with probability one.

To complete the proof, we must furnish matrices $\Omega^{(1)},\ldots,\Omega^{(n)}$ for which
\begin{equation*}
    p(\Omega^{(1)},\ldots,\Omega^{(n)}) \ne 0.   
\end{equation*}
Since $A$ is rank-$p$, $A$ contains $p$ linearly independent columns $A(:,i_1),\ldots,A(:,i_p)$.
The matrix $\Omega = \begin{bmatrix} e^{(i_1)} & \cdots & e^{(i_p)} \end{bmatrix}$ can be realized as a Khatri--Rao product $\Omega = \Omega^{(1)}\odot \cdots \odot \Omega^{(n)}$, where $e^{(i)}$ denotes the standard basis vector with entries $e^{(i)}_j = \delta_{ij}$.
Thus, $p(\Omega^{(1)},\ldots,\Omega^{(n)}) = \det(A(:,\{i_1,\ldots,i_p\})^\conj A(:,\{i_1,\ldots,i_p\})) > 0$, completing the proof. \hfill $\blacksquare$

\section{\texorpdfstring{Proof of \cref{thm:src-exact}}{Proof of Theorem 3}} \label{app:src-exact-proof}

At first glance, it might seem that \cref{thm:src-exact} is a trivial consequence of \cref{thm:khatri-rao}, which guarantees rank-$r$ \QB decomposition is exact with probability one when applied to a matrix of rank at most $r$.
The weakness in this argument is that SRC uses a common set of matrices $\Omega^{(1)},\ldots,\Omega^{(n-1)}$ throughout the algorithm, and the matrix which we applied \QB decomposition at each step depends on previous steps of the algorithm.
The dependencies between the steps of the algorithm could be a potential source of trouble.

Fortunately, this potential concern is not a worry.
At each step of the SRC algorithm, we apply \QB decomposition to a matrix representation $B^{(\ell)}$ of the first $\ell$ sites of the MPO--MPS product
\begin{equation*}
    B^{(\ell)} = \begin{gathered}
    \includegraphics[scale=0.35]{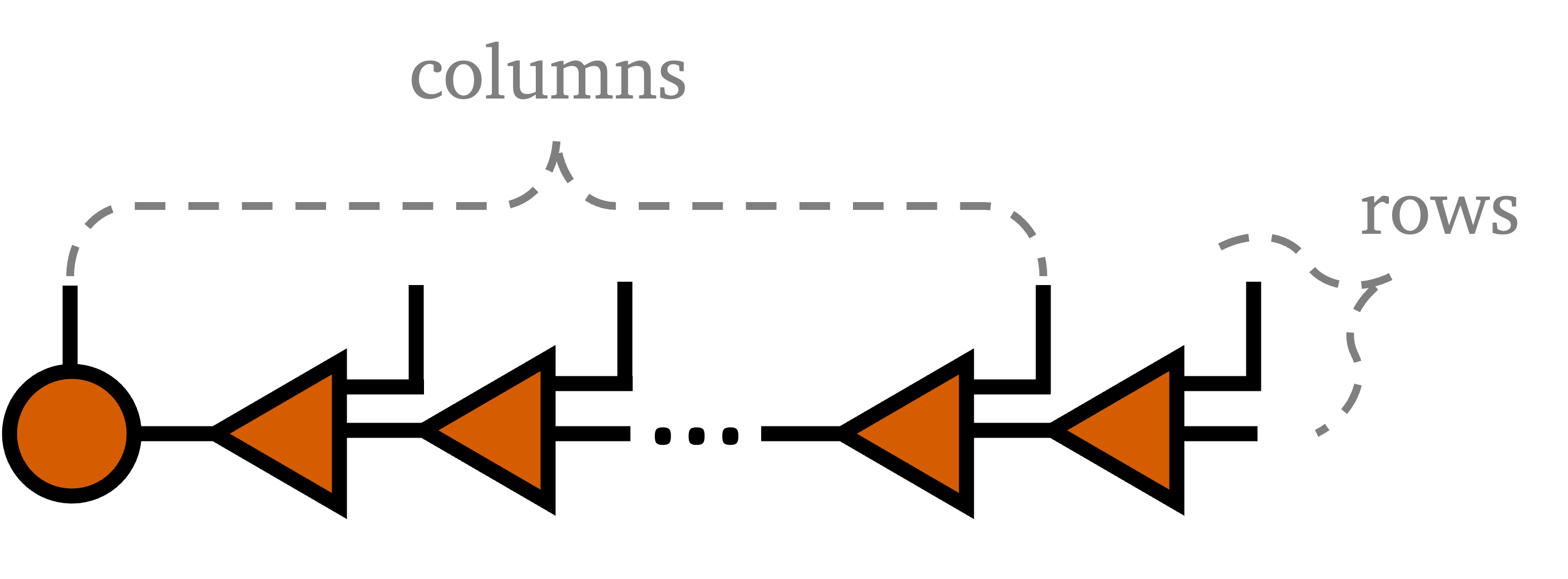}
    \end{gathered}\, \in \complex^{\overline{\chi} d \times d^{\ell-1}}.
\end{equation*}
The right canonical form of an MPS is unique up to unitary transformations along the bond indices.
As such, the $B^{(\ell)}$ matrix that we apply \QB decomposition at each step of the algorithm is fixed independent of the randomness in $\Omega^{(1)},\ldots,\Omega^{(n-1)}$, up to a potential unitary transformation on the left.
The randomized \QB algorithm is unitarily covariant (i.e., if we apply the algorithm to both $A$ and a unitary transformation $U\cdot A$ with the same random test matrix, the outputs will be $\hat{A}$ and $U\cdot \hat{A}$).
Therefore, the randomized \QB decompositions are exact at each step of the algorithm with probability one, so SRC outputs an exact MPS representation of $\ket{\eta} = H\ket{\psi}$. \hfill $\blacksquare$

\section{Implementing SRC with a tolerance} \label{sec:approx}

In \cref{sec:alg}, we presented a version of SRC where the output bond dimension $\overline{\chi}$ is provided as an input to the algorithm. 
This section will describe an adaptive implementation of SRC that chooses the bond dimension adaptively to meet a tolerance.
\Cref{sec:error-estimation} discusses error estimation, \cref{sec:adaptivity} uses the error estimate to implement SRC adaptively, and \cref{app:qr-updating} optimizes the adaptive implementation using QR factorization updating algorithms.

\subsection{Error estimation} \label{sec:error-estimation}

Each step of the randomized MPO--MPS multiplication algorithm involves a randomized \QB approximation of the form
\begin{equation*}
    B^{(j)} \approx \eta^{(j)} (\eta^{(j)})^\conj B^{(j)}.
\end{equation*}
Even the first step of the algorithm has this form, with $B^{(n)} \coloneqq H\ket{\psi}$.
When $H\ket{\psi}$ is exactly representable as an MPS of the specified bond dimension $\overline{\chi}$, this randomized \QB approximation is exact with probability one, but this is typically not the case in practice.
Thus, in order to choose the appropriate bond dimension $\overline{\chi}$, it is important to understand the error of the approximation
\begin{equation*}
    \norm{B^{(j)} - \eta^{(j)} (\eta^{(j)})^\conj B^{(j)}}_{\rm F}.
\end{equation*}
For this purpose, we will employ the leave-one-out error estimator \cite{ET24}.

\subsubsection{Leave-one-out error estimation: Matrix case}

Let us first introduce the leave-one-out error estimator in the matrix context.
To compute a \QB approximation of a matrix $A\in\complex^{M\times N}$, we draw a Gaussian or Khatri--Rao random matrix $\Omega \in \complex^{N\times p}$, apply $A$ to $\Omega$, and compute a \QR factorization
\begin{equation*}
    A\Omega = QR.
\end{equation*}
The \QB approximation of rank $p$ then takes the form
\begin{equation} \label{eq:loo-qr}
    \hat{A} = Q Q^\conj A.
\end{equation}
The root mean-square error of the approximation is
\begin{equation*}
    \mathrm{Err}_p \coloneqq \left(\expect \left[ \norm{\smash{A - \hat{A}}}_{\rm F}^2 \right]\right)^{1/2}.
\end{equation*}
We write a subscript $p$ to indicate the rank of the approximation $p = \rank\hat{A}$.

Surprisingly, the \QR factorization \cref{eq:loo-qr} already contains enough information to compute an estimate of the error $\mathrm{Err}_p$.
First, compute the inverse-adjoint of the matrix $R$:
\begin{equation*}
    G \coloneqq R^{-\conj} = \begin{bmatrix} g_1 & \cdots & g_p \end{bmatrix}.
\end{equation*}
The \emph{leave-one-out estimate for the root-mean-square error} is
\begin{equation*}
    \smash{\hat{\mathrm{Err}}} \coloneqq \left(\frac{1}{p} \sum_{i=1}^p \norm{g_i}^{-2}\right)^{1/2}.
\end{equation*}
For an explanation of the name ``leave-one-out estimate'', intuition for why $\smash{\hat{\mathrm{Err}}}\approx \mathrm{Err}_p$, and empirical evaluation of this estimator for matrix \QB approximation, see \cite{ET24}.
The leave-one-out mean-square error estimate satisfies the following guarantee:

\begin{theorem}[Leave-one-out error estimation] \label{thm:loo}
    Let $\Omega\in\complex^{N\times p}$ be a random matrix with independent columns $\omega_1,\ldots,\omega_p$ satisfying the isotropy condition $\expect[\omega_i^{\vphantom{\conj}}\omega_i^\conj] = \Id$, such as a standard Gaussian matrix or Khatri--Rao product of standard Gaussian matrices. 
    Then, with the prevailing notation,
    \begin{equation*}
        \expect\left[ \smash{\hat{\mathrm{Err}}}^2 \right] = \mathrm{Err}_{p-1}^2 \ge \mathrm{Err}_p^2.
    \end{equation*}
    That is, $\smash{\hat{\mathrm{Err}}}^2$ is a statistically unbiased estimate for the mean-square error $\mathrm{Err}_{p-1}^2$ of the rank-$(p-1)$ randomized \QB approximation.
    Consequently, $\smash{\hat{\mathrm{Err}}}^2$ is an overestimate for $\mathrm{Err}_p^2$, on average.
\end{theorem}

The leave-one-out error estimate is useful because it provides a quick estimate of the error $\norm{\smash{A - \hat{A}}}_{\rm F}$ of randomized \QB approximation.
Indeed, computing this estimate just requires inverting the small $p\times p$ matrix $R^\dagger$ and computing the norms of its columns.
The leave-one-out error estimate only costs a measly $\order(p^3)$ operations to compute, dwarfed by the $\Omega(Mp^2)$ cost of computing and orthogonalizing $A\Omega$.
The estimator is slightly conservative, typically overestimating $\mathrm{MSE}_p$ by a small amount on average.

\subsubsection{Leave-one-out error estimation: SRC algorithm}

The leave-one-out error estimate can directly be used in the randomized MPO--MPS multiplication algorithm to estimate the error at each step of the algorithm.
Simply compute the inverse-adjoint of $R^{(j)}$ and compute the leave-one-out mean-square error estimate:
\begin{equation} \label{eq:err-est}
    G^{(j)} = \left(R^{(j)}\right)^{-\conj} = \begin{bmatrix}
        g_1^{(j)} & \cdots & g_{\overline{\chi}}^{(j)}
    \end{bmatrix}, \quad \smash{\hat{\mathrm{Err}}}^{(j)} \coloneqq \left(\frac{1}{\overline{\chi}} \sum_{i=1}^{\overline{\chi}} \norm{g_i^{(j)}}^{-2}\right)^{1/2}.
\end{equation}
This serves as a drop-in estimate for the error incurred in the current step of the algorithm:
\begin{equation*}
    \smash{\hat{\mathrm{Err}}}^{(j)}\approx \norm{B^{(j)} - \eta^{(j)}\cdot B^{(j-1)}}_{\rm F}.
\end{equation*}

\subsubsection{Norm estimation}

In order to implement MPO--MPS truncation adaptively, we will also need access to an estimate of the \emph{norm} $\norm{B^{(j)}}_{\rm F}$.
This can be achieved by the use of another stochastic estimate
\begin{equation} \label{eq:norm_est}
    \hat{\mathrm{Norm}} \coloneqq \frac{1}{\sqrt{\overline{\chi}}} \norm{Y^{(j)}}_{\rm F} = \frac{1}{\sqrt{\overline{\chi}}} \norm{R^{(j)}}_{\rm F}.
\end{equation}
This is an instance of the Girard--Hutchinson norm estimator, closely related to the Girard--Hutchinson trace estimator \cite[\S4]{MT20a}.
In particular, we have the following standard guarantee \cite[\S4.8]{MT20a}:

\begin{proposition}[Norm estimation]
    Under the hypotheses of \cref{thm:loo}, the squared norm estimate $\smash{\hat{\mathrm{Norm}}}^2$ is a statistically unbiased estimate for $\norm{\smash{B}^{(j)}}^2_{\rm F}$.
    That is,
    \begin{equation*}
        \expect\left[\hat{\mathrm{Norm}}^2\right] = \norm{\smash{B}^{(j)}}^2_{\rm F}.
    \end{equation*}
    Moreover, the variance of the estimator $\hat{\mathrm{Norm}}^2$ decays at a rate $\order(1/\overline{\chi})$.
\end{proposition}
\subsection{Adaptive determination of bond dimension} \label{sec:adaptivity}

Using the leave-one-out error estimate \cref{eq:err-est} and norm estimate \cref{eq:norm_est}, we can adaptively choose the dimension $\overline{\chi}$ adaptively for each bond in the output MPS using a tolerance.
Let $\tau_{\rm abs}$ and $\tau_{\rm rel}$ be absolute and relative tolerances, and let $\overline{\chi}_0$ and $\Delta_\chi$ be a user-specified minimum bond dimension and increment.
Begin each step $j$ of SRC using the minimum bond dimension $\overline{\chi}_0$, obtaining a $j$th site tensor $\eta^{(j)}$.
Then, do the following steps:
\begin{enumerate}
    \item \textbf{Error estimation.} Compute the error estimate $\smash{\hat{\mathrm{Err}}}^{(j)}$ and the norm estimate $\smash{\hat{\mathrm{Norm}}}$.
    \item \textbf{Check tolerance.} Check whether $\smash{\hat{\mathrm{Err}}}^{(j)} \le \tau_{\rm abs} + \tau_{\rm rel} \cdot \smash{\hat{\mathrm{Norm}}}$. 
    If yes, then exist this loop and continue the algorithm; otherwise, proceed to step 3.
    \item \textbf{Increase bond dimension.} Increase $\overline{\chi}$ by the increment $\Delta_\chi$.
    \item \textbf{Update random matrix and intermediate tensors.} If $\overline{\chi}$ is larger than the number of columns in $\Omega^{(1:j-1)}$, then append columns to $\Omega^{(1:j-1)}$ until it has $\overline{\chi}$ in total. 
    Update the intermediate tensors $C^{(1)},\ldots,C^{(j-1)}$.
    \item \textbf{Update randomized \QB approximation.} Compute the additional columns $\Delta_\chi$ of $Y^{(j)}$ using the (possibly updated) intermediate tensor $C^{(j-1)}$.
    Recompute (or update, see \cref{app:qr-updating}) the \QR factorization  of $Y^{(j)}$.
    Go to step 1.
\end{enumerate}

When implemented using \QR updating rather recomputation from scratch, as described in the next section, this adaptive version of the MPO--MPS multiplication algorithm has the same time complexity as reported in \cref{sec:pseudocode-complexity}.
In our experiments in \cref{sec:adaptive_experiments}, we use parameters $\overline{\chi}_0 = 2$, $\Delta_\chi = 3$, and use only a relative tolerance (setting $\tau_{\rm abs} = 0$).
To implement with oversampling, we set the relative tolerance to be $0.1$ times the requested tolerance and run a final truncation with the requested tolerance.

\subsection{Final optimization: Updating the \QR factorization} \label{app:qr-updating}

In the adaptive MPO--MPS multiplication algorithm (\cref{sec:adaptivity}), we continue to accumulate ``columns'' in the 2- or 3-tensor $Y^{(j)}$ until a tolerance is satisfied.
Recomputing a \QR factorization each time we do this is computationally expensive, and we can make the algorithm more efficient by instead updating the \QR factorization to include new columns.
This material is standard but perhaps not well-known, so we provide a brief discussion.

The simplest---but not most numerically stable---way of updating the \QR decomposition is through the block Gram--Schmidt process.
To simplify notation, drop the superscript $Y = Y^{(j)}$ and let $Y'$ be a set of new columns being appended to $Y = QR$.
To update the \QR factorization, first orthogonalize the new columns $Y'$ against $Q$:
\begin{equation*}
    Z \coloneqq Y - QR', \quad R' = Q^\conj Y'.
\end{equation*}
Then, compute a \QR factorization $Z = Q'R''$.
This results in a \QR factorization of the full matrix $Y$ with the appended columns
\begin{equation*}
    \onebytwo{Y}{Y'} = \onebytwo{Q}{Q'} \twobytwo{R}{R'}{0}{R''}.
\end{equation*}
We have suceeded at updating the \QR factorization with the addition of new columns.
A similar procedure based on Householder \QR factorization is more numerically stable and is preferable.
We use the Householder-based procedure in our code.

Another matrix that must be updated to implement the adaptive MPO--MPS multiplication algorithm efficiently is the matrix $G$ used in error estimation \cref{eq:err-est}.
Again using our superscript-less notation, we begin with $G = R^{-\conj}$ and with to update $G$ to 
\begin{equation*}
    G' = \twobytwo{R}{R'}{0}{R''}^{-\conj}.
\end{equation*}
To do so, invoke the standard formula for the inverse of a block two-by-two matrix, obtaining
\begin{equation*}
    G' = \twobytwo{R^{-\conj}}{0}{-(R'')^{-\conj}(R')^\conj R^{-\conj}}{(R'')^{-\conj}} = \twobytwo{G}{0}{-(R'')^{-\conj}(R')^\conj G}{(R'')^{-\conj}}.
\end{equation*}
We recognize the original $G$ matrix in the top left corner of this matrix.
To finish the formula for $G'$, we just need to compute $(R'')^\dagger$ and form the matrix triple product $-((R'')^{-\conj}(R')^\conj) G$.
Since $R''$ is a small $\Delta_\chi \times \Delta_\chi$ matrix, this is efficient to do.

\section{Full operation counts}

Full operation counts for the methods in terms of all five parameters $n$, $d$, $\chi$, $D$, and $\overline{\chi}$ are presented in \cref{tab:Complete_operation_counts}.
For simplified operation counts, see \cref{tab:simplified_operation_counts}.

\begin{table*}[t]
  \centering
  \caption{Operation counts for methods to apply an MPO (bond dimension $D$) to an MPS (bond dimension $\chi$), both of $n$ sites and physical dimension $d$, and truncate the output to bond dimension $\overline{\chi}$.} 
    \label{tab:Complete_operation_counts} 
  \begin{tabular}{ll}\toprule  
    Method  & Operation Count \\ \midrule
    Contract-then-compress \\
    \quad $\bullet$ Basic   & $\order(n \cdot dD^2\chi^2[D\chi + d])$ \\
    \quad $\bullet$ Randomized & $\order(n \cdot dD^2\chi^2[d+\overline{\chi}])$ \\
    Density Matrix  & $\order(n \cdot [dD\chi (D\chi^2 + dD^2\chi + D\chi \overline{\chi} + d\overline{\chi}^2) + d^3\overline{\chi}^3])$ \ENE{Confirmed}\\
    Fitting  & $\order\left(n\cdot[ dD\chi\overline{\chi}^2+d^2D^2\chi^2]\right)$ per iteration\\
    Zip-up   & $\order(n \cdot d^2D\chi\overline{\chi} [D + \overline{\chi}])$ \ENE{Confirmed} \\
    SRC & $\order(n\cdot dD\chi \overline{\chi}(\chi+\overline{\chi}+dD))$ \ENE{Confirmed} 
    \\
    \bottomrule
  \end{tabular}
\end{table*} 

\end{document}